\documentclass{aa}

\usepackage{txfonts}
\RequirePackage{textcomp}  

\usepackage{amsmath}
\usepackage{mathtools}

\DeclareFontFamily{U}{mathx}{\hyphenchar\font45}
\DeclareFontShape{U}{mathx}{m}{n}{<-> mathx10}{}
\DeclareSymbolFont{mathx}{U}{mathx}{m}{n}
\DeclareMathAccent{\widebar}{0}{mathx}{"73}

\usepackage{emmt-unicode}

\usepackage{xspace}

\newcommand*{\etc}{\emph{etc.}\xspace}
\newcommand*{\eg}{\emph{e.g.}\xspace}
\newcommand*{\ie}{\emph{i.e.}\xspace}
\newcommand*{\cf}{\emph{cf.}\xspace}

\usepackage{xcolor}
\usepackage[unicode,colorlinks=true,allcolors=teal]{hyperref}

\usepackage{graphicx}

\usepackage[boxed,noline]{algorithm2e}
\DontPrintSemicolon
\SetKw{And}{and}
\SetKw{Or}{or}
\SetKw{Break}{break}
\SetKw{Not}{not}

\SetKwComment{Comment}{\ensuremath{\vartriangleleft} }{}

\SetCommentSty{CommentFormat}

\usepackage{subcaption}         
\usepackage{lscape}             
\usepackage{placeins}           

\usepackage{array}
\newcommand*{\LargeRowHeight}{\setlength{\extrarowheight}{5pt}}
\newcommand*{\NormalRowHeight}{\setlength{\extrarowheight}{0.5pt}}

\DeclarePairedDelimiterX{\Paren}[1]{(}{)}{#1}
\DeclarePairedDelimiterX{\Brace}[1]{\{}{\}}{#1}
\DeclarePairedDelimiterX{\Brack}[1]{[}{]}{#1}
\DeclarePairedDelimiterX{\Abs}[1]{|}{|}{#1}
\DeclarePairedDelimiterX{\Norm}[1]{\lVert}{\rVert}{#1}
\DeclarePairedDelimiterX{\Avg}[1]{\langle}{\rangle}{#1}
\DeclarePairedDelimiterX{\Round}[1]{\lfloor}{\rceil}{#1}
\DeclarePairedDelimiterX{\Floor}[1]{\lfloor}{\rfloor}{#1}
\DeclarePairedDelimiterX{\Ceil}[1]{\lceil}{\rceil}{#1}
\DeclarePairedDelimiterX{\Inner}[1]{\langle}{\rangle}{#1}
\DeclarePairedDelimiterX{\IntRange}[1]{\llbracket}{\rrbracket}{#1}
\DeclarePairedDelimiterX{\Group}[1]{\lgroup}{\rgroup}{#1}
\DeclarePairedDelimiterXPP{\List}[3]{}{\{}{\}}{_{#2,\ldots,#3}}{#1}

\DeclareMathOperator*{\argmin}{arg\,min}

\DeclareMathOperator{\Var}{Var}
\DeclareMathOperator{\Cov}{Cov}
\newcommand*{\Expect}{\mathbb{E}}
\DeclareMathOperator{\MSE}{MSE}
\DeclareMathOperator{\Median}{median}

\newcommand*{\from}{{:}\;}

\newcommand*{\Tag}[1]{{\mathsf{#1}}}

\newcommand*{\bydef}{\stackrel{{\scriptscriptstyle \text{def}}}{=}}

\newcommand*{\Adj}{\top}
\newcommand*{\T}{^{\Adj}}

\newcommand{\Paragraph}[1]{\medskip\noindent{\sffamily\bfseries #1}}

\newcommand*{\mapping}[1]{\mathcal{#1}}
\newcommand*{\estim}[1]{\widehat{#1}} 
\newcommand*{\mean}[1]{\widebar{#1}}  

\newcommand*{\Ndat}{N_{\mathrm{dat}}} 
\newcommand*{\Nwgt}{N_{\mathrm{wgt}}} 
\newcommand*{\Nact}{N_{\mathrm{act}}} 

\newcommand*{\Dif}{\Tag{dif}} 
\newcommand*{\Obj}{\Tag{ref}} 

\newcommand{\Field}[1]{\ensuremath{\mathtt{#1}}}

\begin{document}
\title{Self-calibration of adaptive optics systems}

\subtitle{Application to the THEMIS solar telescope}

%
%
%
%

\author{Clémentine Béchet\inst{1}\corrauth{clementine.bechet@univ-lyon1.fr}
  \and Éric Thiébaut\inst{1}\email{eric.thiebaut@univ-lyon1.fr}
  \and Michel Tallon\inst{1}\email{mtallon@obs.univ-lyon1.fr}
  \and Isabelle Tallon-Bosc\inst{1}\email{bosc@obs.univ-lyon1.fr}
  \and Bernard Gelly\inst{2} \email{bgelly@themis.iac.es}
  \and Richard Douet\inst{2}\email{richard.douet@cnrs.fr}
  }

  \institute{\inst{1}Université Lyon 1, ENS de Lyon, CNRS, CRAL, UMR 5574, Saint-Genis-Laval, France \\
    \inst{2} THEMIS - INSU-CNRS IRL 2009, Tenerife, Spain}

\date{In review.}

\abstract
{Nowadays most large ground-based telescopes implement adaptive optics (AO) to compensate
  turbulence effects and achieve diffraction limited angular resolution. The so-called
  \emph{interaction matrix} models the effects of deformable mirror (DM) commands as
  perceived by the wavefront sensor (WFS) and is the cornerstone of the AO control.
  Correctly calibrating this matrix is thus of uttermost importance. To perform this task,
  difficulties arise due to the number of degrees of freedom, the non-linearity, and
  temporal evolution of the AO system.}%
{The purpose of this article is to develop calibration methods applicable to any AO system
  under various conditions --- particularly in closed-loop while observing with the AO
  system --- and in real-time to continuously improve and update the AO system model.}%
{As a general approach, an affine approximation of the unknown mapping between the DM
  commands and the WFS measurements is considered. For a nonlinear AO system, we advocate
  that such a model may be preferable provided it can be updated to follow variable
  conditions of operation. Optimal estimators of the parameters of this affine model are
  obtained by fitting, in the weighted least squares sense, WFS data acquired with random
  but known probe commands sent to the DM. Following this framework, several calibration
  methods are being considered depending, in particular, on whether the affine model is
  directly fit to the WFS data or to the differences between successive WFS data frames. For
  all these methods, we derive closed-form expressions of the estimators of the components
  of the model. An advantage of the proposed calibration methods is that they can be applied
  under different conditions: before observing, on an internal source, or on-sky in open- or
  closed-loop. Provided probe commands are small enough to only negligibly impact the PSF at
  the output of the AO system, the latter case is of particular interest as it offers the
  possibility to continuously calibrate the interaction matrix while observing with the AO
  system. By introducing \emph{forgetting factors} to reduce the weight of calibration data
  as they age and thus loose their relevance, we show that the components of the affine
  model can be learned continuously from the sequence of WFS frames using simple recurrence
  rules. The low computational complexity of these rules makes them suitable for real-time
  applications (\ie\ at the same frequency as the AO loop). To theoretically compare the
  accuracy of the various calibration methods, we derive simple expressions for the mean
  squared errors (MSE) of the proposed estimators.}%
{We apply the proposed calibration methods to real telemetry data from the AO system of
  the THEMIS solar telescope. Our results demonstrate the effectiveness of these methods.
  As predicted by our theoretical expressions of the MSE of the estimators, our results
  show that the \emph{simple differences} (SD) method is the method of choice: not only
  does it produce estimators with the least MSE, but it is also very simple to implement
  compared to the \emph{push-pull} (PP) method which is widely used in AO systems.}%
{This work proposes a general approach for developing and comparing calibration methods
  for any AO system. This approach leads to various possible methods, applicable under
  varying conditions (offline and on-sky in closed-loop system), for continuously learning
  the model of the AO system. These calibration methods are therefore suitable for
  nonlinear AO systems and situations in which the behavior of the AO system is variable.}

\keywords{Instrumentation: adaptive optics -- Methods: numerical -- Methods: statistical --
  self-calibration -- generalized interaction matrix -- recursive least-squares}

\maketitle

\nolinenumbers

\section{Methods for adaptive optics calibration}

Adaptive optics (AO) is widely exploited on large telescopes to recover diffraction limited
angular resolution in the scientific path of the instrument in spite of the wavefront
aberrations caused by the atmospheric turbulence \citep{Roddier1999a}. AO systems are
equipped with a wavefront sensing device (WFS) and a correcting device, usually a deformable
mirror (DM), controlled in a servo loop to compensate in real-time, \ie hundreds or
thousands times per second, for the evolution of the residual wavefront aberrations. The
calibration of the relationship between the DM actuators and the WFS measurements is the
very first step in defining the AO control law. Due to the finite number of degrees of
freedom of the system and its linearity (in a good approximation), this is referred to in
the literature as the calibration of the AO \emph{interaction matrix} \citep{Roddier1999a}.
Until the early 2000's, small size AO systems for astronomy were usually calibrated using an
internal source upstream of the DM and the WFS. This was the case for instance for NAOS
\citep{RoussetEtAl2000a} at the Very Large Telescope (VLT), and still the case for SAXO
\citep[SPHERE AO,][]{SauvageEtAl2016a}, the high-contrast imaging adaptive optics system
installed about 15 years later at the VLT. Until then, most of the AO systems were isolated
from the rest of the optical train of the telescope, such that only slight drifts of the
system alignments were expected and thus calibration of the interaction matrix could be
typically executed a few times a month to track these changes \citep{BerdeuEtAl2024a}.

Over the past decade, AO calibration needed to evolve to be applied more frequently or to
become more flexible for a variety of reasons \citep{Kolb2016a}, among which the ones
discussed in the next paragraphs.

With the increased complexity of the new AO systems including adaptive mirrors located in
the optical train of the telescope (e.g. at Large Binocular Telescope, at the VLT Adaptive
Optics Facility), more frequent calibrations are needed \citep{EspositoEtAl2010a} and
sometimes there are no available AO calibration sources even at an intermediate focal plane
\citep{ObertiEtAl2006a, KolbEtAl2012a}. On-sky calibration is then required. A complete
calibration of the interaction matrix on sky is however known to be time-consuming and to
produce noisy estimates. To improve these two aspects, pseudo-synthetic interaction matrices
are often used for AO based on Shack-Hartmann WFS. They avoid calibrating the whole matrix
\citep[\eg,][]{NeichelEtAl2012a, KolbEtAl2012a,HeritierEtAl2018a,BerdeuEtAl2024a}, relying
on a model and only estimating a few parameters of mis-registration (relative shifts,
rotation, magnification, \etc). While non-invasive methods have been developed for the
Adaptive Optics Facility \citep{BechetEtAl2011a}, higher precision can be obtained with
invasive methods, introducing probing commands on the deformable mirror.

On-sky calibration data are also strongly affected by the atmospheric disturbances. The main
strategies developed to face this consist in successively pushing and pulling each actuator
(or correction mode) fast enough to freeze the turbulence \citep{TaylorEtAl2024a} or to use
modulation and demodulation of given modes as made on the LBT \citep{EspositoEtAl2006b}. All
these methods have been studied with particular interest in the perspectives of Giant
Segmented-Mirror Telescopes (GSMTs) and the huge increase of the AO number of degrees of
freedom \citep{HeritierEtAl2018a}. More recently, \citet{LaiEtAl2020a} presented the
DO-CRIME method to calibrate on-sky in working conditions a complete interaction matrix for
a wide-field AO system using random probes applied on the DM. This method does not use the
push and pull technique, but instead a high-pass filter is used to reject low-temporal
frequencies from the measurements, considering the latter as the contribution of the
atmosphere.

Calibration in working conditions is key when non-linearities are present in the system
response, which can be the case with pyramid-based WFS \citep{Ragazzoni1996a} or solar
wavefront sensing \citep{ThiebautEtAl2018a}. We currently note an evolution from the usual
Shack-Hartmann sensors towards the use of the pyramid WFS in many AO systems. On most of
such pyramid-based systems, modal calibration is reported, meaning that the DM actuators are
excited altogether producing large scale deformations as input to calibrate the response
\citep{EspositoEtAl2010a}. As a particular case however, the METIS adaptive optics system
under preparation for the ELT is foreseen to use a zonal pseudo-synthetic calibration
\citep{CorreiaEtAl2022a}, referring to excite the actuators one after the other, creating
local deformations only. The non-linearity of the sensor is a highlighted issue on
pyramid-based AO, with its sensitivity response to deformations of different spatial scales
affected by the working conditions \citep{KorkiakoskiVerinaud2007a}. Therefore it is
required to calibrate the AO interaction matrix according to the working conditions, and to
adapt its knowledge to the variations of those.

In the context of the French THEMIS solar telescope, the adaptive optics system is
relatively small (97-actuator deformable mirror) but the wavefront sensing is made on
solar granulation which is not reproducible by an internal source
\citep{ThiebautEtAl2022a}. Calibration on the sun is then a must. The limited linearity
range of the wavefront sensing on solar granulation \citep{ThiebautEtAl2018a,
  TallonEtAl2022a} is expected to induce variations of the interaction matrix sensitivity.
Other solar telescopes among the world use an internal source, coupled sometimes with a
slide of granulation printed on it, located before the AO correcting mirrors for their
calibration \citep{ShumkoEtAl2014a, Schmidt_et_al_2018}. Even in such calibration
conditions, \citet{Schmidt_et_al_2018} report that the response of the wavefront sensing
depends on the pattern of the target, thus demonstrating that the interaction matrix does
evolve with the observing conditions. AO calibration using random probes was initially
developed for the AO of the THEMIS telescope by \citet{ThiebautEtAl2022a}, but its
implementation was restricted to open-loop.

Taking into account all these considerations, this paper presents a general approach for an
AO system calibration in its current working conditions. To remain as general as possible,
the model of the interaction between DM and WFS is not restricted to be a single interaction
matrix but rather an affine approximation of the nonlinear behavior of these combined
devices. Provided it is updated frequently enough, such a model can follow the evolution of
the behavior with the operating conditions.

The presented method can be applied on an AO bench, with or without introduced wavefront
aberrations, as well as in an open or closed-loop system on sky or on the Sun. In order to
optimize the AO performances, the main driver is to be able to calibrate in conditions as
close as possible to the observing ones. The calibration is made with some known
perturbations added to the control commands. It is thus an invasive method. The amplitude of
the perturbations must be determined by a trade-off between the time (number of frames)
required to reach a given accuracy of the model parameters and the degradation of the AO
performances if perturbations are applied in closed-loop. In THEMIS AO, the perturbations
are first introduced in an open-loop run. The telemetry is used to calibrate a
``\emph{first-guess}'' model to close the loop. The loop is then closed and perturbations
are sent again for a closed-loop calibration this time. The parameters of the model are then
updated and used for the later AO corrections, without perturbations anymore.

In the remainder of this paper, the specific features of the method are analyzed. In
Sect.~\ref{sec:dm-2-wfs}, the approximation of an AO system by an affine model from the
commands to the measurement space is presented. In Sect.~\ref{sec:method} methods to
calibrate such a model are detailed together with their different features. The links with
calibration methods reported on other AO systems are discussed. In Sect.~\ref{sec:errors}, a
theoretical analysis of the errors induced on the interaction matrix and offset estimators
depending on the chosen calibration method is provided. In Sect.~\ref{sec:updating}, rules
are derived to update the parameters of the affine model as new WFS frames become available.
This paves the way to a continuous update of the AO model. Finally, in
Sect.~\ref{sec:results}, a comparison of the results of calibration, using the various
methods under study, is made using THEMIS AO data recorded on the Sun.

\section{The DM-to-WFS model}
\label{sec:dm-2-wfs}

In this section, we describe the model of the wavefront sensor (WFS) response to the
commands sent to the actuators of the deformable mirror (DM). The scope of the present
work is not to consider a specific WFS or DM but to provide a general analysis applicable
to any system.

\subsection{General model}

A general model of a WFS data frame in an adaptive optics system is given by:
\begin{equation}
  \label{eq:general-model}
  d_t ≈ \mapping{S}(w_t, c_{t-ℓ}),
\end{equation}
with $d_t ∈ ℝ^{\Ndat}$ the measurements during $t$-th exposure of the WFS camera (e.g., $t$
is the camera frame number), $w_t ∈ 𝕎$ the incoming wavefront, $c_{t-ℓ} ∈ ℝ^{\Nact}$ the DM
commands computed from the WFS data $d_{t-ℓ}$, and $ℓ$ the lag between the commands and WFS
frame affected by these commands. $\Ndat$ and $\Nact$ are the respective numbers of WFS
measurements and of DM commands elements. The mapping
$\mapping{S} \from 𝕎 \times ℝ^{\Nact} \to ℝ^{\Ndat}$ is the response of the wavefront sensor
to the incident wavefront and applied DM commands during the exposure of the WFS camera. No
matter how accurate the mapping may be, the relationship in Eq.~\eqref{eq:general-model} is
only an approximation due to measurement noise (this is expressed by the $≈$ sign).

Accounting for the times needed to expose, read, process an image from the WFS camera,
compute new commands, and send them to the DM, there is a lag $ℓ$, with $ℓ ⩾ 1$ frame,
before the effects of the commands can be seen by the WFS again. The system lag, or loop
delay, is usually close to 2 frames in AO, with variations around this value depending on
the AO design, the performance goals and the observing conditions \citep{Roddier1999a}.

In Eq.~\eqref{eq:general-model}, we are simply assuming that there is a causal
relationship between the wavefront, the DM commands and the WFS measurements: the mapping
$\mapping{S}$ exists even though it is unknown in practice. This mapping may have any
form, in particular, it may be a nonlinear function of the wavefront and the DM commands.
As an illustration, in the simple case of an AO system for which the sensor response only
depends on the residual wavefront, we could have:
\begin{equation}
  \label{eq:ex-simple-AO-model}
  \mapping{S}(w_t, c_{t-ℓ}) = \mapping{R}(w_t - \mapping{M}(c_{t-ℓ})),
\end{equation}
with $\mapping{M} \from ℝ^{\Nact} \to 𝕎$ the response in the wavefront space of the DM to
a command and $\mapping{R} \from 𝕎 \to ℝ^{\Ndat}$ the WFS response as a function of the
residual wavefront.

For the real-time control (RTC) of the AO system, that is for estimating the best commands
to apply to the DM given the WFS data, a more simple model than the one in
Eq.~\eqref{eq:general-model} must be assumed for several reasons. First, nonlinear models
are not well adapted for quick computations and, hence, real-time applications. Second,
nonlinear models can be very difficult to calibrate for the full domain of their input
variables. In the following, we consider an affine approximation of the mappings
$\mapping{S}$ of Eq.~\eqref{eq:general-model} to overcome these issues.

\subsection{Affine approximation}
\label{sec:affine-approximation}

In simple AO systems, the so-called WFS \emph{reference}, $d^{\Obj}$, represents the
measurements that should be given (on average) by the WFS for a good AO correction. The
task of the AO controller of such systems can be seen as inferring an update $u_{t-ℓ}$ of the
DM commands $c_{t - ℓ}$ that should have been applied during the acquisition of WFS data
$d_t$ so as to measure $d^{\Obj}$ instead. In other words, the controller attempts to
solve:
\begin{equation}
  \label{eq:command-update-problem}
  \mapping{S}(w_t, c_{t-ℓ} + u_{t-ℓ}) ≈ d^{\Obj}
\end{equation}
for $u_{t-ℓ} \in ℝ^{\Nact}$ and knowing the data $d_t$ for which the direct model in
Eq.~\eqref{eq:general-model} holds. The symbol $≈$ is used here to account for the fact
that the problem in Eq.~\eqref{eq:command-update-problem} may not have an exact solution
or, in the case of a nonlinear response, that an exact solution may not be computable in
real time. To find an approximate solution to the problem stated in
Eq.~\eqref{eq:command-update-problem} given the constraints, it is reasonable to
approximate the behavior of $\mapping{S}(w_t, c_{t-ℓ} + u_{t-ℓ})$ with respect to $u_{t-ℓ}$ by
something more simple and more manageable like the following affine model:
\begin{equation}
  \label{eq:affine-approx-of-S-in-u}
  \mapping{S}(w_t, c_{t - ℓ} + u_{t-ℓ}) ≈ G_t \,u_{t-ℓ} + b_t,
\end{equation}
where $G_t ∈ ℝ^{\Ndat × \Nact}$ is the so-called \emph{interaction matrix} and $b_t$ is an
offset term. Setting $u_{t-ℓ} = 0$ in the affine model~\eqref{eq:affine-approx-of-S-in-u}
immediately yields that:
\begin{equation}
  \label{eq:simple-approx-of-b}
  b_t ≈ \mapping{S}(w_t, c_{t - ℓ})
\end{equation}
which, according to Eq.~\eqref{eq:general-model}, can be seen as the noiseless WFS
measurements. It follows that:
\begin{align}
  G_t \,u_{t-ℓ} ≈
  \mapping{S}(w_t, c_{t - ℓ} + u_{t-ℓ}) - \mapping{S}(w_t, c_{t - ℓ}).
  \label{eq:G_t-by-Taylor}
\end{align}
A possibility to identify $G_t$ could be to use first order Taylor series which yields:
\begin{align}
  \label{eq:simple-approx-of-G}
  G_t ≈ \frac{∂\mapping{S}(w_t, c_{t-ℓ})}{∂c}.
\end{align}
Equations~\eqref{eq:simple-approx-of-b} and \eqref{eq:simple-approx-of-G} do provide
expressions for the affine model parameters, but expressed in terms of an unknown mapping
$\mapping{S}(w_t, c_{t - ℓ})$ which leaves unsolved the problem of their calibration.
Anyway, Taylor series yields a tangential approximation of the nonlinear behavior which is
good in $u_{t-ℓ} = 0$, but worsen as the norm of $u_{t-ℓ}$ increases and is thus worse on
average than, say, a Minimum Mean-Square-Error (MMSE) affine approximation. In the
following, we show how to overcome these two issues in practice.

Going back to the control, given $G_t$ and neglecting the WFS noise leads to $b_t ≈ d_t$ and
a possible solution to Eq.~\eqref{eq:command-update-problem} is then provided by:
\begin{equation}
  \label{eq:typical-commands}
  \estim{u}_{t-ℓ} = - G_t^{\dagger} \, \Paren[\big]{d_t - d^{\Obj}}
\end{equation}
with $G_t^{\dagger}$ the pseudo-inverse of $G_t$ or some regularized approximation of it.
Classical AO controllers using an integrator-based control law can be explained by a
similar expression \citep{Roddier1999a}.

In our notation, $G_t$ and $b_t$ have an explicit dependence on the temporal index $t$
because, for a nonlinear WFS response in the DM commands, the affine approximation depends
on the conditions of operation of the AO system. However, if the model is allowed to change
in time, its evolution must be slow enough to be able to calibrate and update the model
parameters by integrating information during a certain amount of time (certainly several WFS
frames).
In the following, we propose to tackle this issue by a general approach
to calibrate an affine quasi-static model derived from Eqs.~\eqref{eq:general-model} and
\eqref{eq:affine-approx-of-S-in-u} with respect to a perturbation $u_{t-ℓ}$ of the commands:
\begin{equation}
  \label{eq:affine-model}
  d_t = G_t \, u_{t-ℓ} + b_t + n_t
\end{equation}
where the term $n_t \in ℝ^{\Ndat}$ is a \emph{nuisance} term that not only accounts for the
measurement noise but also for the approximations of the model. As a consequence, the
nuisance depends on the unknown wavefront $w_t$ and the commands $c_{t-ℓ}$. The set of
perturbations $u_{t-ℓ} \in ℝ^{\Nact}$ is chosen to cover a basis of the possible commands to
be sent to the deformable mirror.

\begin{table}
  \LargeRowHeight
  \centering
  \resizebox{\columnwidth}{!}{ 
    \begin{tabular}{ll}
    \textbf{Symbol} & \textbf{Description}\\
    \hline
    $d_t ∈ ℝ^{\Ndat}$ & Measurements in $t$-th WFS frame\\
    $w_t ∈ 𝕎$         & Incoming wavefront during $t$-th WFS frame\\
    $c_t ∈ ℝ^{\Nact}$ & Commands computed after $t$-th WFS frame\\
    $u_{t-ℓ} \in \mathbb{R}^{\Nact}$ & Delayed probe commands impacting $t$-th WFS frame\\
    $ℓ ⩾ 1$ & Lag between DM commands and WFS data\\
    $G_t ∈ ℝ^{\Ndat × \Nact}$ & Quasi-static interaction matrix \eqref{eq:affine-model}\\
    $b_t = 𝔼_{W_t, Z_t}(d_t - G_t\,u_{t-ℓ})$ & Quasi-static offset \eqref{eq:model-offset}\\
    $\mean{d}_t ∈ ℝ^{\Ndat}$ & WAVG of measurements up to $t$-th WFS frame \eqref{eq:mean(d)_t}\\
    $\mean{u}_t ∈ ℝ^{\Ndat}$ & WAVG of delayed probes up to $t$-th WFS frame \eqref{eq:mean(u)_t} \\
    \end{tabular}
  }
  \NormalRowHeight
  \caption{Main notations for the model of data. ``WAVG'' stands for \emph{weighted average}.}
  \label{tab:notations}
\end{table}

Note that, unless $b_t=0$ ($\forall t$), the model of $d_t$ in Eq.~\eqref{eq:affine-model}
is not linear, but affine, with respect to the perturbation commands $u_{t-ℓ}$. The
wavefront sensor model is always considered relatively to some reference measurements
defined out of the adaptive optics control loop, as are for instance flat-wavefront
reference measurements made on a bench or with an internal source. The linear part of the
model, $G_t$, is often the main focus to design the AO reconstruction and analyze its
quality \citep{KorkiakoskiVerinaud2007a,DeoEtAl2019b,ChambouleyronEtAl2020a}. As shown in
Sec.~\ref{sec:offset-themis}, in practice, the affine term $b_t$ can be non-negligible, may
evolve with the working conditions and, hence, needs to be taken into account.

In the following, we first discuss on how the model can be calibrated and we then propose
different estimators and criteria to estimate $G_t$ and $b_t$.

\section{Calibration methods}
\label{sec:method}

In order to calibrate the components $G_t$ and $b_t$ of the affine model in
Eq.~\eqref{eq:affine-model}, arbitrary \emph{probe} commands, noted $u_{t-ℓ}$, can be
applied to the DM. To keep a general approach, possibly applicable in real-time (\ie, in
open-loop as well as in closed-loop), the probe commands $u_{t-ℓ}$ are added to the control
commands $c_{t-ℓ}$. Note that in open-loop, no control commands are sent for correction
($c_{t-ℓ} = 0, \forall t$) or the control commands are constant ($c_{t-ℓ} = c, \forall t$)
to maintain a given DM shape. In closed-loop, the probing commands must be small, random and
independent of the control commands to avoid visible artifacts in the point spread function
resulting from the AO correction. The response of the system to the control plus the probe
DM commands $\{c_{t'-\ell} + u_{t'-ℓ}\}_{t' ≤ t}$ is observable in the set of WFS data
$\{d_{t'}\}_{t' ≤ t}$. All these constitute \emph{calibration} data available to fit the
behavior of the system.

Considering the number of degrees of freedom of the model and the unknown nuisances (noise
and residual wavefront), faithful estimation of the model parameters can only be carried out
by collecting information from numerous calibration frames. As a consequence, the model
calibrated at time $t$ must be assumed to be \emph{quasi-static} in the sense that for a
large enough number WFS data frames $\{d_{t'}\}_{t' ≤ t}$, the model:
\begin{equation}
  \label{eq:quasi-static-model}
  d_{t'} = G_t \, u_{t'-ℓ} + b_t +n_{t'}
\end{equation}
is supposed to hold. Table~\ref{tab:notations} summarizes the main notations of the
variables for the model and the calibration data.

\subsection{Direct calibration method}
\label{sec:direct-calibration}

Given a set of calibration commands $\Brace{u_{t'-ℓ}}_{t' ≤ t}$ and the corresponding WFS
measurements $\Brace{d_{t'}}_{t' ≤ t}$, the parameters $G_t$ and $b_t$ of the quasi-static
affine model can be estimated by means of weighted least squares:
\begin{equation}
  \label{eq:WLS:general}
  \estim{G}^\Tag{D}_t, \estim{b}^\Tag{D}_t
  = \argmin_{G_t, b_t} \sum_{t'=ℓ+1}^{t} \Norm{d_{t'} - G_t \, u_{t'-ℓ} - b_t}^2_{Q_{t,t'}}
\end{equation}
where $Q_{t,t'} ∈ ℝ^{\Nact × \Nact}$ are symmetric non-negative weighting matrices and
$\Norm{x}^{2}_{Q} \bydef x\T\,Q\,x$ denotes the weighted squared norm of $x$. We call this
approach the \emph{direct} calibration method and denote the resulting estimators with a
letter $\Tag{D}$. There are multiple reasons to introduce the weighting matrices $Q_{t,t'}$:
accounting for the covariance of the \emph{nuisance} terms $n_{t'}$ and, possibly, reducing
the importance of calibration data $d_{t'}$ as the time difference $t - t'$ increases. The
latter is to relax the \emph{quasi-static} model assumption which becomes increasingly
invalid with the age of the calibration data.

We note that minimizing the least squares criterion in Eq.~\eqref{eq:WLS:general} is
supported by assuming that the nuisance term be \emph{centered} in the sense that:
\begin{equation}
  \label{eq:centered-nuisance}
  𝔼(n_{t'}) = 0
\end{equation}
where the expectation $𝔼(\ldots)$ is taken with respect to the wavefront and the noise. As
a consequence, the offset term is given by:
\begin{equation}
  \label{eq:model-offset}
  b_t = 𝔼(d_t - G_t\,u_{t-ℓ})
\end{equation}
which is readily the expectation of the data without perturbation. It is thus characteristic
of the working conditions of the system.

The criterion in the right-hand side of Eq.~\eqref{eq:WLS:general} is quadratic in $G_t$ and
in $b_t$ which warrants that a unique solution exists under mild conditions. For example,
the estimator of the offset $b_t$ knowing the interaction matrix $\estim{G}^\Tag{D}_t$ has
the following simple closed-form expression:
\begin{equation}
  \label{eq:b^D:general}
  \estim{b}^\Tag{D}_t =
  \Paren*{\sum\nolimits_{t'=ℓ+1}^{t} Q_{t,t'}}^{-1}\,\sum\nolimits_{t'=ℓ+1}^{t}
  Q_{t,t'}\,\Paren{ d_{t'} - \estim{G}^\Tag{D}_t \, u_{t'-ℓ}},
\end{equation}
provided $\sum_{t'=ℓ+1}^{t} Q_{t,t'}$ is invertible. Unfortunately, even though a closed
form-expression of $\estim{G}^\Tag{D}_t$ can also be obtained\footnote{such an
  expression could be given with Kronecker products of matrices} it would require
integrating 4-index tensors that, in practice, would be too large to be stored and
inverted. In order to obtain a more manageable solution, we impose that the weighting
matrices $Q_{t,t'}$ be all proportional to the same symmetric positive, hence invertible,
matrix $Q$:
\begin{equation}
  \label{eq:Q(t,t')=rho(t,t')*Q}
  Q_{t,t'} = ρ_{t,t'}\,Q,
\end{equation}
where $ρ_{t,t'} ≥ 0$ are non-negative scalar weights. The matrix $Q$ implements a
\emph{metric} that can account for the uneven quality of the WFS measurements and their
correlations whereas the scalar weights $ρ_{t,t'}$ can be used to reduce the importance
of the data frames as time passes. With this choice of weighting matrices, $Q$ is factorized
out of the denominator and numerator of $\estim{b}^\Tag{D}_t$ in Eq.~\eqref{eq:b^D:general}
which simplifies to:
\begin{equation}
  \label{eq:b^D_t}
  \estim{b}^\Tag{D}_t = \mean{d}_t - \estim{G}^\Tag{D}_t \, \mean{u}_t
\end{equation}
with:
\begin{align}
  \label{eq:mean(d)_t}
  \mean{d}_t
  &= \frac{1}{η_t}\,\sum\nolimits_{t'=ℓ+1}^{t} ρ_{t,t'}\,d_{t'}\\
  \label{eq:mean(u)_t}
  \mean{u}_t
  &= \frac{1}{η_t}\,\sum\nolimits_{t'=ℓ+1}^{t} ρ_{t,t'}\,u_{t'-ℓ}
\end{align}
the weighted averages of the data and of the delayed probe commands, and:
\begin{equation}
  \label{eq:sum-of-weights}
  η_t = \sum\nolimits_{t'=ℓ+1}^{t} ρ_{t,t'}
\end{equation}
the sum of weights. The required conditions for the existence and uniqueness of the
solution in Eq.~\eqref{eq:b^D_t} are that $Q$ be invertible and that $η_t > 0$ holds.
Substituting the estimator $\estim{b}^\Tag{D}_t$ in the weighted least squares criterion
and introducing
\begin{align}
  \label{eq:x^D_(t,t')}
  x^\Tag{D}_{t,t'} &= u_{t'-ℓ} - \mean{u}_t\\
  \label{eq:y^D_(t,t')}
  y^\Tag{D}_{t,t'} &= d_{t'} - \mean{d}_t
\end{align}
the recentered probe commands and WFS data, the \emph{direct} estimator of the interaction
matrix writes:
\begin{align}
  \estim{G}^\Tag{D}_t
  \label{eq:G^D_t:WLS}
  &= \argmin_{G_t} \sum\nolimits_{t'=ℓ+1}^{t} ρ_{t,t'}\,\Norm{y^\Tag{D}_{t,t'} - G_t \, x^\Tag{D}_{t,t'}}^2_Q\\
  \label{eq:G^D_t}
  &= B^\Tag{D}_t\,\Paren[\big]{A^\Tag{D}_t}^{-1}
\end{align}
with:
\begin{align}
  \label{eq:A^D_t}
  A^\Tag{D}_t &= \sum\nolimits_{t'=ℓ+1}^{t}
                ρ_{t,t'}\,x^\Tag{D}_{t,t'}\,\Paren[\big]{x^\Tag{D}_{t,t'}}\T,\\
  \label{eq:B^D_t}
  B^\Tag{D}_t &= \sum\nolimits_{t'=ℓ+1}^{t}
                ρ_{t,t'}\,y^\Tag{D}_{t,t'}\,\Paren[\big]{x^\Tag{D}_{t,t'}}\T.
\end{align}
The required condition for the existence and uniqueness of the solution in
Eq.~\eqref{eq:G^D_t} is that $A^\Tag{D}_t$ be invertible. This condition holds, in
practice as soon as the probes cover a given basis of the commands space. Interestingly,
neither $\estim{b}^\Tag{D}_t$ in Eq.~\eqref{eq:b^D_t} nor $\estim{G}^\Tag{D}_t$ in
Eq.~\eqref{eq:G^D_t} depends on the matrix $Q$. In other words, $\estim{G}^\Tag{D}_t$ and
$\estim{b}^\Tag{D}_t$ are the best estimators in the sense that they minimize the
quadratic error of the assumed affine model whatever the chosen metric. Hence, taking into
account non-uniform variance and correlations in the WFS data would not change the
expressions of these estimators. Although expression~\eqref{eq:G^D_t} is not new, the
literature of calibration in AO usually replaces the covariance matrices here by the
product of matrices built from gathering in their columns the sets of used data and probe
vectors \citep[see for instance,][]{KasperFedrigo2004a,KolbEtAl2012a,MeimonEtAl2015a,HeritierEtAl2018a}. The
set of probe vectors are often chosen to be Hadamard or Karhuenen-Loeve modes, less
frequently random modes \citep{LaiEtAl2020a}. The link between the two writings for least-squares AO
calibration has already been evidenced \citep{BechetTallon2012a} but the present derivation
goes beyond, explicitly handling the affine term and a possibly non-homogeneous metric $Q$.

\subsection{Calibration based on successive differences}

The accuracy of the estimators $\estim{b}^\Tag{D}_t$ and $\estim{G}^\Tag{D}_t$ in
Eqs.~\eqref{eq:b^D_t} and \eqref{eq:G^D_t} strongly depends on the variance of the nuisance
terms $n_{t'} = d_{t'} - G_t \, u_{t'-ℓ} - b_t$, that is the difference between the WFS data
and their model. In this variance, the contribution of the noise is unavoidable while the
contribution of the wavefront residuals depends on the quality of the AO correction and may
be the dominant contribution. The AO loop is usually so fast that the wavefront residuals do
not change very much between successive frames. Their contribution can thus be strongly
reduced\footnote{to the cost of doubling the variance of the noise contribution} if the
calibration is performed on the \emph{differences} between successive WFS frames:
\begin{align}
  \label{eq:y^dif_(t')}
  y^\Tag{dif}_{t'} \bydef d_{t'} - d_{t'-1} ≈ G_t\,x^\Tag{dif}_{t'}
\end{align}
for some known $x^\Tag{dif}_{t'}$ based on the probe commands. As detailed further, there
are several possible calibration methods based on the differences between successive WFS
frames. These methods differ in the sequence of probe commands and in the chosen
expression for $x^\Tag{dif}_{t'}$.

Following the \emph{direct} calibration method, the interaction matrix can be estimated by
minimizing the weighted squared error:
\begin{align}
  \estim{G}^\Tag{dif}_{t}
  \label{eq:G^dif_t:WLS}
  &= \argmin_{G_t} \sum\nolimits_{t'=ℓ+2}^{t}
    ρ_{t,t'}\,\Norm{y^\Tag{dif}_{t'} - G_t \, x^\Tag{dif}_{t'}}^2_Q\\
  \label{eq:G^dif_t}
  &= B^\Tag{dif}_t\,\Paren[\big]{A^\Tag{dif}_t}^{-1}
\end{align}
with:
\begin{align}
  \label{eq:A^dif_t}
  A^\Tag{dif}_t &= \sum\nolimits_{t'=ℓ+2}^{t}
                  ρ_{t,t'}\,x^\Tag{dif}_{t'}\,\Paren[\big]{x^\Tag{dif}_{t'}}\T,\\
  \label{eq:B^dif_t}
  B^\Tag{dif}_t &= \sum\nolimits_{t'=ℓ+2}^{t}
                  ρ_{t,t'}\,y^\Tag{dif}_{t'}\,\Paren[\big]{x^\Tag{dif}_{t'}}\T.
\end{align}

In the right-hand side of Eq.~\eqref{eq:y^dif_(t')}, the model of the successive
differences data $y^\Tag{dif}_{t'}$ only depends on the interaction matrix $G_t$ because
the offset $b_t$ has been suppressed by taking the difference. However, the offset term of
the affine model can be estimated by applying Eq.~\eqref{eq:b^D_t} but with
$\estim{G}^\Tag{dif}_{t}$ instead of $\estim{G}^\Tag{D}_t$:
\begin{equation}
  \label{eq:b^dif_t}
  \estim{b}^\Tag{dif}_t = \mean{d}_t - \estim{G}^\Tag{dif}_t \, \mean{u}_t.
\end{equation}

We now list a number of possible calibration methods that can be described by these
generic equations for processing the successive differences of WFS data frames. Analytic
expressions of the propagation of errors and experimental tests are provided in futher
sections to compare all proposed calibration strategies.

\subsubsection{Push-pull calibration}
\label{sec:push-pull-calibration}

The so-called \emph{push-pull} calibration \citep[\eg,][]{EspositoEtAl2010a} is a common
strategy to reduce the contribution of the turbulent wavefront to the measurements in AO
calibration. The method consists in applying a probe command and its opposite in two
successive frames and in calibrating the model based on the differences of each pair of
such frames. In our formalism, the \emph{push-pull} strategy amounts to taking
$u_{t'-ℓ} = -u_{t'-ℓ-1}$ for all even calibration frame index $t'$. To calibrate the
quasi-static interaction matrix $G_t$, Eqs.~\eqref{eq:G^dif_t:WLS} to \eqref{eq:b^dif_t}
can be applied except that the successive differences can only be taken every two frames.
As a consequence, in these equations, all sums over $t'$ must be done for $t'$ even and
replacing $x^\Tag{dif}_{t'}$ by:
\begin{equation}
  \label{eq:x^PP_(t')}
  x^\Tag{PP}_{t'} = u_{t'-ℓ} - u_{t'-ℓ-1} = 2\,u_{t'-ℓ}.
\end{equation}
For further analysis, one should note here that, provided probes are centered, \emph{i.e.}
$\Expect({u}_{t})=0$, and $\rho_{t',t}=\rho_{t'-1,t}$ for all even $t'$ and $t$,
$A^\Tag{dif}_t$ and $B^\Tag{dif}_t$ from Eqs.~\eqref{eq:A^dif_t}-\eqref{eq:B^dif_t} can be
expanded to demonstrate that $A^\Tag{dif}_t=2 A^\Tag{D}_{t}$ and
$ B^\Tag{dif}_{t}=2B^\Tag{D}_{t}$ in Eqs.~\eqref{eq:A^D_t}-\eqref{eq:B^D_t}. The
\emph{direct} (D) method applied on a push-pull calibration sequence is therefore expected
to provide the same estimate of interaction matrix as the \emph{push-pull} method.
Hereinafter, $\Tag{PP}$ is used to label quantities obtained by the \emph{push-pull} method.

\subsubsection{Other successive differences}
\label{sec:successive-differences}

The \emph{push-pull} strategy is one of the possible means to reduce the wavefront
contribution to the nuisance in the calibrated quantities. We also consider simply sending
independent probe commands in successive frames. Then, there are three different obvious
choices for $x^\Tag{dif}_{t'}$ for which the approximation in Eq.~\eqref{eq:y^dif_(t')}
holds and that Eqs.~\eqref{eq:G^dif_t:WLS} to \eqref{eq:b^dif_t} are applicable to
calibrate the interaction matrix:
\begin{align}
  \label{eq:x^SD_(t')}
  x^\Tag{SD}_{t'} &= u_{t'-ℓ} - u_{t'-ℓ-1},\\
  \label{eq:x^FD_(t')}
  x^\Tag{FD}_{t'} &= u_{t'-ℓ},\\
  \label{eq:x^BD_(t')}
  x^\Tag{BD}_{t'} &= -u_{t'-ℓ-1}.
\end{align}
We call these approaches the \emph{simple differences}, \emph{forward differences}, and
\emph{backward differences} methods and denote the resulting quantities with respective
labels $\Tag{SD}$, $\Tag{FD}$, and $\Tag{BD}$.

\subsection{Generic formalism}
\label{sec:generic}

\begin{table*}[t]
  \LargeRowHeight
  \centering
  \begin{tabular}{rlll}
    \hline
                         & \multicolumn{1}{c}{Calibration data} & \multicolumn{1}{c}{Calibration probe} & \multicolumn{1}{c}{Calibration nuisance} \\
    Calibration strategy & \multicolumn{1}{c}{($y_{t,t'}$)} &                                                               \multicolumn{1}{c}{($x_{t,t'}$)} &  \multicolumn{1}{c}{($z_{t,t'}$)} \\
    \hline
    Direct ($t' ⩾ ℓ+1$) & $y^\Tag{D}_{t,t'} = d_{t'} - \mean{d}_t$ & $x^\Tag{D}_{t,t'} = u_{t'-ℓ} - \mean{u}_t$ & $z^\Tag{D}_{t,t'} = n_{t'} - \mean{n}_t ≈ n_{t'}$\\
    Simple Differences ($t' ⩾ ℓ+2$) & $y^\Tag{SD}_{t,t'} = d_{t'} - d_{t'-1}$ & $x^\Tag{SD}_{t,t'} = u_{t'-ℓ} - u_{t'-ℓ-1}$ & $z^\Tag{SD}_{t,t'} = n_{t'} - n_{t'-1}$ \\
    Forward Differences ($t' ⩾ ℓ+2$) & $y^\Tag{FD}_{t,t'} = d_{t'} - d_{t'-1}$ & $x^\Tag{FD}_{t,t'} = u_{t'-ℓ}$ & $z^\Tag{FD}_{t,t'} = n_{t'} - n_{t'-1} - G_t\,u_{t'-ℓ-1}$\\
    Backward Differences ($t' ⩾ ℓ+2$) & $y^\Tag{BD}_{t,t'} = d_{t'} - d_{t'-1}$ & $x^\Tag{BD}_{t,t'} = -u_{t'-ℓ-1}$ & $z^\Tag{BD}_{t,t'} = n_{t'} - n_{t'-1} + G_t\,u_{t'-ℓ}$\\
    Push-Pull ($t$ and $t'$ even with $t' ⩾ ℓ+2$) & $y^\Tag{PP}_{t,t'} = d_{t'} - d_{t'-1}$ & $x^\Tag{PP}_{t,t'} = u_{t'-ℓ} - u_{t'-ℓ-1} = 2\,u_{t'-ℓ}$ & $z^\Tag{PP}_{t,t'} = n_{t'} - n_{t'-1}$ \\
    \hline
  \end{tabular}
  \NormalRowHeight
  \caption{Calibration probes and data for the different calibration strategies. $G_t$ is the quasi-static interaction matrix for the sequence of
  calibration data. Known
    quantities are $d_{t'}$ and $u_{t'-ℓ}$, the WFS data and the corresponding probe commands
    in frame $t'$ of the calibration sequence. The nuisance terms $n_{t'}$ are unknown
    random quantities accounting for the noise, the residual wavefront, and the model
    approximation in the WFS data. The overbar denote weighted average of a quantity for the
    calibration sequence.}
  \label{tab:general:eqns}
\end{table*}

All calibration methods under consideration can be described by a common formalism stated
in this section. This formalism will be useful to study the propagation of errors in the
estimators and the practical implementation of these estimators for real-time
computations.

In the considered calibration methods, the estimator of the quasi-static interaction
matrix is the result of a weighted least squares fit:
\begin{align}
  \label{eq:G_t:WLS}
  \estim{G}_t
  &= \argmin_{G_t} \sum\nolimits_{t'=t'_\Tag{min}}^{t}
    ρ_{t,t'}\,\Norm[\big]{y_{t,t'} - G_t\,x_{t,t'}}^2_Q\\
  \label{eq:G_t}
  &= B_t \, A_t^{-1},
\end{align}
with $Q$ a symmetric positive definite matrix, $ρ_{t,t'}$ non-negative scalar weights, and
$t'_\Tag{min} = ℓ+1$ for the \emph{direct} method while $t'_\Tag{min} = ℓ+2$ for the
\emph{differential} ones. The expressions of the calibration probes and data, $x_{t,t'}$
and $y_{t,t'}$, depend on the calibration method as summarized by
Table~\ref{tab:general:eqns}. The matrices $A_t$ and $B_t$ are given by:
\begin{align}
  \label{eq:A_t}
  A_t &= \sum\nolimits_{t'=t'_\Tag{min}}^{t} ρ_{t,t'}\,x_{t,t'}\,x_{t,t'}\T,\\
  \label{eq:B_t}
  B_t &= \sum\nolimits_{t'=t'_\Tag{min}}^{t} ρ_{t,t'}\,y_{t,t'}\,x_{t,t'}\T.
\end{align}
The equations of this generic formalism apply to all calibration methods except that, for
the \emph{push-pull} calibration method, $t$ must be even and sums must be restricted to
even values of $t'$.

The offset term of the affine model is estimated as:
\begin{equation}
  \label{eq:b_t}
  \estim{b}_t = \mean{d}_t - \estim{G}_t \, \mean{u}_t.
\end{equation}
where $\mean{d}_t$ and $\mean{u}_t$ are given in Eqs.~\eqref{eq:mean(d)_t} and
\eqref{eq:mean(u)_t}. It is worth noting the close connection of this expression of the
estimator $\estim{b}_t$ with the definition of $b_t$ in Eq.\eqref{eq:model-offset}.

\section{Errors on the estimators}
\label{sec:errors}

This section provides a theoretical comparison of the errors in the model estimators
depending on the chosen calibration method.

\subsection{Statistics of the probe commands}
\label{sec:probe-statistics}

We restrict our study to random probe commands $u_{t'-ℓ}$ that are centered and independent,
this avoids noticeable artifacts in closed-loop calibration which is an important
objective of the proposed methods. To be more specific about the statistics of the probe
commands and using $𝔼_U(\ldots)$ to denote expectation with respect to the probe commands,
we will assume that random independent probes sequence follow the properties, for all $t' ≤ t$ and $t'' ≤ t$:
\begin{align}
  \label{eq:probe-expectation}
  𝔼_U(u_{t'-ℓ}) &= 0,\\
  \label{eq:probe-covariance}
  𝔼_U(u_{t'-ℓ}[j]\,u_{t''-ℓ}[j']) &= δ_{j,j'}\,δ_{t',t''}\,ϵ^2_j,
\end{align}
with $ϵ^2_j$ the variance of the $j$-th coefficient in the probe commands. The first of
these properties states that the probe commands are centered. The second one states that
they are independent for different indices in the command vector (indices may refer to
actuators or modes depending if zonal or modal commands are sent) and/or frames but their
variance $ϵ^2_i$ may only depend on the index $i$, not on the frame index $t'$. Note that in a push-pull sequence of probes, the above properties hold except for the case $t''=t'-1$ with $t'$ even, since by construction,
\begin{equation}
    \label{eq:probe-covariance-pp}
  𝔼_U(u_{t'-ℓ}[j]\,u_{t'-1-ℓ}[j']) = - δ_{j,j'} ϵ^2_j.
\end{equation}

\subsection{Calibration errors for the interaction matrix}
\label{sec:G_t:errors}

In the above generic formalism, the quasi-static interaction matrix $G_t$ is calibrated
from a sequence of centered or differential WFS data frames modeled by:
\begin{equation}
  \label{eq:y_(t,t')}
  y_{t,t'} = G_t\,x_{t,t'} + z_{t,t'}
\end{equation}
where $z_{t,t'}$ is a nuisance term accounting for the WFS data noise, for the residual
wavefront, and for the approximations by the quasi-static affine model in
Eq.~\eqref{eq:quasi-static-model}. The expressions of the nuisance terms $z_{t,t'}$ depend
on the calibration method and are given in Table~\ref{tab:general:eqns}.

Using Eqs.~\eqref{eq:G_t}--\eqref{eq:B_t} and \eqref{eq:y_(t,t')} and assuming $A_t$ is
invertible, the estimation error for the quasi-static interaction matrix can be rewritten as
follows:
\begin{align}
  \label{eq:G-error-def}
  δG_t &\bydef \estim{G}_t - G_t \\
  \label{eq:G-error-alt}
       &= \Paren[\Big]{\sum\nolimits_{t'=t'_\Tag{min}}^{t}
         ρ_{t,t'}\,z_{t,t'}\,x_{t,t'}\T}\,A_t^{-1}\\
  \label{eq:G-error}
       &= \sum\nolimits_{t'=t'_\Tag{min}}^{t}
         ρ_{t,t'}\,z_{t,t'}\,v_{t,t'}\T
\end{align}
where $v_{t,t'} = A_t^{-1}\,x_{t,t'}$ are known vectors that only depend on the probe
commands. Although it is more general due to the scalar weights $ρ_{t,t'}$, the right-hand
side of Eq.~\eqref{eq:G-error-alt} is consistent with expressions of the errors found in
the AO literature \citep[\eg,][]{MeimonEtAl2015a,HeritierEtAl2018a}. The expression in
Eq.~\eqref{eq:G-error} shows that the statistics of the nuisances $z_{t,t'}$ drives the
errors in the estimator $\estim{G}_t$ of $G_t$. For example, since $v_{t,t'}$ is known,
the estimator $\estim{G}_t$ is unbiased provided the nuisances $z_{t,t'}$ are centered,
\ie $𝔼\Paren{z_{t,t'}} = 0 \Rightarrow 𝔼\Paren{δG_t} = 0$ where expectation is taken
with respect to the realizations of the noise and of the wavefront.

To simplify Eq.~\eqref{eq:G-error}, we first derive an approximate expression of the known
terms $v_{t,t'}$. As previously noted, the calibration of the quasi-static model
parameters requires that Eq.~\eqref{eq:quasi-static-model} holds for a long enough
sequence of WFS frames such that $t' ≤ t$ and $ρ_{t,t'} > 0$. Under this condition, the
known matrix $A_t$, defined in Eq.~\eqref{eq:A_t}, can be approximated by its expectation
with respect to the realizations of the random probe commands $u_{t'-ℓ}$:
\begin{equation}
  \label{eq:A_t=E(A_t)}
  A_t ≈ 𝔼_{U}\Paren*{A_t}
  = \sum\nolimits_{t'=t'_\Tag{min}}^{t} ρ_{t,t'}\,𝔼_{U}\Paren[\big]{x_{t,t'}\,x_{t,t'}\T}.
\end{equation}
Considering the possible expressions for $x_{t,t'}$ given in Table~\ref{tab:general:eqns}
and properties \eqref{eq:probe-expectation}, \eqref{eq:probe-covariance} and \eqref{eq:probe-covariance-pp} of the probe
commands $u_{t'-ℓ}$, the vectors $x_{t,t'}$ are independent\footnote{neglecting the tiny
  correlation between $u_{t'-ℓ}$ and $\mean{u}_t$ in the \emph{direct} method} and centered.
It follows that:
\begin{equation}
  𝔼_{U}\Paren[\big]{x_{t,t'}[j]\,x_{t,t'}[j']} = δ_{j,j'}\,ω_j
\end{equation}
with
\begin{equation}
  \label{eq:omega_j}
  ω_j \bydef 𝔼_{U}\Paren[\big]{x_{t,t'}[j]^2} = \Var_{U}\Paren[\big]{x_{t,t'}[j]} > 0
\end{equation}
which does not depend on the frame indices $t$ or $t'$. Finally, the entries of $A_t$ can
be approximated by:
\begin{equation}
  \label{eq:A_t[j,j']}
  A_t[j,j'] ≈ δ_{j,j'}\,η_t\,ω_j
\end{equation}
with $η_t = \sum_{t'=t'_\Tag{min}}^{t} ρ_{t,t'}$, as defined in
Eq.~\eqref{eq:sum-of-weights}. This shows that, for a long enough calibration sequence,
$A_t$ tends to be a diagonal matrix which is invertible as all its diagonal entries are
positive. Now, for a long enough calibration sequence, the correlation between $x_{t,t'}$
and $A_t$ can be neglected to obtain the following approximation for the entries of
$v_{t,t'}$:
\begin{equation}
  \label{eq:v_(t,t')[j]}
  v_{t,t'}[j] ≈ \frac{x_{t,t'}[j]}{η_t\,ω_j}.
\end{equation}
With this approximation, Eq.~\eqref{eq:G-error} becomes:
\begin{align}
  \label{eq:G-error-approx}
  δG_t[i,j] &≈\sum\nolimits_{t'=t'_\Tag{min}}^{t}
              \frac{ρ_{t,t'}}{η_t\,ω_j}\,z_{t,t'}[i]\,x_{t,t'}[j].
\end{align}
Finally, the \emph{mean squared error} (MSE) of the estimator of the quasi-static
interaction matrix is simply given by:
\begin{align}
  \label{eq:MSE(G)}
  \MSE\Paren[\big]{\estim{G}_t[i,j]}
  &\bydef 𝔼\Paren[\big]{δG_t[i,j]^2}\notag\\
  &≈ \sum\nolimits_{t'=t'_\Tag{min}}^{t}
    \frac{ρ^2_{t,t'}}{η^2_t\,ω^2_j}\,𝔼_Z\Paren[\big]{z^2_{t,t'}[i]}
    \,\underbrace{𝔼_U\Paren[\big]{x^2_{t,t'}[j]}}_{\displaystyle ω_j}\notag\\
  &≈ \frac{χ_i}{ξ_t\,ω_j},
\end{align}
with\footnote{according to the quasi-static hypothesis, the statistics of the nuisance
  $z_{t,t'}[i]$ is stationary and, hence, its variance may only depend on the measurement index
  $i$}
\begin{align}
  \label{eq:chi_i}
  χ_i &= 𝔼_Z\Paren[\big]{z^2_{t,t'}[i]} = \Var_Z\Paren[\big]{z_{t,t'}[i]},\\
  \label{eq:effective-number-of-frames}
  ξ_t &= \frac{η_t^2}{\sum_{t'=t'_\Tag{min}}^{t} ρ^2_{t,t'}}
       = \frac{\Paren*{\sum_{t'=t'_\Tag{min}}^{t} ρ_{t,t'}}^2}{\sum_{t'=t'_\Tag{min}}^{t} ρ^2_{t,t'}}.
\end{align}

Taking scalar weights $ρ_{t,t'}$ all equal to the same positive constant yields
$ξ_t = t-t'_{\Tag{min}}+1$, the number of WFS frames used from the sequence. Hence, $ξ_t$ can
be seen as an \emph{effective number of calibration frames}. The value of $ξ_t$ is
nearly\footnote{differential methods other than the \emph{push-pull} one consider one less
  frame than in the \emph{direct} method} the same for all considered calibration methods
except the \emph{push-pull} method which only considers differential WFS frames with an
even index $t'$:
\begin{align}
  \label{eq:xi^D_t}
  ξ^\Tag{D}_t
  &\bydef \frac{\Paren[\Big]{\sum_{t'=ℓ+1}^{t} ρ_{t,t'}}^2}{\sum_{t'=ℓ+1}^{t} ρ^2_{t,t'}}\\
  \label{eq:xi^SD_t}
  ξ^\Tag{SD}_t
  &= ξ^\Tag{FD}_t = ξ^\Tag{BD}_t
  \bydef \frac{\Paren[\Big]{\sum_{t'=ℓ+2}^{t} ρ_{t,t'}}^2}{\sum_{t'=ℓ+2}^{t} ρ^2_{t,t'}}
  ≈ ξ^\Tag{D}_t\\
  \label{eq:xi^PP_t}
  ξ^\Tag{PP}_t
  &\bydef \frac{\Paren[\Big]{\sum_{\substack{t'=ℓ+2\\t'\text{ even}}}^{t} ρ_{t,t'}}^2}
  {\sum_{\substack{t'=ℓ+2\\t'\text{ even}}}^{t} ρ^2_{t,t'}}
  ≈ \frac{\Paren[\Big]{\frac{1}{2}\,\sum_{t'=ℓ+1}^{t} ρ_{t,t'}}^2}
  {\frac{1}{2}\,\sum_{t'=ℓ+1}^{t} ρ^2_{t,t'}} = \frac{ξ^\Tag{D}_t}{2}.
\end{align}

For clarity reasons, the derivations of simple approximations of the variances $ω_j$ and
$χ_i$ for each calibration method are gathered in Appendix~\ref{sec:variances}. The
results are used to summarize below the MSEs of the estimator of the quasi-static
interaction matrix for the different calibration methods, using Eq.~\eqref{eq:MSE(G)}.

\Paragraph{Direct method.} Using the variances $ω^\Tag{D}_j$ and $χ^\Tag{D}_i$ given in
Eqs.~\eqref{eq:omega^D_j} and \eqref{eq:chi^D_i}
:
\begin{equation}
  \label{eq:MSE(G^D)}
  \MSE\Paren[\big]{\estim{G}^\Tag{D}_t[i,j]}
  ≈ \frac{χ^\Tag{D}_i}{ξ^\Tag{D}_t\,ω^\Tag{D}_j}
  = \frac{σ^2_i}{ξ^\Tag{D}_t\,ϵ^2_j}
\end{equation}
with $σ^2_i = \Var\Paren[\big]{n_{t'}[i]}$ the variance of the nuisance term in the WFS
data (\cf\ Eq.~\eqref{eq:quasi-static-model}) and where $ε_j$ is introduced in
Eq.~\eqref{eq:probe-covariance}.

\Paragraph{Simple differences method.} With $ξ^\Tag{SD}_t ≈ ξ^\Tag{D}_t$ and using the variances
$ω^\Tag{SD}_j$ and $χ^\Tag{SD}_i$ given in Eqs.~\eqref{eq:omega^SD_j} and
\eqref{eq:chi^SD_i}
:
\begin{equation}
  \label{eq:MSE(G^SD)}
  \MSE\Paren[\big]{\estim{G}^\Tag{SD}_t[i,j]}
  ≈ \frac{χ^\Tag{SD}_i}{ξ^\Tag{SD}_t\,ω^\Tag{SD}_j}
  = \frac{(1 - ζ_i)\,σ^2_i}{ξ^\Tag{D}_t\,ϵ^2_j}
\end{equation}
with $ζ_i = \Cov\Paren[\big]{n_{t'}[i], n_{t'-1}[i]}/σ^2_i$ the coefficient of correlation
for successive nuisances in the $i$-th WFS data.

\Paragraph{Forward differences method.} With $ξ^\Tag{FD}_t ≈ ξ^\Tag{D}_t$ and using the variances
$ω^\Tag{FD}_j$ and $χ^\Tag{FD}_i$ given in Eqs.~\eqref{eq:omega^FD_j} and
\eqref{eq:chi^FD_i}
:
\begin{equation}
  \label{eq:MSE(G^FD)}
  \MSE\Paren[\big]{\estim{G}^\Tag{FD}_t[i,j]}
  ≈ \frac{χ^\Tag{FD}_i}{ξ^\Tag{FD}_t\,ω^\Tag{FD}_j}
  = \frac{2\,(1 - ζ_i)\,σ^2_i}{ξ^\Tag{D}_t\,ϵ^2_j}.
\end{equation}

\Paragraph{Backward differences method.} Since $ξ^\Tag{BD}_t ≈ ξ^\Tag{D}_t$ and using the
variances $ω^\Tag{BD}_j$ and $χ^\Tag{BD}_i$ given in Eqs.~\eqref{eq:omega^BD_j} and
\eqref{eq:chi^BD_i}
:
\begin{equation}
  \label{eq:MSE(G^BD)}
  \MSE\Paren[\big]{\estim{G}^\Tag{BD}_t[i,j]}
  ≈ \frac{χ^\Tag{BD}_i}{ξ^\Tag{BD}_t\,ω^\Tag{BD}_j}
  = \frac{2\,(1 - ζ_i)\,σ^2_i}{ξ^\Tag{D}_t\,ϵ^2_j}.
\end{equation}
Not surprisingly, this result is the same as for the \emph{forward differences} method.

\Paragraph{Push-pull method.} With $ξ^\Tag{PP}_t ≈ ξ^\Tag{D}_t/2$ and using the variances
$ω^\Tag{PP}_j$ and $χ^\Tag{PP}_i$ given in Eqs.~\eqref{eq:omega^PP_j} and
\eqref{eq:chi^PP_i}
:
\begin{equation}
  \label{eq:MSE(G^PP)}
  \MSE\Paren[\big]{\estim{G}^\Tag{PP}_t[i,j]}
  ≈ \frac{χ^\Tag{PP}_i}{ξ^\Tag{PP}_t\,ω^\Tag{PP}_j}
  = \frac{(1 - ζ_i)\,σ^2_i}{ξ^\Tag{D}_t\,ϵ^2_j}
\end{equation}

The variance $σ_i^2$ of the nuisance $n_{t'}[i]$ and its correlation $ζ_i$ between
successive frames strongly depend on the conditions of calibration. For example, if
calibration is performed on an internal source so that the turbulence is negligible, the
nuisance $n_{t'}$ in the WFS data is only due to measurement noise. As another example, if
calibration is performed on sky and in closed-loop, the contribution of the turbulent
wavefront should be much reduced compared to calibration in open-loop. Nevertheless, for
similar conditions and according to the above expressions, the \emph{push-pull} (PP) and
\emph{simple differences} (SD) methods appear to be equivalent in terms of MSE, although SD
is easier to implement in practice. On the sky, the correlation $ζ_i$ between successive
frames becomes significant due to the contribution of the turbulence in open-loop and of the
residual wavefront in closed-loop. In this case, the \emph{push-pull} (PP) and \emph{simple
  differences} (SD) methods are expected to be better than the \emph{direct} (D) method. As
already noted, the \emph{forward differences} (FD) and \emph{backward differences} (BD) are
almost identical and, due to the fact that only one-half of the known probe commands are
taken into account, they yield interaction matrix estimators whose MSE is twice worse than
that obtained by the \emph{push-pull} (PP) and \emph{simple differences} (SD) methods. To
summarize, whatever the conditions, the \emph{push-pull} (PP) and \emph{simple differences}
(SD) methods are expected to provide the best estimators. It is worth noting that this
analysis of errors is general and valid for both zonal or modal calibration. These
conclusions, based on approximate expressions of the MSEs, must be confirmed by tests in
different conditions. In Section~\ref{sec:results}, we present results based on the
telemetry acquired on the AO system of the THEMIS solar telescope.

\subsection{Calibration errors for the offset term}
\label{sec:b_t:errors}

Whatever the calibration method, our estimator $\estim{b}_t$ of the offset term of the
affine model is given by Eq.~\eqref{eq:b_t}. Using Eqs.~\eqref{eq:quasi-static-model} and
\eqref{eq:mean(d)_t}--\eqref{eq:sum-of-weights}, the weighted average of the WFS data can
be re-expressed as:
\begin{equation}
  \label{eq:d-stationary-alt}
  \mean{d}_t = G_t\,\mean{u}_t + b_t + \mean{n}_t
\end{equation}
with $\mean{n}_t \bydef (1/η_t)\,\sum_{t'=ℓ+1}^{t} ρ_{t,t'}\,n_t'$ the weighted average of the
nuisance term in Eq.~\eqref{eq:quasi-static-model}. Inserting this expression of
$\mean{d}_t$ in that of the offset estimator in Eq.~\eqref{eq:b_t} yields
the following expression of the error of this estimator:
\begin{equation}
  \label{eq:b-error}
  δb_t \bydef \estim{b}_t - b_t = \mean{n}_t - δG_t\,\mean{u}_t
\end{equation}
with $δG_t$ the error of the estimator of the interaction matrix defined in
Eq.~\eqref{eq:G-error-def} and $\mean{u}_t$ the weighted average of the probe commands
defined in \eqref{eq:mean(u)_t}. Hence, the estimator of the offset is unbiased provided
$𝔼\Paren[\big]{\mean{n}_t} = 0$ and $𝔼\Paren[\big]{δG_t\,\mean{u}_t} = 0$ which is the
case in practice.

Noting that, for a long enough calibration sequence, the mean probe commands $\mean{u}_t$
and $x_{t,t'}$ are negligibly correlated and that $\mean{n}_t ≈ 0$ and $\mean{u}_t ≈ 0$,
the MSE of the offset estimator can be approximated by:
\begin{align}
  \label{eq:MSE(b_t)}
  \MSE\Paren[\big]{\estim{b}_t[i]}
  &\bydef 𝔼\Paren[\big]{δb_t[i]^2}\\
  &≈ 𝔼\Paren[\big]{\mean{n}_t[i]^2}
    + \sum_j 𝔼\Paren[\big]{δG_t[i,j]^2}\,𝔼\Paren[\big]{\mean{u}_t[j]^2}\\
  &≈ 
    \frac{σ_i^2}{ξ_t} + \frac{χ_i}{ξ_t^2}\,\sum_j \frac{ϵ_j^2}{ω_j}
\end{align}
Indeed, $𝔼\Paren[\big]{\mean{n}_t[i]^2} = σ_i^2/ξ_t$ and
$𝔼\Paren[\big]{\mean{u}_t[j]^2} = ϵ_j^2/ξ_t$ straightforwardly follow from the definitions
of $\mean{n}_t$ and $\mean{u}_t$, from the statistics of $n_{t'}$ and $u_{t'-ℓ}$, and from
the definition of $ξ_t$, the effective number of calibration frames, in
Eq.~\eqref{eq:effective-number-of-frames}. As shown in Appendix~\ref{sec:variances}, the
ratio $ϵ_j^2/ω_j$ is a small constant which depends on the calibration method but not on
$ϵ_j$. Using the results of Appendix~\ref{sec:variances}, it can be found that:
\begin{align}
  \label{eq:MSE(b^D_t)}
  \MSE\Paren[\big]{\estim{b}^\Tag{D}_t[i]}
  & ≈ \frac{σ_i^2}{ξ^\Tag{D}_t}\,\Paren*{1 + \frac{\Nact}{ξ^\Tag{D}_t}},\\
  \label{eq:MSE(b^SD_t)}
  \MSE\Paren[\big]{\estim{b}^\Tag{SD}_t[i]}
  & ≈ \frac{σ_i^2}{ξ^\Tag{D}_t}\,
    \Paren*{1 + \Paren[\big]{1 - ζ_i}\,\frac{\Nact}{ξ^\Tag{D}_t}},\\
  \label{eq:MSE(b^FD_t)}
  \MSE\Paren[\big]{\estim{b}^\Tag{FD}_t[i]}
  & = \MSE\Paren[\big]{\estim{b}^\Tag{BD}_t[i]}\notag\\
  & ≈ \frac{σ_i^2}{ξ^\Tag{D}_t}\,
    \Paren*{1 + 2\,\Paren[\big]{1 - ζ_i}\,\frac{\Nact}{ξ^\Tag{D}_t}},\\
  \label{eq:MSE(b^PP_t)}
  \MSE\Paren[\big]{\estim{b}^\Tag{PP}_t[i]}
  & ≈ \frac{2\,σ_i^2}{ξ^\Tag{D}_t}\,
    \Paren*{1 + \Paren[\big]{1 - ζ_i}\,\frac{\Nact}{ξ^\Tag{D}_t}}.
\end{align}
For the offset term, the \emph{simple differences} (SD) method is the most accurate:
compared to the \emph{direct} (D) method, it has the ability to reduce the part of the
nuisance correlated between successive frames and, compared to the \emph{push-pull} (PP)
method, it uses twice as much frames. The \emph{forward differences} (FD) and \emph{backward
  differences} (BD) are less accurate than the \emph{simple differences} (SD) method to
reduce the variance of correlated terms. As concerns the offset term, the \emph{simple
  differences} (SD) method is therefore the method of choice. However, when the effective
number of calibration frames is much larger than the degrees of freedom in the commands,
that is to say when $ξ^\Tag{D}_t \gg \Nact$, all methods but the \emph{push-pull} (PP) one
tend to be equally good. Again, the conclusions are valid for both zonal or modal
calibration and probing.

\section{Simple updating rules}
\label{sec:updating}

In this section, we are interested in deriving rules to compute the quantities
($\mean{d}_t$, $\mean{u}_t$, $A_t$ and $B_t$) needed by our estimators of the affine model
components. For frequent updating of the model, these computations should be of low
computational complexity. We show here that a suitable choice of the weights $ρ_{t,t'}$
leads to simple rank-1 updates for the integrated matrices $A_t$ and $B_t$ but also for
$A_t^{-1}$ and for the estimator $\estim{G}_t$ of the interaction matrix.

\subsection{Weights based on forgetting factors}
\label{sec:forgetting-factors}

In the weighted least squares, the weights shall be set to $ρ_{t,t'} = 0$ for all $t$ and
$t'$ when the needed WFS data frame(s) or DM probe command(s) are missing for calibration
frame $t'$; otherwise, $ρ_{t,t'} > 0$ shall hold. Looking at Eq.~\eqref{eq:G_t:WLS}, the
weighted least squares estimator $\estim{G}_t$ is unchanged if, for all $t' ≤ t$, the
weights $ρ_{t,t'}$ are scaled by a positive factor that may only depend on $t$. This
invariance also holds for the estimator $\estim{b}_t$ given by Eq.~\eqref{eq:b_t} and for
the weighted averages, $\mean{d}_t$, $\mean{u}_t$, and $\mean{n}_t$. Hence, with no loss of
generality, we choose to impose that:
\begin{equation}
  \label{eq::rho(t,t)}
  ρ_{t,t} \equiv  \begin{cases}
    1 & \text{if $t$-th calibration frame is valid,}\\
    0 & \text{otherwise.}\\
  \end{cases}
\end{equation}
Furthermore, for the scalar weights to decrease with the age of the calibration data, we
assume the following recurrence:
\begin{equation}
  \label{eq:rho-recurrence}
  ρ_{t,t'} = γ_t\,ρ_{t-1,t'} \quad (\forall t, t' ≤ t - 1)
\end{equation}
for chosen \emph{forgetting factors} $γ_t \in [0,1]$ ($\forall t$). It follows from this
recurrence that the sum of weights defined in Eq.~\eqref{eq:sum-of-weights} are initialized
and updated by:
\begin{equation}
  η_t = \begin{cases}
    0 & \text{if $t < 1$,}\\
    ρ_{t,t} + γ_t\,η_{t-1} & \text{if $t ≥ 1$.}\\
  \end{cases}
  \label{eq:eta-recurrence}
\end{equation}

\subsection{Updating rules for direct calibration}

While $η_t = 0$, no valid calibration frames are available. For the first valid calibration
frame $t = t₁$ (\ie, when $η_{t₁} > 0$ and $η_{t₁-1} = 0$), the needed quantities are given by:
\begin{subequations}
  \label{eq:initial-quantities}
  \begin{align}
    \label{eq:mean(d)_t:init}
    \mean{d}_{t₁} &= d_{t₁},\\
    \label{eq:mean(u)_t:init}
    \mean{u}_{t₁} &= u_{t₁ - ℓ},\\
    \label{eq:A^D_t:init}
    A^\Tag{D}_{t₁} &= 0,\\
    \label{eq:B^D_t:init}
    B^\Tag{D}_{t₁} &= 0.
  \end{align}
\end{subequations}

For subsequent calibration frames ($t > t₁$ and thus $η_t > 0$) and accounting for the
forgetting factors, $\mean{d}_t$ and $\mean{u}_t$ respectively defined in
Eqs.~\eqref{eq:mean(d)_t} and \eqref{eq:mean(u)_t} can be updated by:
\begin{align}
  \mean{d}_t &= α_t\,d_t + \Paren*{1 - α_t}\,\mean{d}_{t-1},
  \label{eq:mean(d)_t:update}\\
  \mean{u}_t &= α_t\,u_{t-ℓ} + \Paren*{1 - α_t}\,\mean{u}_{t-1},
  \label{eq:mean(u)_t:update}
\end{align}
where:
\begin{equation}
  \label{eq:alpha}
  α_t = ρ_{t,t}/η_t.
\end{equation}

In Appendix~\ref{sec:updating-cross-products}, we derive the updating rules for
$A^\Tag{D}_t$ and $B^\Tag{D}_t$ defined in Eqs.~\eqref{eq:A^D_t} and \eqref{eq:B^D_t} when
the scalar weights are driven by forgetting factors. These rules, given by
Eqs.~\eqref{eq:A^D_t:update:simplified} and \eqref{eq:B^D_t:update:simplified}, are recalled
here for convenience:
\begin{align}
  A^\Tag{D}_t
  &= γ_t\,A^\Tag{D}_{t-1} + ρ_{t,t}\,(1 - α_t)\,
    \Paren{u_{t-ℓ} - \mean{u}_{t-1}}\,
    \Paren{u_{t-ℓ} - \mean{u}_{t-1}}\T,
    \label{eq:A^D_t:update}\\
  B^\Tag{D}_t
  &= γ_t\,B^\Tag{D}_{t-1} + ρ_{t,t}\,(1 - α_t)\,
    \Paren{d_t - \mean{d}_{t-1}}\,
    \Paren{u_{t-ℓ} - \mean{u}_{t-1}}\T.
    \label{eq:B^D_t:update}
\end{align}

By avoiding summation over $t' ≤ t$, all these updating rules considerably reduce the
computational burden to obtain the quantities $η_t$, $\mean{d}_t$, $\mean{u}_t$,
$A^\Tag{D}_t$, and $B^\Tag{D}_t$. All these rules are applicable for any $t ≥ 1$, if all
updated quantities are initially set to zero. With these rules, it is not needed to
explicitly apply Eqs.~\eqref{eq:initial-quantities}.

With the added flexibility introduced by the forgetting factors, the updating rules of
$\mean{d}_t$ and $\mean{u}_t$ in Eqs.~\eqref{eq:mean(d)_t:update} and
\eqref{eq:mean(u)_t:update} and the rank-1 updating rules of $A^\Tag{D}_t$ and $B^\Tag{D}_t$
in Eqs.~\eqref{eq:A^D_t:update} and \eqref{eq:B^D_t:update} generalize the equations derived
by \citet{Welford-1962-corrected_sums} for computing the sample mean and variance in online
statistics\footnote{The results of \citet{Welford-1962-corrected_sums} are retrieved by
  taking $γ_t \equiv 1$ which yields $η_t = t$, that is the number of samples, and
  $α_t = 1/t$.}. Moreover, \citet{Chan+1983-computing_sample_variance} have shown that such
one-pass updating rules are robust to catastrophic rounding errors provided they involve
\emph{shifted quantities}\footnote{that is variables approximately centered by subtracting
  an estimate of their mean} as it is the case in the formulae that we obtained.

\subsection{Updating rules for differential calibration}

We now consider updating the matrices $A^\Tag{dif}_t$ and $B^\Tag{dif}_t$ needed by
\emph{differential} calibration methods as defined by Eqs.~\eqref{eq:A^dif_t} and
\eqref{eq:B^dif_t} but assuming that the scalar weights are driven by forgetting factors.
These matrices are the weighted averages of the cross products of $x^\Dif_{t'}$ and
$y^\Dif_{t'}$ (for $t' ≤ t$). Since these terms do not depend on $t$, the updating rules are
much easier to derive compared to the \emph{direct} calibration method and we find:
\begin{align}
  A^\Dif_t
  &= γ_t\,A^\Dif_{t-1} + ρ_{t,t}\,x^\Dif_t\,\Paren[\big]{x^\Dif_t}\T,
    \label{eq:A_t^dif:update}\\
  B^\Dif_t
  &= γ_t\,B^\Dif_{t-1} + ρ_{t,t}\,y^\Dif_t\,\Paren[\big]{x^\Dif_t}\T,
    \label{eq:B_t^dif:update}
\end{align}
where $ρ_{t,t} \in \Brace{0,1}$ is given by Eq.~\eqref{eq::rho(t,t)}. In addition to this
rule, for the \emph{push-pull} (PP) method, $ρ_{t,t} = 0$ if $u_{t-ℓ} \not= -u_{t-ℓ-1}$.

\subsection{Updating the interaction matrix}
\label{sec:updating-imat}

As a result of using forgetting factors to set the calibration weights and whatever the
calibration method, the two matrices $A_t$ and $B_t$ can be updated by the following simple
recurrences:
\begin{align}
  A_t &= γ_t\,A_{t-1} + μ_t\,x_t\,x_t\T, \label{eq:A_t:update}\\
  B_t &= γ_t\,B_{t-1} + μ_t\,y_t\,x_t\T. \label{eq:B_t:update}
\end{align}
with
\begin{align}
  \label{eq:mu_t}
  μ_t &=
  \begin{cases}
    ρ_{t,t}\,(1 - α_t) & \text{for the \emph{direct} method}\\
    ρ_{t,t} & \text{for \emph{differential} methods}\\
  \end{cases}\\
  \label{eq:x_t}
  x_t &=
  \begin{cases}
    u_{t-ℓ} - \mean{u}_{t-1} & \text{for the \emph{direct} method}\\
    x^\Dif_t & \text{for \emph{differential} methods}\\
  \end{cases}\\
  \label{eq:y_t}
  y_t &=
  \begin{cases}
    d_t - \mean{d}_{t-1} & \text{for the \emph{direct} method}\\
    y^\Dif_t & \text{for \emph{differential} methods.}\\
  \end{cases}
\end{align}

Thanks to these simple recurrences, updating $A_t$ and $B_t$ has a constant computational
cost\footnote{$\sim4\,\Nact^2$ and $\sim4\,\Nact\,\Ndat$ operations to update $A_t$ and
  $B_t$ respectively} whatever the length of the calibration sequence. However, for
real-time applications, the bottleneck is the computation\footnote{$\sim(1/2)\,\Nact^3$
  operations for the Cholesky decomposition of $A_t$ plus $\sim2\,\Ndat\,\Nact^2$
  operations to compute the matrix product} of $\estim{G}_t = B_t\,A_t^{-1}$, according to
Eq.~\eqref{eq:G_t}, which prevents to continuously update the interaction matrix except
for the smallest AO systems\footnote{for a $\sim 70\,\text{Gflops}$ CPU, updating the
  interaction matrix at $\sim 1\,\text{kHz}$ is only possible if the AO system has at most
  $\sim 400$ degrees of freedom which corresponds to a $20 \times 20$ DM}. Fortunately,
the rule in Eq.~\eqref{eq:A_t:update} is a rank-1 update and the numerical complexity of
updating the inverse of $A_t$ can be much reduced by applying the
Sherman--Morisson--Woodbury matrix identity:
\begin{equation}
  \label{eq:Sherman-Morisson-Woodbury}
  (A + U\,C\,V)^{-1} = A^{-1}
  - A^{-1}\,U\,(C^{-1} + V\,A^{-1}\,U)^{-1}\,V\,A^{-1}
\end{equation}
which holds for any invertible matrix $A$ and matrices $C$, $U$, and $V$ of suitable
sizes. Applying the Sherman--Morisson--Woodbury identity\footnote{with the substitutions
  $A \to γ_t\,A_{t-1}$, $C \to μ_t\,I_{1×1}$, $U \to x_t$, and $V \to x_t\T$ where
  $I_{1×1}$ denotes the $1×1$ matrix identity}, the inverse of $A_t$ can also be obtained
by a (scaled) rank-1 update:
\begin{equation}
  \label{eq:A_t^{-1}:update}
  A_t^{-1} = \frac{1}{γ_t}\,\Paren*{A_{t-1}^{-1} - λ_t\,v_t\,v_t\T}
\end{equation}
with:
\begin{align}
  \label{eq:v_t}
  v_t &= A_{t-1}^{-1}\,x_t, \\
  \label{eq:lambda_t}
  λ_t & = \frac{μ_t}{θ_t\,μ_t + γ_t}, \\
  \label{eq:alpha_t}
  θ_t &= x_t\T\,v_t.
\end{align}
Combining the rank-1 updating rules for $A_t^{-1}$ and $B_t$, given by
Eq.~\eqref{eq:A_t^{-1}:update} and Eq.~\eqref{eq:B_t:update}, and simplifying yields a
simple rank-1 updating formula for the estimator of the interaction matrix:
\begin{align}
  \estim{G}_t = \estim{G}_{t-1} + λ_t\,e_t\,v_t\T,
  \label{eq:G_t:update}
\end{align}
with:
\begin{align}
  \label{eq:e_t}
  e_t
  & = y_t - B_{t-1}\,v_t
    = y_t - \estim{G}_{t-1}\,x_t.
\end{align}
Similar updating formulae are found in the method of \emph{Recursive Least Squares}
\citep[RLS,][]{Hayes1996a} where $λ_t\,v_t$ in Eq.~\eqref{eq:G_t:update} is known as the
\emph{gain} while $e_t$ in Eq.~\eqref{eq:e_t} is known as the \emph{error}.

\begin{algorithm}
  \caption{Update interaction matrix $\estim{G}_t$.}
  \label{alg:update-G}
  \KwIn{%
    $A_{t-1}^{-1} \in ℝ^{\Nact × \Nact}$,
    $\estim{G}_{t-1} \in ℝ^{\Ndat × \Nact}$,
    $y_t \in ℝ^{\Ndat}$, $x_t \in ℝ^{\Nact}$,
    $γ_t \in [0, 1]$, and $μ_t ≥ 0$.}

  \KwOut{Updated matrices $A_t^{-1}$ and $\estim{G}_t$.}

  \BlankLine
  $v_t = A_{t-1}^{-1}\,x_t$ \Comment*{see Eq.~\eqref{eq:v_t}}
  $θ_t = x_t\T\,v_t$ \Comment*{see Eq.~\eqref{eq:alpha_t}}
  $λ_t = \frac{μ_t}{θ_t\,μ_t + γ_t}$ \Comment*{see Eq.~\eqref{eq:lambda_t}}
  $A_t^{-1} = (A_{t-1}^{-1} - λ_t\,v_t\,v_t\T)/γ_t$ \Comment*{see Eq.~\eqref{eq:A_t^{-1}:update}}
  $e_t = y_t - \estim{G}_{t-1}\,x_t$ \Comment*{see Eq.~\eqref{eq:e_t}}
  $\estim{G}_t = \estim{G}_{t-1} + λ_t\,e_t\,v_t\T$ \Comment*{see Eq.~\eqref{eq:G_t:update}}
\end{algorithm}

Algorithm~\ref{alg:update-G} summarizes the updating of the estimator $\estim{G}_t$ of the
quasi-static interaction matrix. It is worth noting that there are no needs to store and
update $A_t$ and $B_t$, only $A_t^{-1}$ and $\estim{G}_t$ are required. Moreover,
$A_t^{-1}$ and $\estim{G}_t$ may be updated in-place to reduce storage. Looking at
Eq.~\eqref{eq:e_t}, there are 2 possibilities to compute the \emph{error} $e_t$, we choose
the one which does not require the matrix $B_{t-1}$ to avoid storing and updating it.
Another motivation for this choice is that using the expression that involves the current
estimate of the interaction matrix may help to auto-correct numerical errors and avoid
their propagation. The computational burden of Algorithm~\ref{alg:update-G} is
$\sim (5\,\Nact + 4\,\Ndat)\,\Nact$ operations to update $\estim{G}_t$ and $A_t^{-1}$.
With the same $\sim 70\,\text{Gflops}$ CPU as before, all these can be done at
$\sim 1\,\text{kHz}$ for AO systems with more than $\sim 2300$ degrees of freedom which
corresponds to a $\sim 50 \times 50$ DM.

%
%
%

The recurrence implemented by Algorithm~\ref{alg:update-G} can be initialized by a null
interaction matrix and by the expectation of the mean cross product of the calibration
probes, that is $\estim{G}_0 = 0$ and $A_0 = 𝔼_{U}\Paren[\big]{x_{t,t'}\,x_{t,t'}\T}$. The
latter is diagonal, see Eqs.~\eqref{eq:A_t=E(A_t)} to \eqref{eq:omega_j}, and trivial to
invert. In Appendix~\ref{sec:update-model}, we provide Algorithm~\ref{alg:update-model} to
update all model parameters for any of the calibration methods under consideration.

\section{Validation with the THEMIS AO system}
\label{sec:results}

This section presents the results of applying the proposed calibration methods to the AO
system of THEMIS, a 90-cm solar telescope located at the Observatorio del Teide in Tenerife
Canary Island \citep[Spain,][]{Gelly_2016}. THEMIS AO system is composed of a $10\times 10$
Shack-Hartmann wavefront sensor with 76 active sub-apertures of 10-arcsecond field-of-view
each and a $11\times 11$ deformable mirror with 97 active actuators. The wavefront sensing
is based on a \emph{self-referenced generalized linearized matched filter}
\citep{ThiebautEtAl2018a} applied to the images acquired by WFS camera. THEMIS AO is able to
work on the Sun granulation as well as on a small resolved object like Mercury.

In the current version of the THEMIS AO control system \citep{ThiebautEtAl2022a}, only the
calibrated interaction matrix is used. The offset term of the affine model is so far
neglected to compute the AO correction, but it shall be considered in later improvements of
the control. Since 2024, the interaction matrix calibration can be done at the AO loop
framerate (up to $1\,250\,\text{Hz}$), directly on the Sun granulation, by the various
methods presented in this paper. We use centered independent random binomial probe commands
$u_{t'-ℓ}$ of uniform variance $ϵ^2 > 0$ for all actuators and a zonal command vector.
Properties in Eq.~\eqref{eq:probe-expectation}--\eqref{eq:probe-covariance-pp} therefore
hold with $u_{t'-ℓ}[j] = +ϵ$ or $-ϵ$ each with the same probability and are used to build
either a \emph{push-pull} (see Sec.~\ref{sec:push-pull-calibration}) or a \emph{push-only}
sequence, \ie with completely independent successive probe commands. For all presented
results except in Appendix~\ref{sec:working-conditions}, the forgetting factor is $γ_t = 1$
whatever $t$, \ie\ all acquired frames have the same importance which amounts to assuming
that the conditions are stable during the $60\,\text{s}$ (\ie, $60\,000$ frames) of
acquisition of each calibration sequence.

\subsection{Open-loop calibration}
\label{sec:open-loop-results}

\begin{figure*}
  \centering
  \includegraphics[width=0.55\textwidth]{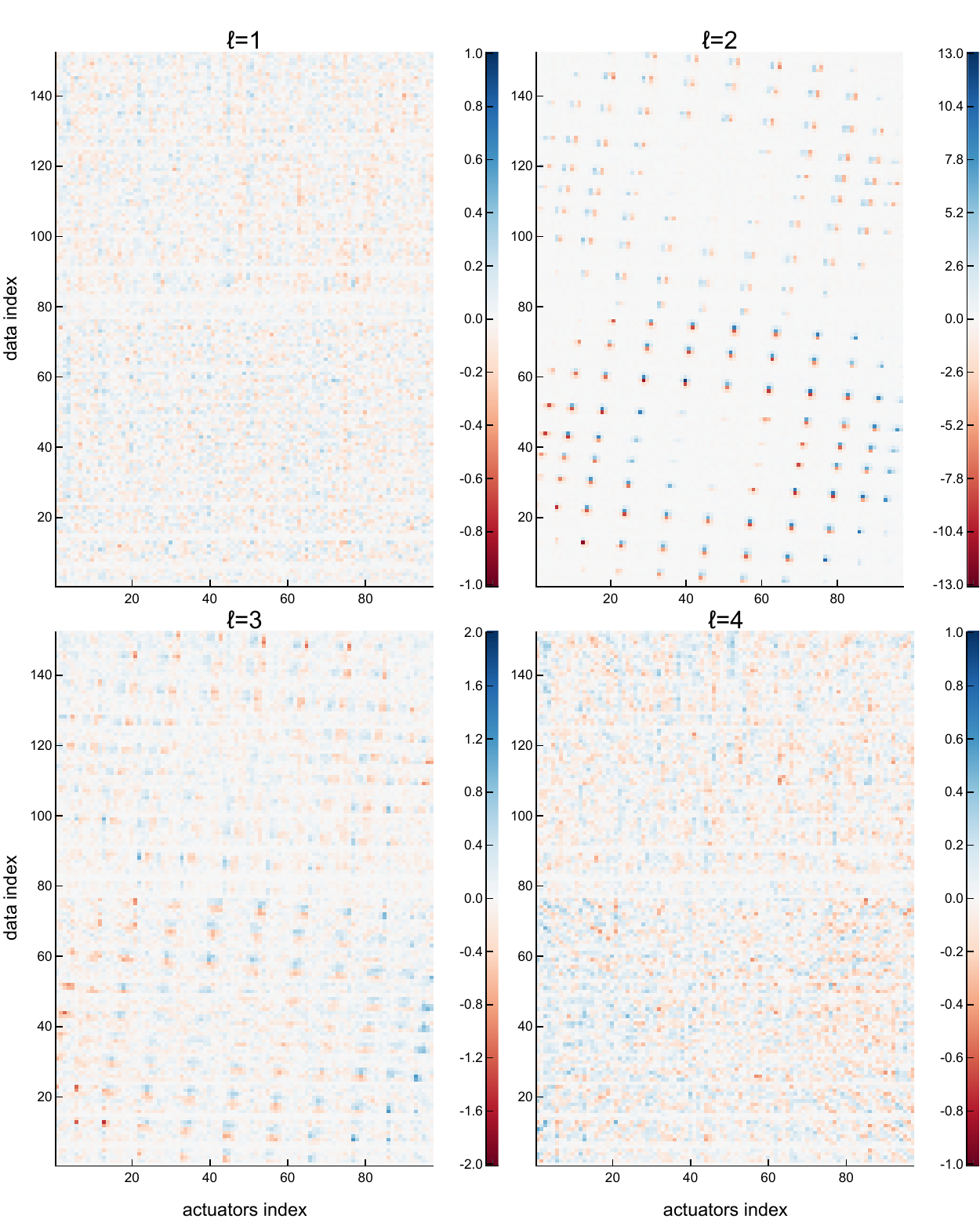}
  \caption[OL Direct vs lag]{\label{fig:G-OL-direct-vs-lag} Interaction matrix
    $\estim{G}_{60\,000}$ for various assumed lags $\ell$ and estimated on-sky by the
    open-loop \emph{direct} (D) calibration method with probe commands of 5\% of the
    maximum allowed for the DM commands and after recording 60\,000 WFS frames. Vertical
    axis is for measured wavefront slopes (with $\Ndat = 2×76$) and horizontal axis is for
    DM actuators (with $\Nact = 97$).}
\end{figure*}

We first present the calibration on the Sun in open-loop and based on $60\,000$ recorded
DM probes and WFS frames at 1\,kHz (60\,s). Two sets of such telemetry are recorded: one in
\emph{push-pull} and the other in \emph{push-only} mode. For each set of telemetry, the
estimated interaction matrix is analyzed depending on the chosen calibration method. In
open-loop calibration, the amplitude of the probe commands is chosen to be $ϵ\simeq5\%$ of
the maximum allowed for the DM commands, which corresponds to $\approx 250$\,nm.

\subsubsection{Estimation of the lag}
\label{sec:lag-estimation}

\begin{figure}
  \centering
  \includegraphics[width=0.9\columnwidth, height=7cm]{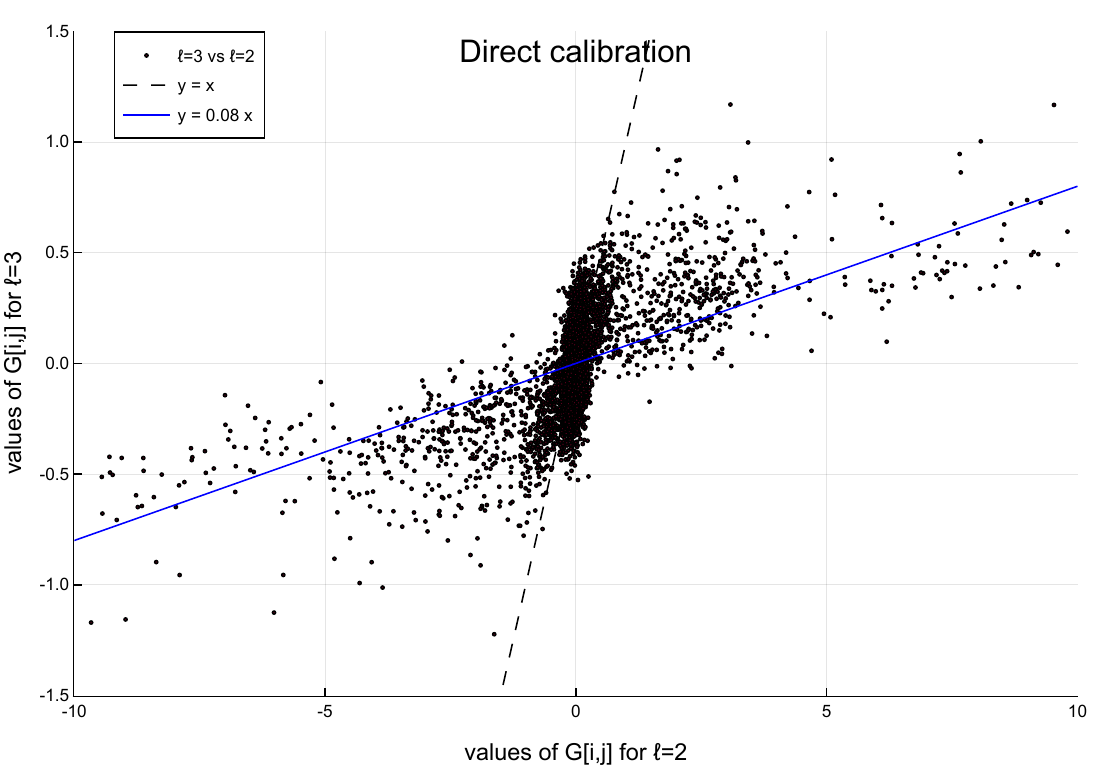}
  \caption[OL Direct l2 vs l3]{\label{fig:G-OL-direct-l2-vs-l3} Values of each element of
    the matrix $\estim{G}_{60\,000}^{\ell=3}$ is plotted as a function of the value of the
    same element in $\estim{G}_{60\,000}^{\ell=2}$, hence for two values of the lag $ℓ$.
    Direct calibration is applied. The black and blue lines stand for $y=x$ and $y=0.08\,x$
    respectively.}
\end{figure}
To determine the lag $\ell$ of the THEMIS AO system which is not exactly known, we applied
the calibration method assuming different integer lag values.
Figure~\ref{fig:G-OL-direct-vs-lag} shows the interaction matrix $\estim{G}_{60\,000}$ as
estimated by the \emph{direct} (D) calibration method for $\ell \in \Brace{1,2,3,4}$. The
coefficients of the estimated interaction matrix are most significant for a lag value
$\ell=2$ which confirms that, as many existing AO systems, THEMIS AO has a lag of
approximately 2 frames. For $\ell=1$ and $\ell=4$, the estimated coefficients of the
interaction matrix are non-significant and dominated by the noise. For $\ell=3$, weakly
significant coefficients can be observed. Fig.~\ref{fig:G-OL-direct-l2-vs-l3} plots the
coefficients of the calibrated interaction matrix with a lag $\ell = 3$ versus those with
a lag $\ell = 2$. We observe that the most significant coefficients of the
matrix for which we assume $\ell=3$ are approximately 8\% of their counterpart in the
matrix for which we assume $\ell=2$. This reveals that the wavefront sensor is actually
sensitive to two successive commands and thus the delay is not exactly 2 frames.
%
An extension of the present work to account for fractional delays and with a multi-frame
model is currently under study and will be the subject of another publication. As
mentioned by \citet{EspositoEtAl2006b}, we observe that the estimate of the lag impacts
the estimation of the interaction matrix on sky.
%
In the remaining, we compare the interaction matrices obtained by the different
calibration methods assuming a lag value $\ell=2$, thus rounding the delay to 2 frames.

\subsubsection{Comparison of calibration methods in open-loop}
\label{sec:open-loop-comparison}


\begin{figure}[!]
  \centering
  \includegraphics[width=60mm]{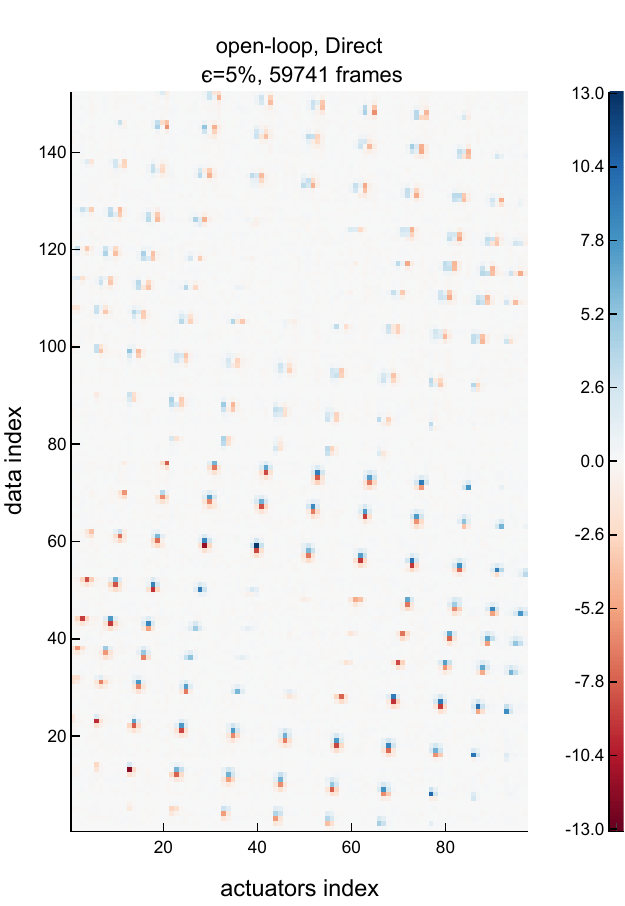}%
  \\
  \centering
  \includegraphics[width=80mm]{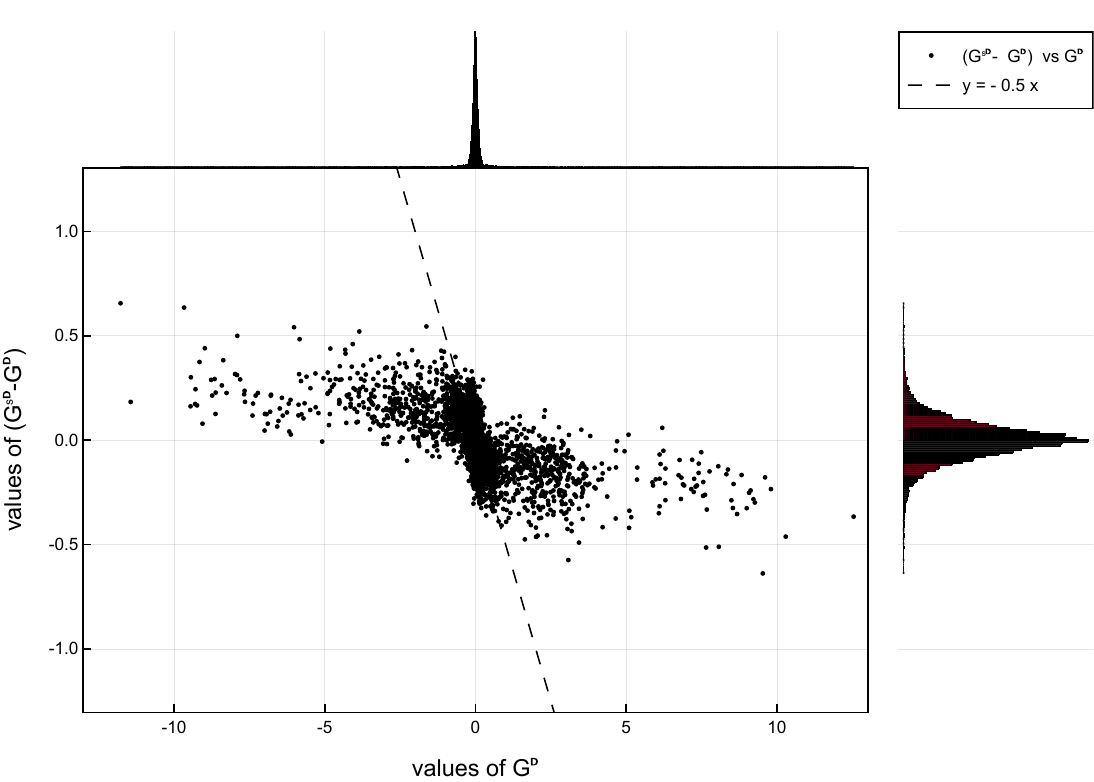}
  \caption{Comparison of interaction matrices calibrated in open-loop on a sequence of
    \emph{push-only} probe commands with $5\%$ amplitude. Top: Interaction matrix
    $\estim{G}_{t}^{\Tag{D}}$ calibrated by the \emph{direct} (D) method. Bottom:
    Coefficients of $\estim{G}_{t}^{\Tag{SD}} - \estim{G}_{t}^{\Tag{D}}$ \emph{vs.} those of
    $\estim{G}_{t}^{\Tag{D}}$ with $\estim{G}_{t}^{\Tag{SD}}$ calibrated by the \emph{simple
      difference} (SD) method. Calibration sequence counts $60\,000$ frames and results in
    $59\,741$ and $59\,394$ calibration frames respectively for the D and SD methods.}
  \label{fig:open-loop:G:maps-p}
\end{figure}

\begin{figure}[!]
  \centering
  \includegraphics[width=60mm]{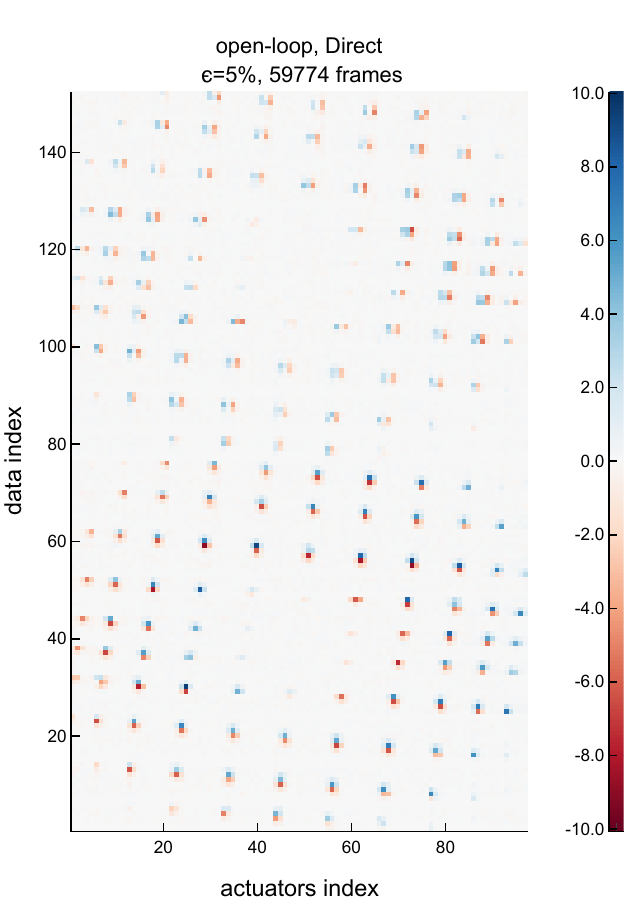}%
  \\
  \centering
  \includegraphics[width=80mm]{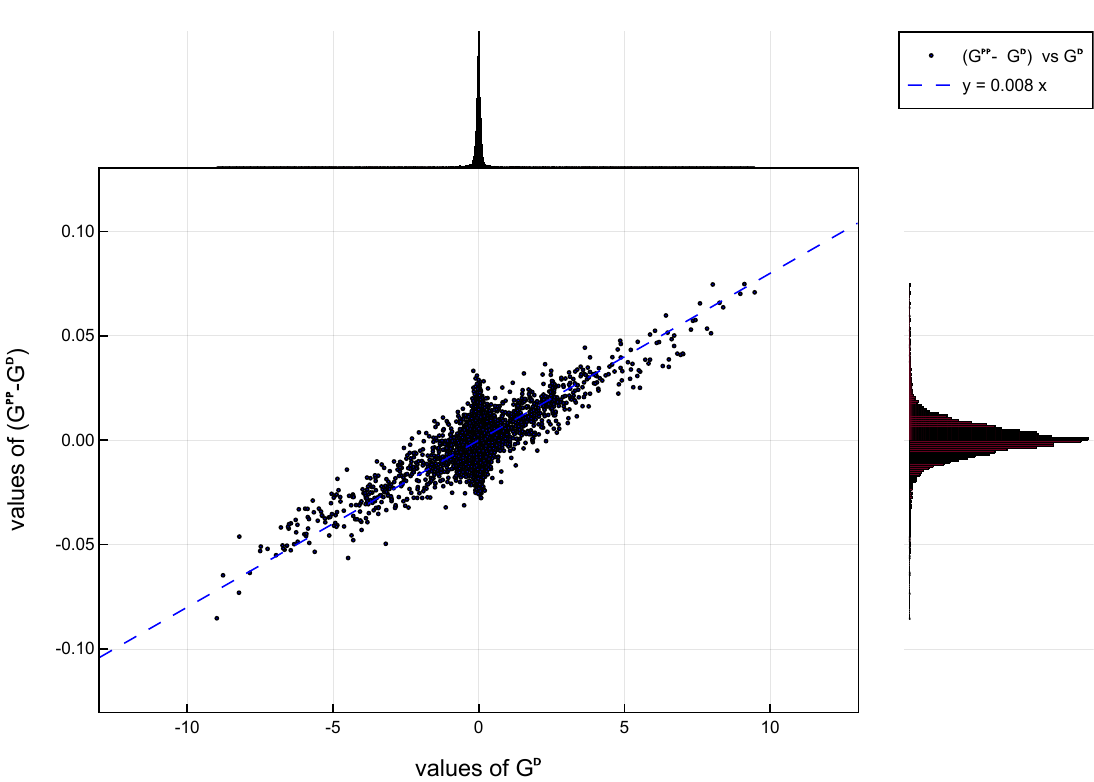}
  \\
  \centering
  \includegraphics[width=80mm]{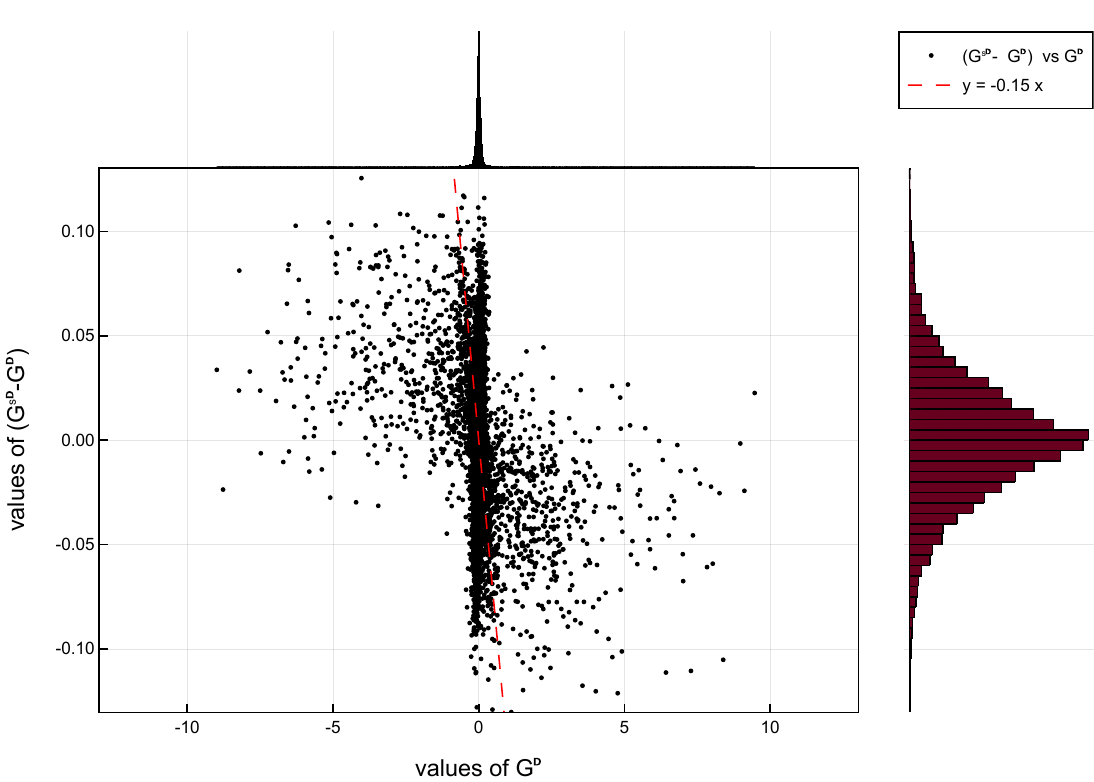}
  \caption{Comparison of interaction matrices calibrated in open-loop on a sequence of
    \emph{push-pull} probe commands with $5\%$ amplitude. Top: Interaction matrix
    $\estim{G}_{t}^{\Tag{D}}$ calibrated by the \emph{direct} (D) method. Middle:
    Coefficients of $\estim{G}_{t}^{\Tag{PP}} - \estim{G}_{t}^{\Tag{D}}$ \emph{vs.} those of
    $\estim{G}_{t}^{\Tag{D}}$ with $\estim{G}_{t}^{\Tag{PP}}$ calibrated by the
    \emph{push-pull} (PP) method. Bottom: Coefficients of
    $\estim{G}_{t}^{\Tag{SD}} - \estim{G}_{t}^{\Tag{D}}$ \emph{vs.} those of
    $\estim{G}_{t}^{\Tag{D}}$ with $\estim{G}_{t}^{\Tag{SD}}$ calibrated by the \emph{simple
      difference} (SD) method. Calibration sequence counts $60\,000$ frames and results in
    $59\,774$ and $29\,565$ calibration frames respectively for the D and PP methods.}
  \label{fig:open-loop:G:maps-pp}
\end{figure}

The top of Fig.~\ref{fig:open-loop:G:maps-p} shows the interaction matrix calibrated by
the \emph{direct} (D) method on $60\,000$ frames of telemetry taken in \emph{push-only}
mode. The interaction matrix obtained by the \emph{simple differences} (SD) on the same
sequence is not visually different and thus not plotted. To compare them quantitatively
instead, the difference between the coefficients of the two matrices
$(\estim{G}_{t}^{\Tag{SD}} -\estim{G}_{t}^{\Tag{D}})$ is plotted as a function of the
coefficients of $\estim{G}_{t}^{\Tag{D}} $, at the bottom of
Fig.~\ref{fig:open-loop:G:maps-p}. We first note that the discrepancies between the
matrices are at most one order of magnitude smaller than the maximum coefficients of the
reference matrix. The position of the points with respect to the horizontal axis reveals
that the significant coefficients of the matrix obtained by the \emph{direct} method tend
to be slightly larger in amplitude than the ones in the matrix obtained by \emph{simple
  differences}. The tilted orientation of the central part of the cloud evidences that the
background part of the matrix is less noisy for the \emph{simple difference} method, since
these coefficients expected to be zero have almost systematically smaller amplitudes. The
actual numbers of calibration frames are $59\,741$ and $59\,394$ respectively for
\emph{direct} and \emph{simple differences} methods. They are both slightly smaller than
$60\,000$ due to lost frames (the task in charge of recording the telemetry having a low
priority in the AO control system). Although not displayed here, the finally estimated
matrices with \emph{forward differences} (FD) and \emph{backward differences} (BD) are
also similar to the matrix in top of Fig.~\ref{fig:open-loop:G:maps-p}.

The interaction matrix displayed on top of Fig.~\ref{fig:open-loop:G:maps-pp} is obtained
from the \emph{push-pull} sequence of calibration recorded a couple of minutes later. This
interaction matrix is estimated applying the \emph{direct} (D) method on this
\emph{push-pull} calibration telemetry. We remind here that the \emph{direct} (D) method
estimates the same matrix as the \emph{push-pull} method when applied on a \emph{push-pull}
sequence (see Sec.~\ref{sec:push-pull-calibration}). In practice here,
Algorithm~\ref{alg:update-model} has been used so the estimated matrices may differ because
the update of $G$ is applied only every two frames for the \emph{push-pull} method. The
difference in the obtained coefficients is plotted in blue as a function of the coefficients
of the matrix obtained with the \emph{direct} method on the same sequence in the middle
graph of Fig.~\ref{fig:open-loop:G:maps-pp}. The differences appear negligible: their
amplitude is less than 1\% of the maximum amplitude.

In this experiment, we finally also consider applying the \emph{simple differences} method
to the \emph{push-pull} sequence of calibration, even if no prediction of the
mean-squared-error (MSE) of the interaction matrix estimation has been derived for it. The
coefficients of the obtained matrix are again similar to the ones of the \emph{direct}
method, and their discrepancy with respect to it are represented in the bottom plot of
Fig.~\ref{fig:open-loop:G:maps-pp}. The differences in amplitude appear larger than for
the \emph{push-pull} method, but smaller than in Fig.~\ref{fig:open-loop:G:maps-p}.
Considering the position of the points with respect to the horizontal axis, we can observe
about the significant coefficients in the matrix (not the noisy background), that
\emph{push-pull} tends to increase their amplitude compared to the \emph{direct} method,
while the \emph{simple differences} method tends to reduce it. The differences remains
however lower than 1\% of the maximum amplitude.

These results confirm that \emph{direct} (D), \emph{simple difference} (SD) and
\emph{push-pull} (PP) methods estimate similar interaction matrices when applied on a same
calibration sequence. The small anti-correlation of the points in
Fig.~\ref{fig:open-loop:G:maps-p} reveals that the estimation errors are smaller with the
\emph{simple differences} (SD) method. The amplitude of the significant coefficients are
however different when applied on the \emph{push-pull} calibration sequence with respect to
the ones obtained in the \emph{push-only} one (see top of Figs.~\ref{fig:open-loop:G:maps-p}
and \ref{fig:open-loop:G:maps-pp}). This is due to different working conditions with a
non-linear wavefront sensor. This is confirmed by the analysis in
Appendix~\ref{sec:working-conditions}, evidencing significant evolution of the working
conditions below the minute scale.

\begin{figure}
  \centering \includegraphics[width=0.9\columnwidth]{%
    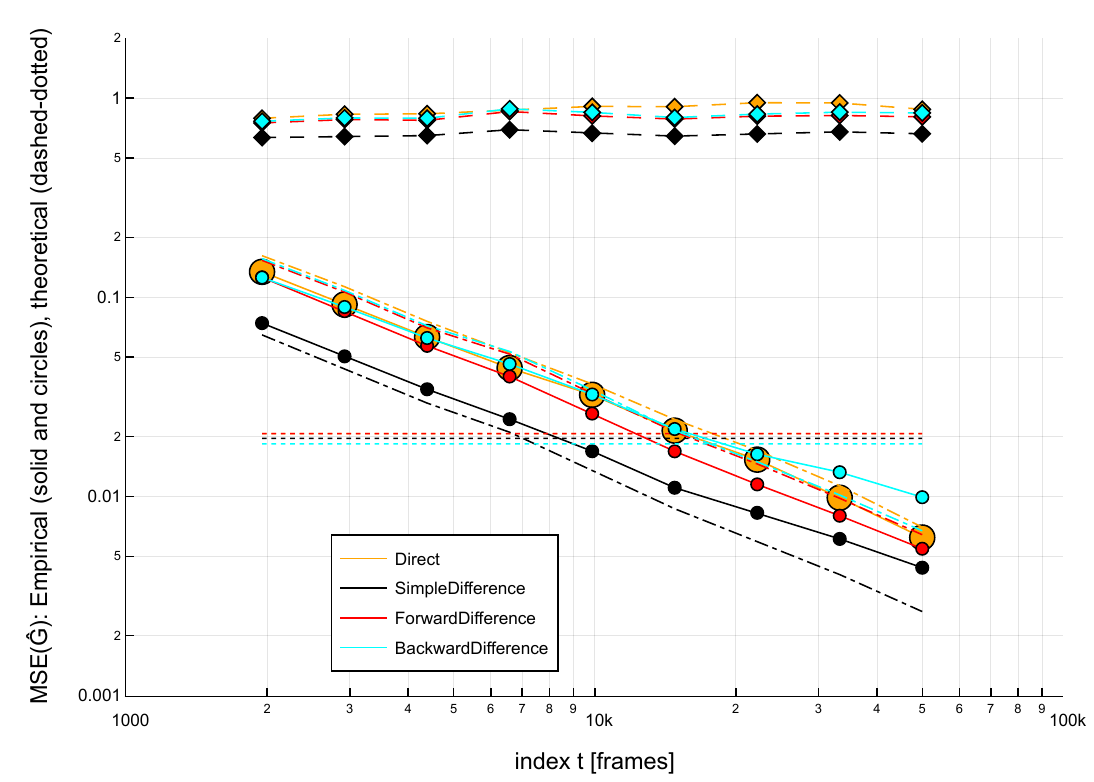}
  \caption{MSEs of the interaction matrix calibrated on-sky from open-loop
    \emph{push-only} calibration sequence as a function of its length $t$ (in frames).
    Methods: \emph{direct} (D in orange), \emph{simple differences} (SD in blue),
    \emph{forward differences} (FD in red) and \emph{backward differences} (BD in cyan).
    Solid lines with circles: empirical MSE of $\estim{G}_t$. Circles for \emph{direct}
    (D) are larger just to be noticeable. Dashed lines with diamonds: empirical variance
    of the calibration nuisance, $\estim{χ}_t$, given by Eq.~\eqref{eq:estim(chi)_t}.
    Dashed-dotted lines: MSE computed according to the theoretical expression in
    Eq.~\eqref{eq:MSE(G)}. Dotted lines: required precision of $2\%$ of the magnitude of
    the most significant coefficients of $\estim{G}_t$. The calibration data sequence is
    the same as for Fig.~\ref{fig:open-loop:G:maps-p}.}
  \label{fig:open-loop-MSEs-p}
\end{figure}

\begin{figure}
  \centering
  \centering \includegraphics[width=0.9\columnwidth]{%
    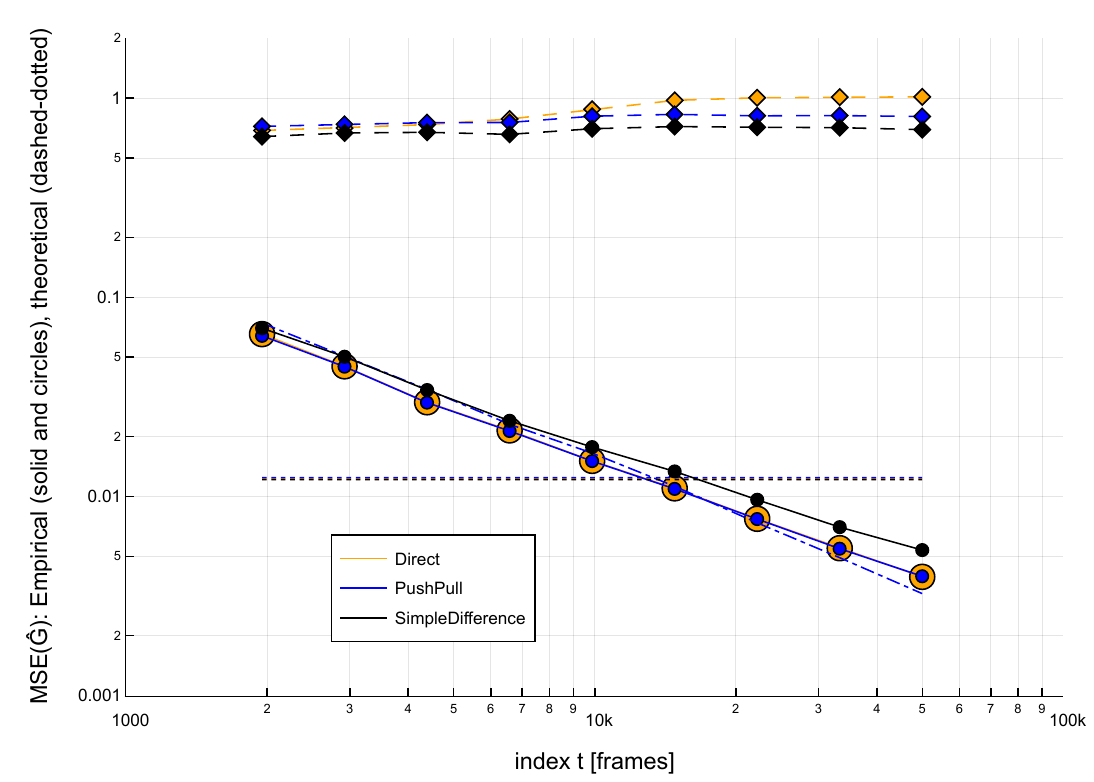} 
  \caption{Same as in Fig.~\ref{fig:open-loop-MSEs-p} but in open-loop and for a sequence of
    \emph{push-pull} probe commands, to compare the \emph{direct} (D in orange),
    \emph{push-pull} (PP in blue), and \emph{simple differences} (SD in black) applied on
    such \emph{push-pull} sequence. The calibration data are the same as for
    Fig.~\ref{fig:open-loop:G:maps-pp}.}
  \label{fig:open-loop-MSEs-pp}
\end{figure}

In order to compare the accuracy of the estimators $\estim{G}_t$ by the different
calibration methods and assert the validity of our theoretical predictions in
Sect.~\ref{sec:errors}, we need to estimate the \emph{mean squared errors} (MSEs) of these
estimators and the corresponding variance $χ_i$ of the calibration nuisance $z_{t,t'}[i]$.
The latter is to evaluate the theoretical MSEs according to Eq.~\eqref{eq:MSE(G)}. To
empirically estimate the MSE, we take advantage of the sparsity of the interaction matrix
$G_t$. Indeed, for an AO system with a Shack-Hartmann WFS and a DM with compact influence
functions, most coefficients of the zonal interaction matrix should be equal to zero. Hence,
an empirical estimator of the MSE of $\estim{G}_t$ (averaged over its coefficients) is given
by the variance of the centered Gaussian that best fit the histogram of the least
significant coefficients of $\estim{G}_t$, say those in the range $[-5,+5]$ here. The values
of this empirical MSE for the different methods in open-loop and as a function of the length
$t$ of the calibration sequence are represented as solid lines with filled circles in
Figs.~\ref{fig:open-loop-MSEs-p} and \ref{fig:open-loop-MSEs-pp}. To estimate empirically
the variance $χ_i$ of the calibration nuisance $z_{t,t'}[i]$, we use interaction matrix
$\estim{G}_\Tag{last}$ calibrated on the complete sequence (hence, with the best SNR under
stable conditions) to approximate the nuisance by:
\begin{equation}
  \label{eq:estim(z)_(t,t')}
  \estim{z}_{t,t'} = y_{t,t'} - \estim{G}_\Tag{last}\,x_{t,t'}
\end{equation}
which follows from Eq.~\eqref{eq:y_(t,t')}, and its variance by:
\begin{equation}
  \label{eq:estim(chi)_t}
  \estim{χ}_t = \Median_{i}\Paren[\big]{\,\estim{z}_{t,t'}[i]^2}
\end{equation}
according to Eq.~\eqref{eq:chi_i} except that the median averaging is taken for index $i$
for a sequence of length $t$, which assumes a stochastic behavior of the nuisance and
mitigates the influence of Shack-Hartmann measurements located close to the edges of the
pupil. The value of $\estim{χ}_t$ as a function of the sequence length $t$ and for the
different calibration methods is plotted as dashed lines with diamonds in
Figs.~\ref{fig:open-loop-MSEs-p} and \ref{fig:open-loop-MSEs-pp}. Finally, the
theoretical MSEs according to Eq.~\eqref{eq:MSE(G)} and for $χ_i = \estim{χ}_t$ are
plotted with dashed-dotted lines in the same figures.

We now analyze the trends observed in Figs.~\ref{fig:open-loop-MSEs-p} (\emph{push-only}
sequence) and \ref{fig:open-loop-MSEs-pp} (\emph{push-pull} sequence) for on-sky calibration
in open-loop. Considering the MSE of the estimators of the quasi-static interaction matrix
(solid lines with circles), the \emph{simple differences} (SD, in black) and
\emph{push-pull} (PP, in blue) are the best methods with similar MSEs, while the
\emph{forward differences} (FD, in red), and \emph{backward differences} (BD, in cyan) are
significantly worse than the \emph{simple differences} (SD) by a factor of $≈2$, as
predicted by our formulae. The \emph{simple differences} (SD) also performs significantly
better, by a factor of $≈2$, than the \emph{direct} method when applied on a
\emph{push-only} sequence. The choice of a \emph{push-pull} excitation however makes the
\emph{direct} method perform as well as the \emph{push-pull} method (as predicted), and
similarly to the \emph{simple differences} (SD).

Combining the expression in Eq.~\eqref{eq:MSE(G)} and our
empirical estimate, $\estim{χ}_t$, of the variance of the calibration nuisance, the
theoretical MSEs (dashed-dotted lines) well approximate the empirical ones (solid lines
with circles) which validates the hypotheses and approximations leading to
Eq.~\eqref{eq:MSE(G)} as well as our empirical estimate of the variance of the calibration
nuisance. The equivalence, in terms of the MSE of the interaction matrix, of the
\emph{push-pull} (PP) and \emph{simple differences} (SD) methods is predicted by the
theoretical expressions in Eqs.~\eqref{eq:MSE(G^PP)} and
\eqref{eq:MSE(G^SD)}. 
In Fig.~\ref{fig:open-loop-MSEs-p}, a
ratio of $≈2$ between the MSEs (solid lines with circles) by the \emph{direct} (D) method
and the \emph{simple differences} (SD) method can be observed.
According to Eqs.~\eqref{eq:MSE(G^D)} and \eqref{eq:MSE(G^PP)} or \eqref{eq:MSE(G^SD)},
this means that the coefficient of correlation between successive nuisances in the WFS
frames is $ζ_i ≈ 50\%$ for this experiment.

To conclude, for on-sky open-loop calibration of the model, the \emph{push-pull} (PP) and
\emph{simple differences} (SD) calibration methods provide interaction matrix estimators
with the least MSEs. In practice and with a $5\%$ probe excitation, these methods require
about $10\,000$ WFS frames ($10\,\text{s}$ at the AO loop frequency of $1\,\text{kHz}$) to
achieve a precision of $2\%$ of the magnitude of the most significant coefficients of the
interaction matrix\footnote{given by the median of the maximal absolute value of
  $\estim{G}_t$ for each measurement component $i$:
  $\Median_i\Paren*{\max_j \Abs[\big]{\estim{G}_t[i,j]}}$} (the dotted line in
Figs.~\ref{fig:open-loop-MSEs-p} and \ref{fig:open-loop-MSEs-pp}). These conclusions do
not depend on the number of degrees of freedom of the AO system, and thus apply to
larger telescopes.

\subsection{Closed-loop calibration}
\label{sec:results-cl}

\begin{figure}[t]
  \centering
  \includegraphics[width=60mm]{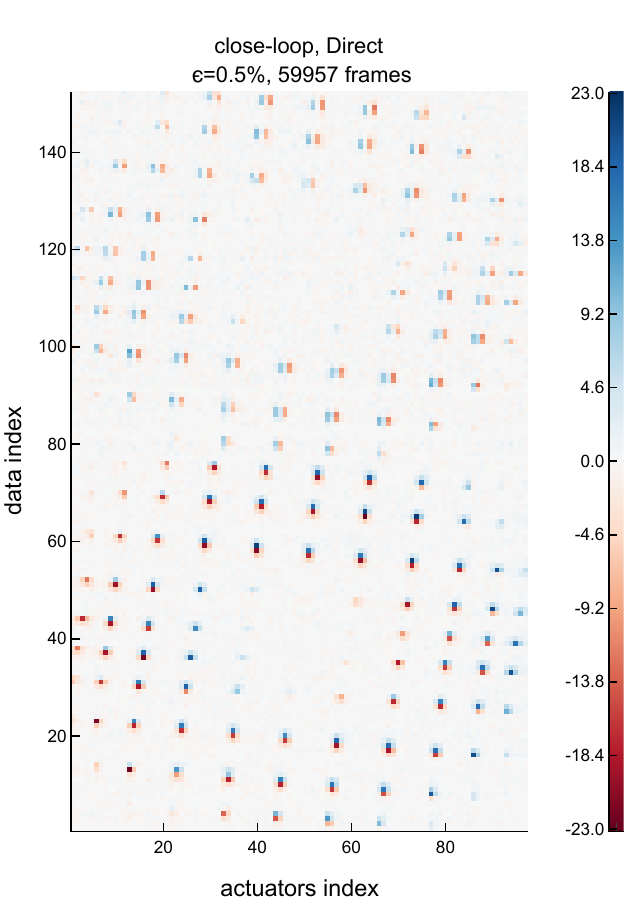}
  \\
  \centering
  \includegraphics[width=80mm]{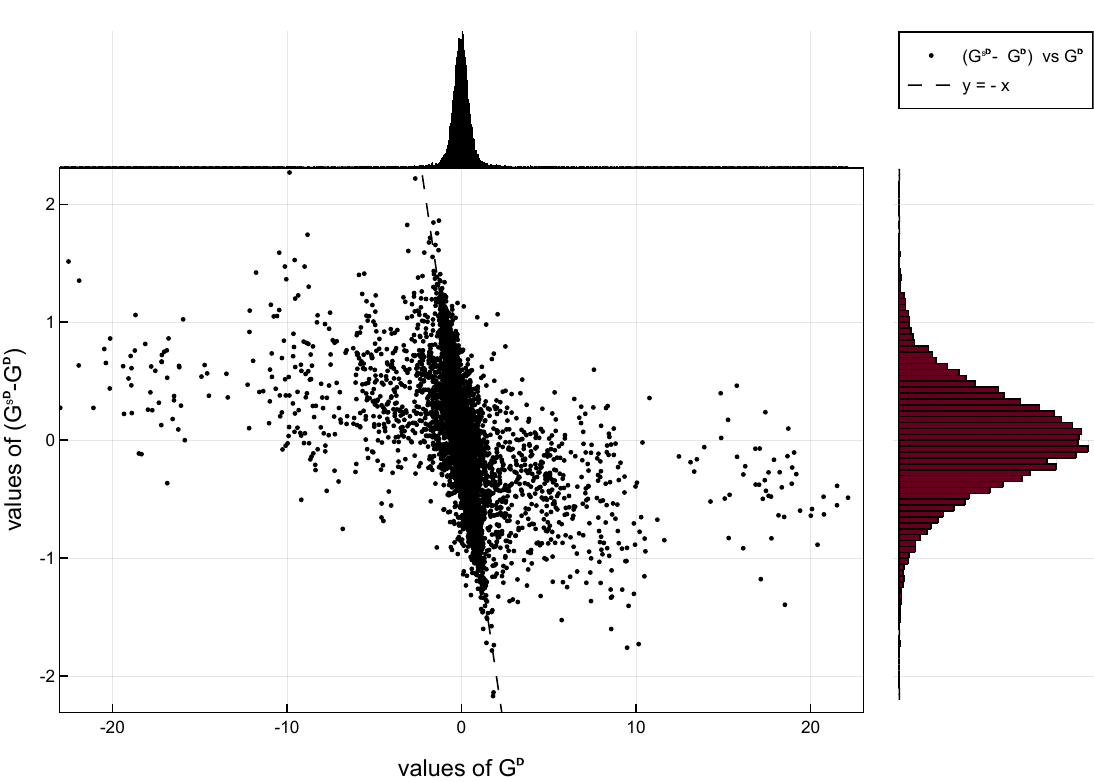}
  \caption{Comparison of interaction matrices calibrated on-sky in open-loop on a sequence
    of \emph{push-only} probe commands with $0.5\%$ amplitude. Top: Interaction matrix
    $\estim{G}_{t}^{\Tag{D}}$ calibrated by the \emph{direct} (D) method. Bottom:
    Coefficients of $\estim{G}_{t}^{\Tag{SD}} - \estim{G}_{t}^{\Tag{D}}$ \emph{vs.} those of
    $\estim{G}_{t}^{\Tag{D}}$ with $\estim{G}_{t}^{\Tag{SD}}$ calibrated by the \emph{simple
      difference} (SD) method. Calibration sequence counts $60\,000$ frames and results in
    $59\,957$ and $59\,905$ calibration frames respectively for the D and SD methods.}
  \label{fig:closed-loop:G:maps-p}
\end{figure}

\begin{figure}
  \centering
  \includegraphics[width=60mm]{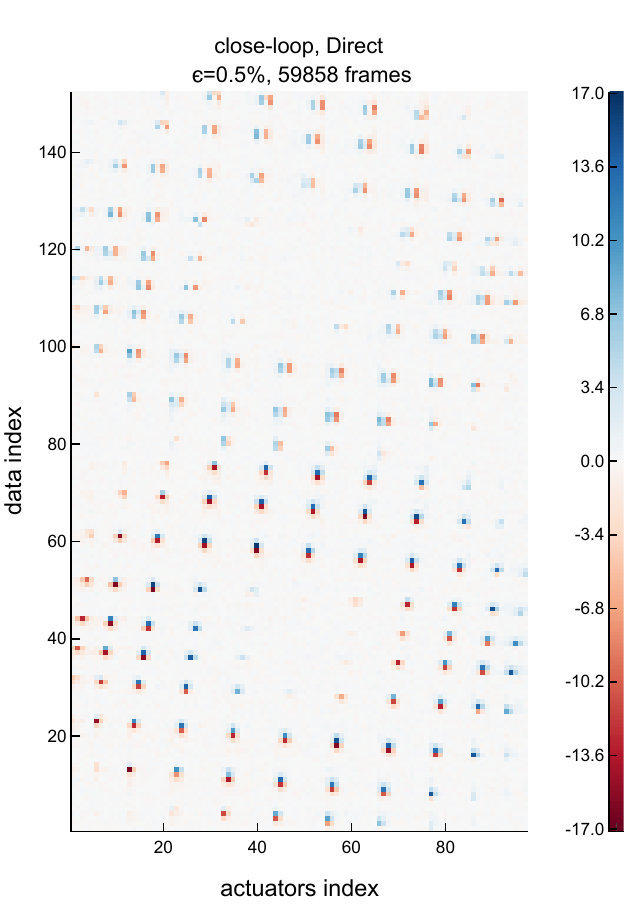}
  \\
  \centering
  \includegraphics[width=80mm]{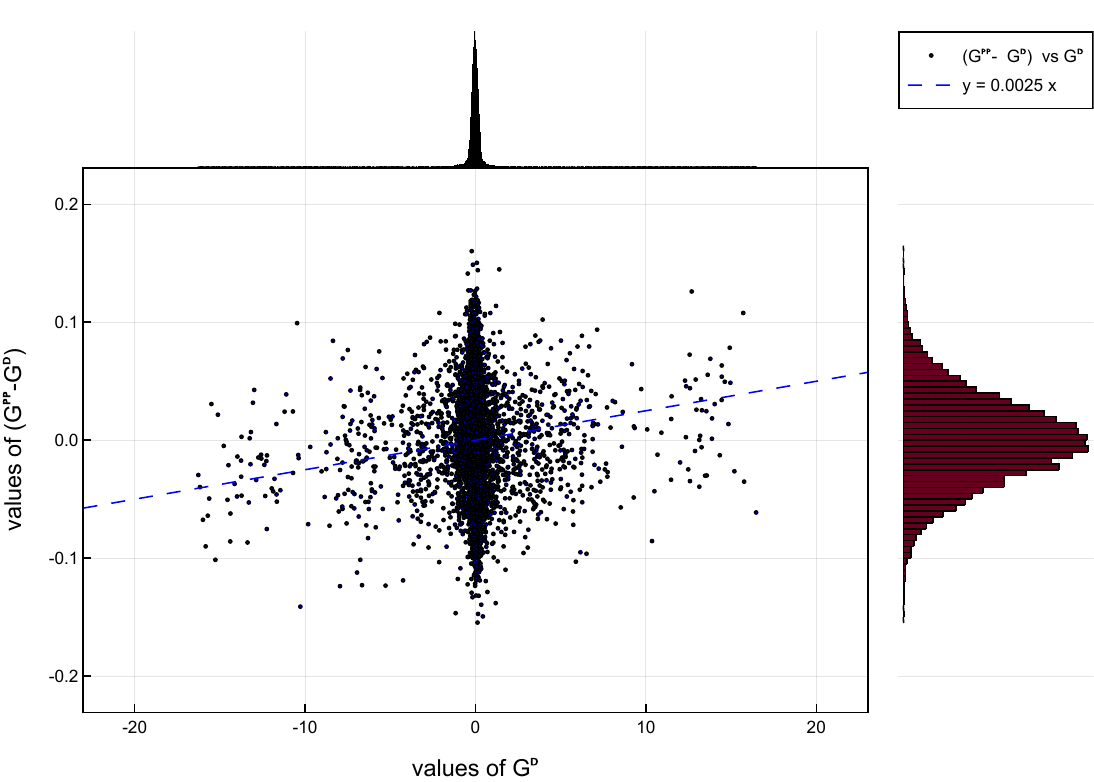}
  \\
  \centering
  \includegraphics[width=80mm]{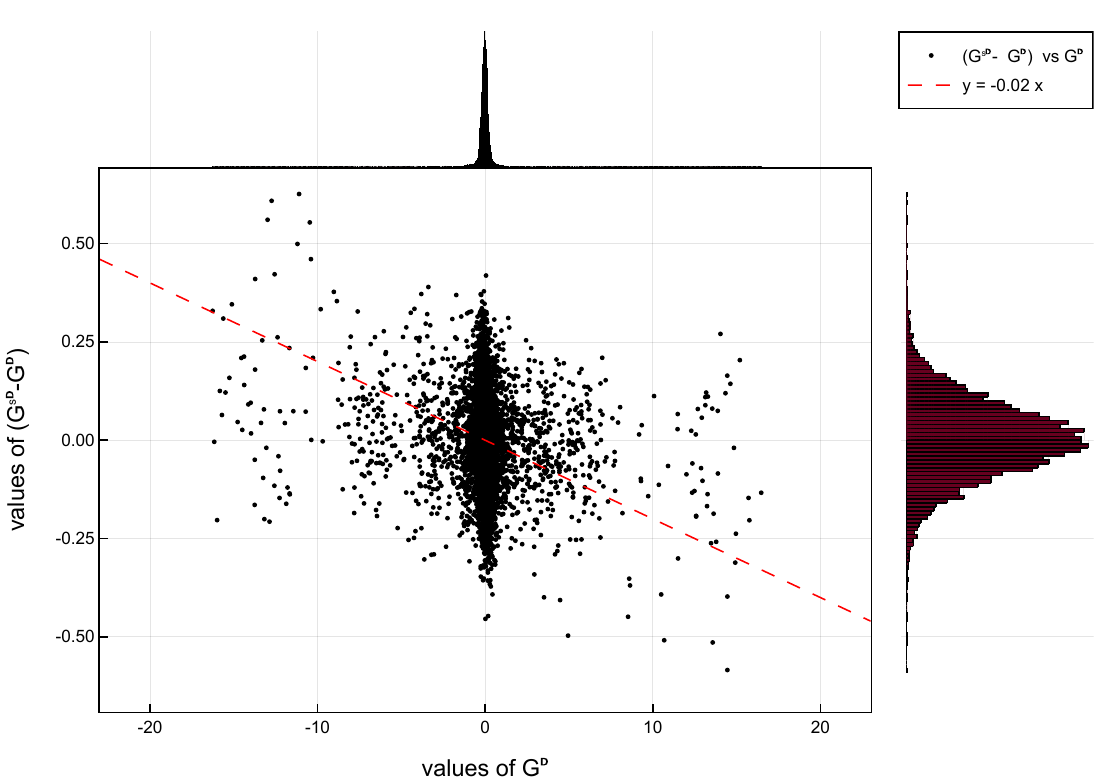}
  \caption[CL PP vs D]{Same as Fig.~\ref{fig:closed-loop:G:maps-p} but for a sequence of
    calibration in closed-loop by \emph{push-pull} probe commands with $0.5\%$ amplitude.
    Top: Interaction matrix $\estim{G}_{t}^{\Tag{D}}$ calibrated by the \emph{direct} (D)
    method. Middle: Coefficients of $\estim{G}_{t}^{\Tag{PP}} - \estim{G}_{t}^{\Tag{D}}$
    \emph{vs.} those of $\estim{G}_{t}^{\Tag{D}}$ with $\estim{G}_{t}^{\Tag{PP}}$ calibrated
    by the \emph{push-pull} (PP) method. Bottom: Coefficients of
    $\estim{G}_{t}^{\Tag{SD}} - \estim{G}_{t}^{\Tag{D}}$ \emph{vs.} those of
    $\estim{G}_{t}^{\Tag{D}}$ with $\estim{G}_{t}^{\Tag{SD}}$ calibrated by the \emph{simple
      difference} (SD) method.}
  \label{fig:closed-loop:G:maps-pp}
\end{figure}

\begin{figure}
  \centering
  \includegraphics[width=\columnwidth]{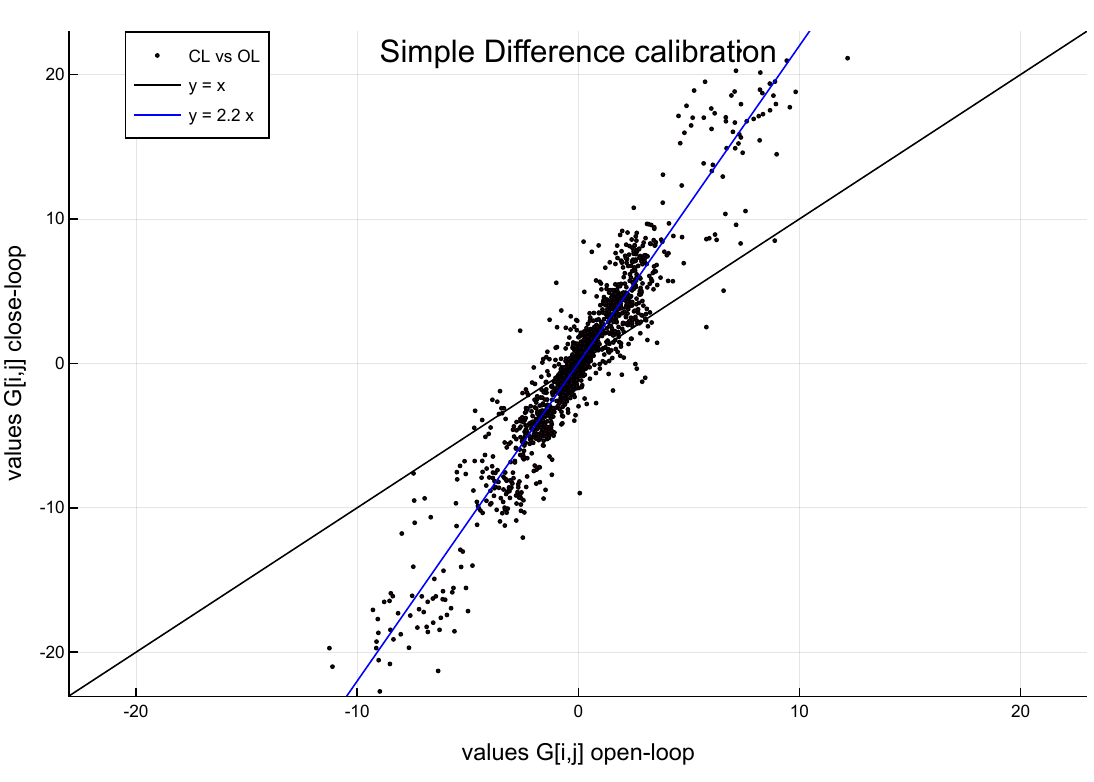}
  \caption[CL SimpleDiff]{Coefficients of the closed-loop interaction matrix \emph{vs.}
    the ones in open-loop. In both cases, calibration has been done by the \emph{simple
      differences} method on sequences of $60\,000$ frames of \emph{push-only} telemetry.
    A mean factor of 2.2 has been fitted (blue line) between the coefficients of the
    matrices in closed- and open-loop.}
  \label{fig:G-SD-CL-vs-OL}
\end{figure}

In THEMIS AO, the initial calibration is made in open-loop on the Sun to obtain an
interaction matrix suitable to close the loop. Once the AO loop is closed with this
open-loop calibration, a closed-loop calibration can be performed, adding probe commands
at every loop to the AO correction commands. Closed-loop calibration aims at following the
system response in these new conditions. To that end, telemetry sequences of $60\,000$
frames are again recorded in closed-loop, with probes of amplitude $ϵ = 0.5\%$ of the
maximum allowed for the DM commands, which corresponds to $\approx 25$\,nm. This excitation
is $10$ times smaller than that used in open-loop, in order to avoid degrading the AO
correction. In this section, we analyze again two calibration sequences: one taken in
\emph{push-only} mode (\ie, with independent successive random probe commands) and one in
\emph{push-pull} mode. An analysis similar to the one presented in
Sec.~\ref{sec:lag-estimation}, but for these sequences of closed-loop calibration data,
leads to similar results concerning the lag, \ie $\ell\simeq 2$.

Figures~\ref{fig:closed-loop:G:maps-p} and~\ref{fig:closed-loop:G:maps-pp} shows the
interaction matrices obtained respectively on the \emph{push-only} sequence and on the
\emph{push-pull} sequence of telemetry. Interaction matrices obtained on both sequences of
calibration, with various methods --- \emph{direct} (D), \emph{simple differences} (SD) and
\emph{push-pull} (PP, only on \emph{push-pull} sequence) --- are compared using the full
length of the telemetry. In the bottom graphs of Figs~\ref{fig:closed-loop:G:maps-p} and
\ref{fig:closed-loop:G:maps-pp}, we still observe that \emph{simple differences} (SD)
coefficients tend to have smaller amplitudes than their counterparts in the matrix
calibrated with the \emph{direct} (D) method.

Comparing Figs.~\ref{fig:open-loop:G:maps-p} and \ref{fig:closed-loop:G:maps-p}, the
interaction matrix appears to have higher amplitude coefficients in closed-loop than in
open-loop. This is confirmed by Fig.~\ref{fig:G-SD-CL-vs-OL} which plots the coefficients
of the interaction matrix calibrated by the \emph{simple differences} (SD) method in
closed-loop versus the ones in open-loop, each one taken on the \emph{push-only}
sequences. In this figure, the fit (blue line) corresponds to closed-loop coefficients
equal to $2.2$ times the corresponding coefficients in open-loop. This significant change
of the system response in closed-loop compared to open-loop clearly demonstrates that the
AO system is nonlinear and that open-loop calibration provides an estimate of the model
far from its working conditions in closed-loop (see Appendix~\ref{sec:working-conditions}
for more details). The nonlinear behavior is not unexpected considering the WFS sensing
method \citep{ThiebautEtAl2018a}, but such a large factor has very important consequences
for the AO control of THEMIS. Indeed, previously, the interaction matrix of the THEMIS AO
system was calibrated only in open-loop and had to be multiplied by a \emph{fudge factor}
between 2 to 4 to be able to close the loop. With the possibility to measure the
interaction matrix in closed-loop, the updated AO control system can work without
\emph{fudge factor}, evidencing a better calibration of the system around its working
conditions. This was one of the motivations for us to develop these new calibration
methods which are promising to improve the AO correction and control in nonlinear systems.
Furthermore, it paves the way for the AO control to take into account the variations of
the interaction matrix sensitivity, following the subminute scale changes
 evidenced in Fig.~\ref{fig:gain-vs-time}.

\begin{figure}
  \centering \includegraphics[width=0.9\columnwidth]{%
    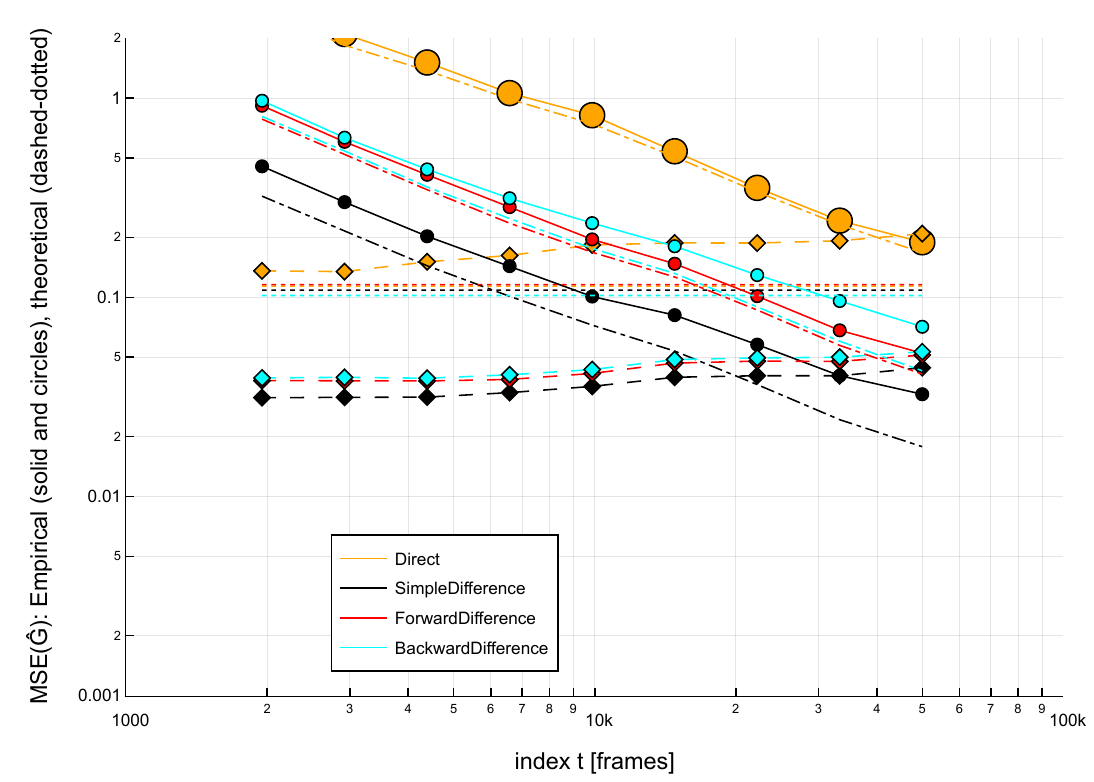}
  \caption{Same as in Fig.~\ref{fig:open-loop-MSEs-p} but for on-sky closed-loop calibration.}
  \label{fig:closed-loop-MSEs-p}
\end{figure}

\begin{figure}
  \centering \includegraphics[width=0.9\columnwidth]{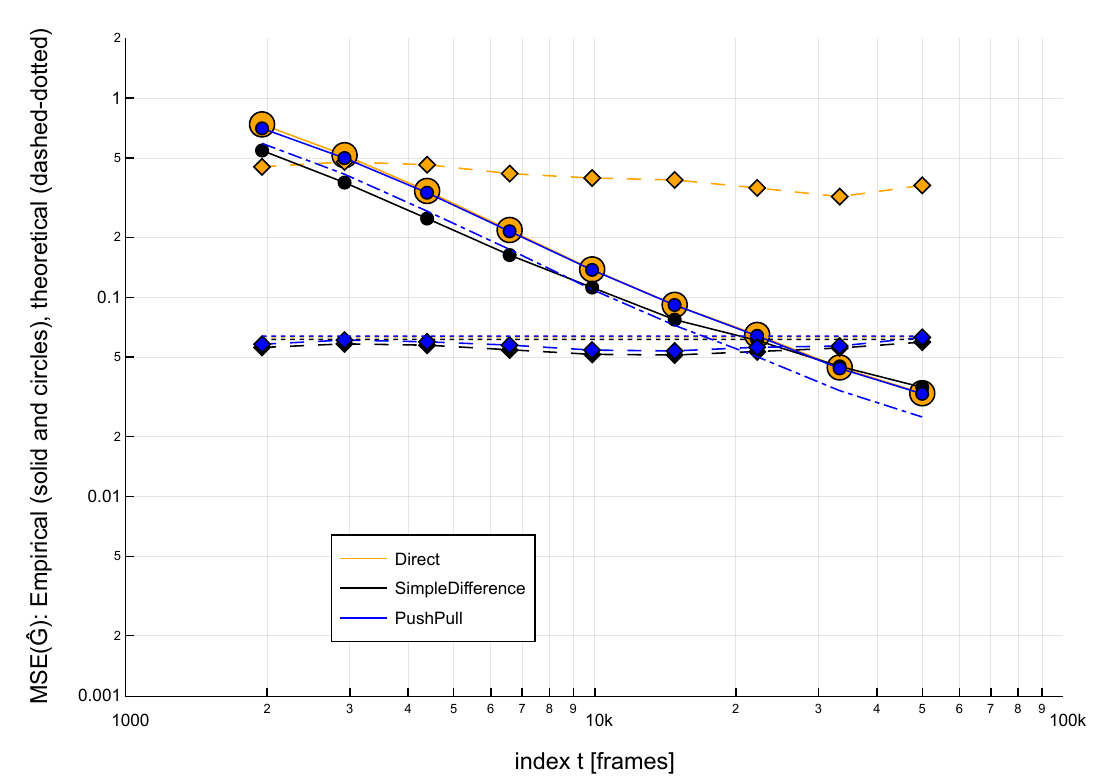}
  \caption{Same as in Fig.~\ref{fig:open-loop-MSEs-pp} but for on-sky closed-loop
    calibration.}
  \label{fig:closed-loop-MSEs-pp}
\end{figure}

Figures~\ref{fig:closed-loop-MSEs-p} and \ref{fig:closed-loop-MSEs-pp} present similar
results as Figs.~\ref{fig:open-loop-MSEs-p}--\ref{fig:open-loop-MSEs-pp} but for
closed-loop calibration. It can be first noticed that the variance of the calibration
nuisance (the dashed lines with diamond) are reduced by a factor between 5 and 10 in
closed-loop compared to open-loop. This reflects that the contribution of the turbulence
is strongly attenuated by the AO correction. This reduction helps improve the MSE of the
calibrated interaction matrix; however, to avoid disrupting the AO system, the amplitude
of the calibration probe commands was reduced by a factor of 10 ($100$ for $ϵ_j^2$). In
such configurations, for the same number of frames in the calibration sequences, the MSE
is always lower in open-loop than in closed-loop. Furthermore, for the \emph{push-only}
sequence and the \emph{simple differences} (SD) method, about $10\,000$ frames of
telemetry are again needed ($10\,\text{s}$ at the AO loop frequency of $1\,\text{kHz}$) to
reach a relative precision of $2\%$ of the magnitude of the most significant coefficients
of $\estim{G}_t$ (indicated by the dotted lines in these figures).

Comparing the MSE achieved by the different calibration methods (the solid lines with
circles in Figs.~\ref{fig:closed-loop-MSEs-p}--\ref{fig:closed-loop-MSEs-pp}) it appears
that, in closed-loop, the \emph{simple differences} (SD) is always the best method being
slightly better than the \emph{push-pull} (PP) one for a moderate number of calibration
frames. As for open-loop calibration, the \emph{direct} (D), \emph{forward differences}
(FD), and \emph{backward differences} (BD) on a \emph{push-only} sequence are significantly
worse than the \emph{simple differences} (SD) and \emph{push-pull} (PP) methods.

These results confirm that \emph{simple differences} (SD) is the method of choice,
regardless of whether calibration is performed in open- or closed-loop, and allowing either
\emph{push-only} or \emph{push-pull} excitation sequences.

\subsection{Offset term}
\label{sec:offset-themis}

\begin{figure}
  \centering
  \includegraphics[width=0.48\columnwidth]{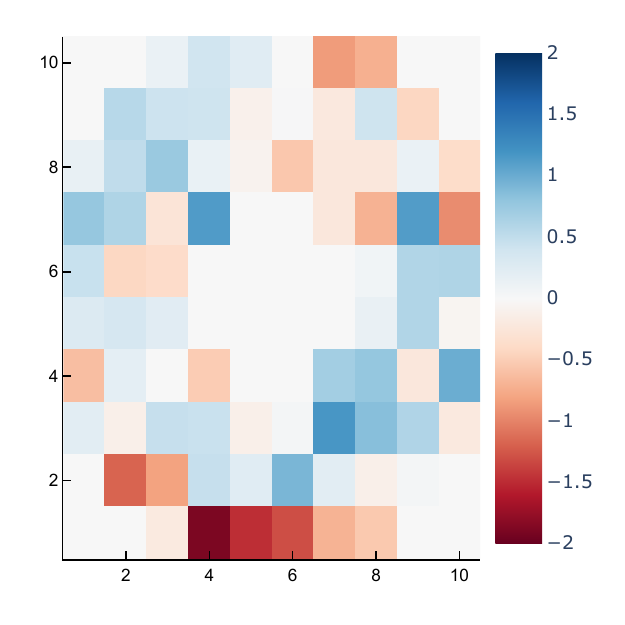}
  \includegraphics[width=0.48\columnwidth]{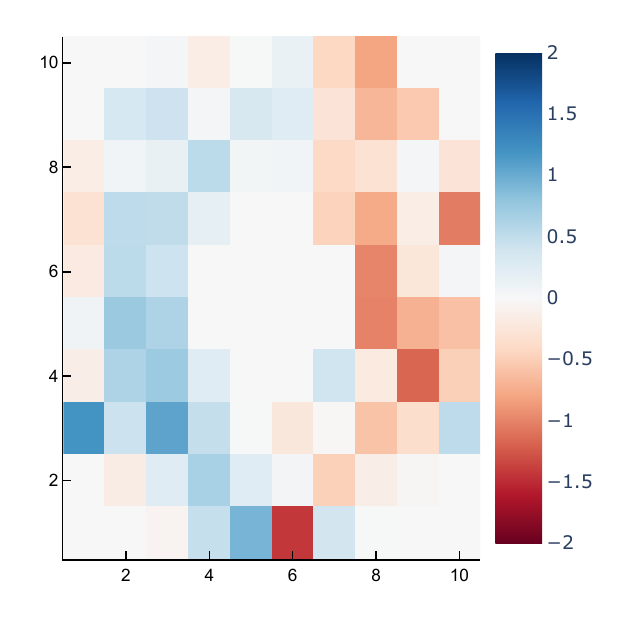}
  \includegraphics[width=0.48\columnwidth]{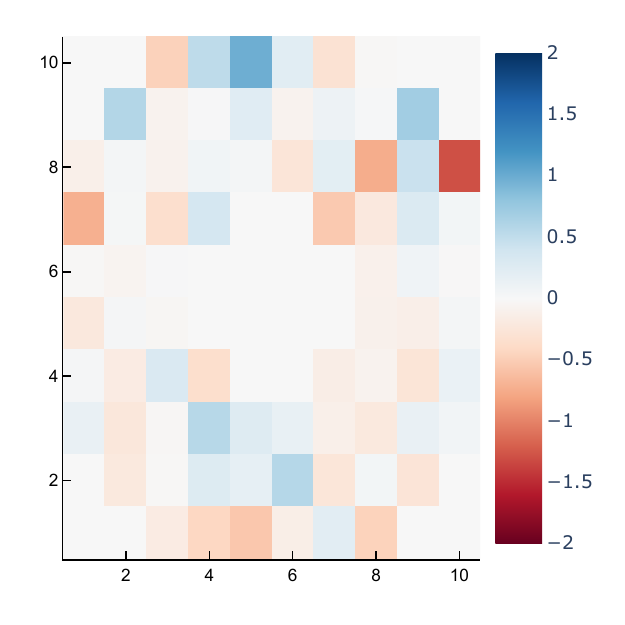}
  \includegraphics[width=0.48\columnwidth]{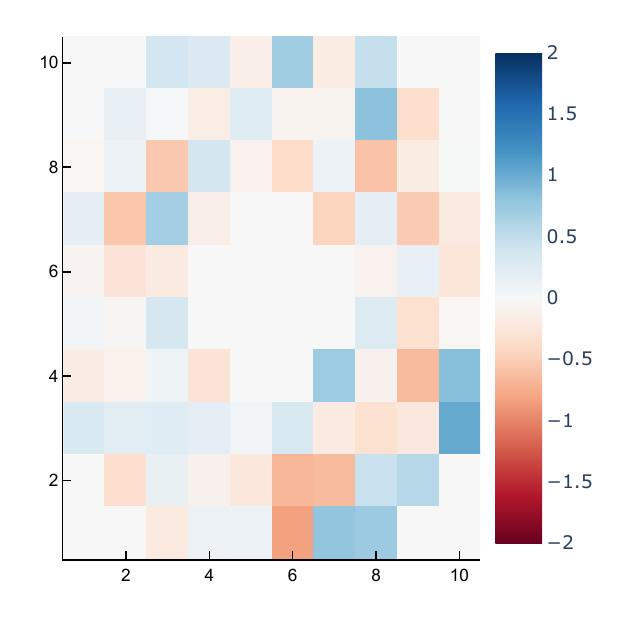}
  \caption{Offset vector $b$ obtained applying \emph{simple differences} (SD) method represented as two maps ($x-$ (left) and $y-$ (right) slopes measurements) of the 76 active Shack-Hartmann wavefront sensor. Top row: from open-loop data. Bottom row: from closed-loop data.}
  \label{fig:offset-term-maps-sd}
\end{figure}

\begin{figure}
  \centering
  \includegraphics[width=0.48\columnwidth]{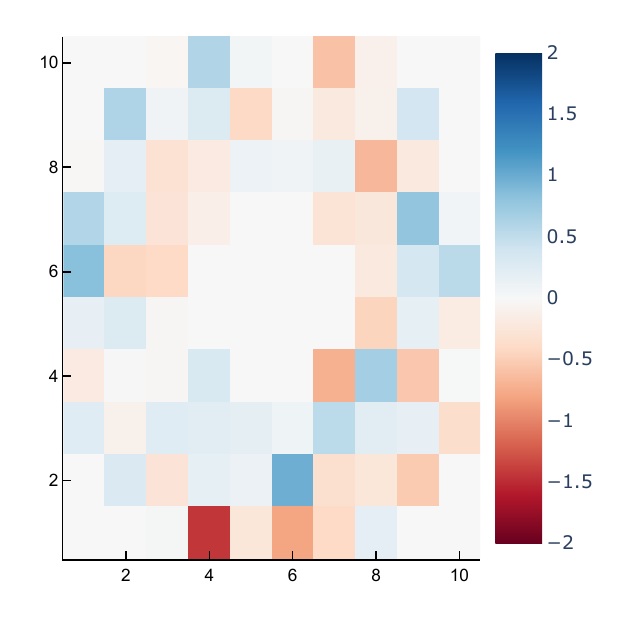}
  \includegraphics[width=0.48\columnwidth]{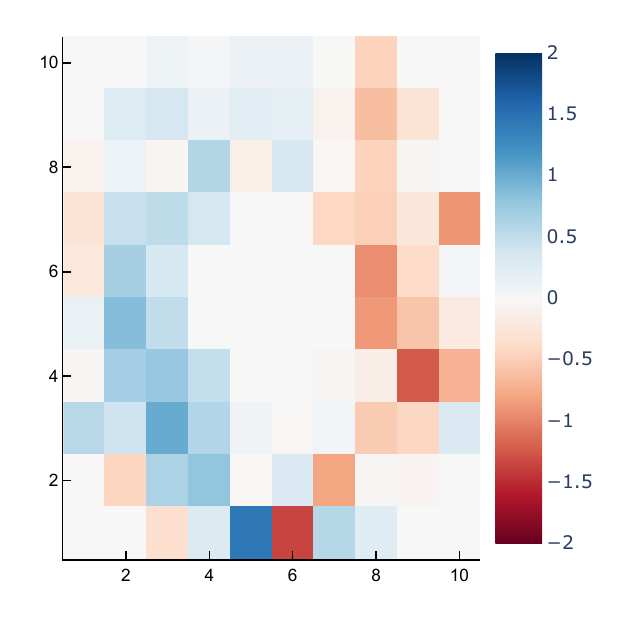}
  \includegraphics[width=0.48\columnwidth]{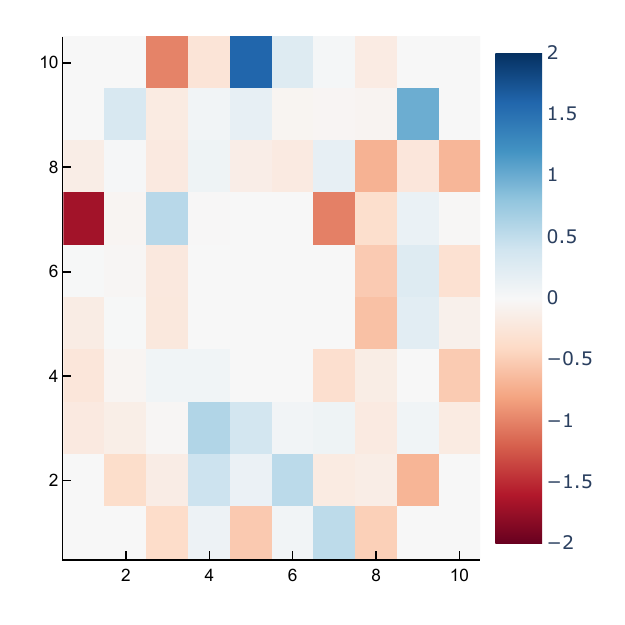}
  \includegraphics[width=0.48\columnwidth]{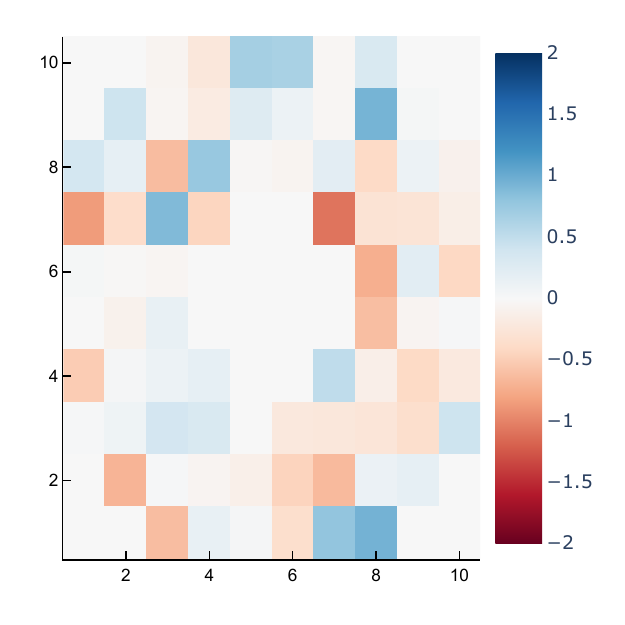}
  \caption{Same as Fig.~\ref{fig:offset-term-maps-sd} but for \emph{push-pull} (PP) method.}
  \label{fig:offset-term-maps-pp}
\end{figure}

The offset term $b$ can be represented as a vector of $x-$ and $y-$slopes (gradients)
measured over the telescope aperture and characteristic of the working conditions. In order
to analyze its values, we have chosen to reorder its elements in two corresponding maps
(left and right) in Figs~\ref{fig:offset-term-maps-sd} and~\ref{fig:offset-term-maps-pp}.

In Figure~\ref{fig:offset-term-maps-sd}, we illustrate the offset terms obtained using
\emph{simple differences} (SD) calibration on open-loop (top row) and closed-loop (bottom
row) \emph{push-only} probe commands. The offset terms obtained from the push-only
calibration frames are not significantly distinguishable whatever the calibration method
applied to compute it (D, SD, FD, or BD), thus we only represent it here for \emph{simple
differences} (SD). The \emph{push-pull} sequence of data leads however to another offset
estimate, represented in Fig.~\ref{fig:offset-term-maps-pp} and obtained with
\emph{push-pull} (PP) method.

Both calibration sequences (\emph{push-only} and \emph{push-pull}) of data have been
recorded just a few minutes apart. It can be observed, particularly visible on open-loop
$y$-slopes map (top right), that offset terms have a common structure in the two
estimates, evidencing some quasi-static aberration. It is also clear that the amplitude of
the offset terms are, on average, larger in open-loop, than in closed-loop. This is
consistent with the fact that the closed-loop control drives the system towards null
residual measurements, and thus correct this aberration.

Until now, THEMIS AO correction does not use the estimated $b$ terms in its control law.
The results presented here show however that this offset has a significant amplitude with
respect to the uncertainties given in Eqs.~\eqref{eq:MSE(b^D_t)}-\eqref{eq:MSE(b^PP_t)},
and a study of how this offset estimate could improve the THEMIS AO correction is under
study.


\section{Conclusions}
\label{sec:conclusions}

The calibration of adaptive optics (AO) system has been revisited through a more general
affine model to approximate the behavior of the system. This model is parameterized by an
interaction matrix $G$ and an offset term $b$. In this framework, a family of unbiased
estimators, $\estim{G}$ and $\estim{b}$, has been derived based on weighted least-squares
fit of calibration data obtained by adding small random perturbations to the commands sent
to the deformable mirror (DM). Beyond the classical \emph{push-pull} method, these
estimators offer new calibration strategies --- referred as \emph{direct}, \emph{simple
  differences}, \emph{forward differences}, and \emph{backward differences} --- that relax
the requirement for a specific sequence of push and pull probing commands.

All the proposed calibration strategies are applicable regardless of whether the AO control
system is active and have been validated on empirical telemetry collected on THEMIS AO
system on the sky and in open- or closed-loop. Our results show in particular that reliable
calibration can be achieved from about $10\,000$ frames of WFS data acquired in closed-loop
with probes amplitude as small as $0.5\%$ of the maximum allowed for the DM commands, which
corresponds to $\approx 25$\,nm.

Theoretical analysis of the calibration errors as well as empirical results based on the
telemetry of the THEMIS AO system evidence the superiority of the \emph{simple differences}
(SD) and \emph{push-pull} (PP) methods in terms of accuracy. The \emph{simple differences}
(SD) method therefore appears to be the method of choice since it is easier to implement.

All the estimators follow a similar formalism and have closed-form expressions. By
introducing \emph{forgetting factors} to mitigate the impact of calibration data over time,
we were able to establish computationally efficient rules to update the estimators, without
resorting to matrix inversions and using only simple operations such as rank-1 updates at
each new acquired WFS frame. This, combined with the fact that the proposed calibration
methods are applicable in closed-loop while observing, makes them suitable for continuously
updating the system model in real-time (that is to say at the same frequency as the control
loop) to account for variable working conditions. Such a capability is of importance for AO
systems with nonlinear devices such as the pyramid WFS. Optimizing the integration time, the
\emph{forgetting factor}, and the amplitude of the perturbations remain to be studied. An
extension of this work to account for the fractional delay is also under work.

Finally, the amplitudes of the offset term $b$ are not negligible in the THEMIS AO system.
It is thus foreseen to integrate into THEMIS AO a continuous update of the model with the
proposed method and to study how the estimation of the offset term can improve the AO
control.

\section*{Acknowledgments}

This work received government funding administered by the French National Research Agency
under the France 2030 program, grant number 22-EXOR-0017 (UPCAO, \emph{Unsupervised AO
  control}), part of the PEPR \emph{Origins: from planets to life}. It was also supported by
the Action Spécifique Haute Résolution Angulaire (ASHRA) of CNRS/INSU co-funded by CNES, by
the European Commission’s FP7 Capacities Programme under Grant Agreement number 312495, and
the Centre National de la Recherche Scientifique. The THEMIS solar telescope operations are
funded by the Centre National de la Recherche Scientifique Terre \& Univers (INSU).

\bibliographystyle{aa}
\bibliography{AO-self-calibration}


\appendix

\nolinenumbers

\section{Estimation of calibration errors}
\label{sec:variances}

In this Appendix, we derive expressions of the variances, $ω_j$ and $χ_i$, defined in
Eqs.~\eqref{eq:omega_j} and \eqref{eq:chi_i} and involved in the mean squared error of the
estimator of the quasi-static interaction matrix given in Eq.~\eqref{eq:MSE(G)}:
\begin{displaymath}
  𝔼\Paren[\big]{δG_t[i,j]^2} ≈ \frac{χ_i}{ξ_t\,ω_j}.
\end{displaymath}
These variances depend on the considered calibration method and if the probes are sent as a
push-pull sequence or not. We examine each case in turn.

\subsection{Push-only sequence of probes}
\label{sec:variances-push-only}

When all the successive probes commands are mutually independent, we refer to a push-only
sequence of probes, by contrast with  a push-pull sequence. All the calibration
methods except the push-pull can be considered in turn on such sequence.

\Paragraph{Direct method.} In the \emph{direct} calibration method, the interaction matrix
is calibrated on the recentered WFS data:
\begin{align}
  y^\Tag{D}_{t,t'}
  &\bydef d_{t'} - \mean{d}_t = G_t\,x^\Tag{D}_{t,t'} + z^\Tag{D}_{t,t'},\\
  x^\Tag{D}_{t,t'} &= u_{t'-ℓ} - \mean{u}_t,\\
  z^\Tag{D}_{t,t'} &= n_{t'} - \mean{n}_t,\\
  \intertext{with}
  \mean{n}_t &\bydef \frac{1}{η_t}\,\sum_{t'=ℓ+1}^{t} ρ_{t,t'}\,n_{t'},
\end{align}
the weighted average of the nuisance term appearing in Eq.~\eqref{eq:quasi-static-model}.
The variance $ω_j$ for the \emph{direct} method can be expanded and approximated as follows:
\begin{align}
  ω^\Tag{D}_j
  &\bydef \Var_U\Paren[\big]{x^\Tag{D}_{t,t'}[j]} \notag\\
  &= \Var\Paren[\big]{u_{t'-ℓ}[j]} + \Var\Paren[\big]{\mean{u}_t[j]}
    - 2\,\Cov\Paren[\big]{u_{t'-ℓ}[j], \mean{u}_t[j]}\notag\\
  &≈ ϵ^2_j,
    \label{eq:omega^D_j}
\end{align}
where $ϵ^2_j \bydef \Var\Paren[\big]{u_{t'-ℓ}[j]}$ is the variance of the $j$-th actuator in
the probe commands and first appears in Eq.~\eqref{eq:probe-covariance}. The above
approximation follows from that, for a long enough calibration sequence, the variance of the
weighted mean of a quantity is much smaller than the variance of any instance of this
quantity in the sequence. In other words,
$\Var\Paren[\big]{\mean{u}_t[j]} \ll \Var\Paren[\big]{u_{t'-ℓ}[j]}$. For the same reasons,
$\Var\Paren[\big]{\mean{n}_t[i]} \ll \Var\Paren[\big]{n_{t'}[i]}$ should hold and the
variance $χ_i$ for the \emph{direct} method can be expanded and approximated as follows:
\begin{align}
  χ^\Tag{D}_i
  &\bydef \Var\Paren[\big]{z^\Tag{D}_{t,t'}[i]}\notag\\
  &= \Var\Paren[\big]{n_{t'}[i]} + \Var\Paren[\big]{\mean{n}_t[i]}
  - 2\,\Cov\Paren[\big]{n_{t'}[i], \mean{n}_t[i]}\notag\\
  &≈ σ^2_i
    \label{eq:chi^D_i}
\end{align}
where:
\begin{equation}
  \label{eq:sigma^2_i}
  σ^2_i \bydef \Var\Paren[\big]{n_{t'}[i]}
\end{equation}
is the variance of the nuisance term appearing in Eq.~\eqref{eq:quasi-static-model} and
accounting for the noise, for the residual wavefront, and for the approximation by the
affine model. It follows from our hypotheses of the stationarity in the sequence of
calibration data, that this variance does not depend on the frame index $t'$ as reflected
by the notation. In practice, this variance depends on the calibration conditions, \eg\
open- or closed-loop, calibration on an internal source, \etc

\Paragraph{Simple differences method.} In the \emph{simple differences} method, the
interaction matrix is calibrated on the differences between successive WFS frames
expressed as functions of the differences between successive probe commands:
\begin{align}
  y^\Tag{SD}_{t,t'}
  &\bydef d_{t'} - d_{t'-1} = G_t\,x^\Tag{SD}_{t,t'} + z^\Tag{SD}_{t,t'},\\
  x^\Tag{SD}_{t,t'} &= u_{t'-ℓ} - u_{t'-ℓ-1}\\
  z^\Tag{SD}_{t,t'} &= n_{t'} - n_{t'-1}.
\end{align}
Hence,
\begin{align}
  ω^\Tag{SD}_j
  &\bydef \Var_U\Paren[\big]{x^\Tag{SD}_{t,t'}[j]} \notag\\
  &= \Var\Paren[\big]{u_{t'-ℓ}[j]} + \Var\Paren[\big]{u_{t'-ℓ-1}[j]} = 2\,ϵ^2_j,
    \label{eq:omega^SD_j}\\
  χ^\Tag{SD}_i
  &\bydef \Var\Paren[\big]{z^\Tag{SD}_{t,t'}[i]}\notag\\
  &= \Var\Paren[\big]{n_{t'}[i]} + \Var\Paren[\big]{n_{t'-1}[i]}
  - 2\,\Cov\Paren[\big]{n_{t'}[i], n_{t'-1}[i]}\notag\\
  &= 2\,(1 - ζ_i)\,σ^2_i.
    \label{eq:chi^SD_i}
\end{align}

\Paragraph{Forward differences method.} In the \emph{forward differences} method, the
interaction matrix is calibrated on the differences between successive WFS frames as
functions of the last probe commands:
\begin{align}
  y^\Tag{FD}_{t,t'}
  &\bydef d_{t'} - d_{t'-1} = G_t\,x^\Tag{FD}_{t,t'} + z^\Tag{FD}_{t,t'},\\
  x^\Tag{FD}_{t,t'} &= u_{t'-ℓ},\\
  z^\Tag{FD}_{t,t'} &= n_{t'} - n_{t'-1} - G_t\,u_{t'-ℓ-1}.
\end{align}
Hence,
\begin{align}
  ω^\Tag{FD}_j
  &\bydef \Var_U\Paren[\big]{x^\Tag{FD}_{t,t'}[j]}
  = \Var\Paren[\big]{u_{t'-ℓ}[j]} = ϵ^2_j,
    \label{eq:omega^FD_j}\\
  χ^\Tag{FD}_i
  &\bydef \Var\Paren[\big]{z^\Tag{FD}_{t,t'}[i]}
  = \Var\Paren[\big]{z^\Tag{SD}_{t,t'}[i] - \Paren{G_t\,u_{t'-ℓ-1}}[i]}\notag\\
  &≈ 2\,(1 - ζ_i)\,σ^2_i.
    \label{eq:chi^FD_i}
\end{align}
For the last right-hand side, we have omitted a term whose contribution in the MSE, as
given by Eq.~\eqref{eq:MSE(G)}, is decreasing as the effective number of calibration
frames. Hence, our expression in Eq.~\eqref{eq:chi^FD_i} holds for a large enough number
of calibration frames.

\Paragraph{Backward differences method.} The \emph{backward differences} method is the
same as the \emph{forward differences} one except for the considered probe commands:
\begin{align}
  y^\Tag{BD}_{t,t'}
  &\bydef d_{t'} - d_{t'-1} = G_t\,x^\Tag{BD}_{t,t'} + z^\Tag{BD}_{t,t'},\\
  x^\Tag{BD}_{t,t'} &= -u_{t'-ℓ-1},\\
  z^\Tag{BD}_{t,t'} &= n_{t'} - n_{t'-1} + G_t\,u_{t'-ℓ}. \label{eq:nuisance^BD}
\end{align}
This method being very similar to the previous one, we directly give the results under
the same assumptions:
\begin{align}
  ω^\Tag{BD}_j
  &\bydef \Var_U\Paren[\big]{x^\Tag{BD}_{t,t'}[j]}
  = \Var\Paren[\big]{-u_{t'-ℓ-1}[j]} = ϵ^2_j,
    \label{eq:omega^BD_j}\\
  χ^\Tag{BD}_i
  &\bydef \Var\Paren[\big]{z^\Tag{BD}_{t,t'}[i]} ≈ 2\,(1 - ζ_i)\,σ^2_i.
    \label{eq:chi^BD_i}
\end{align}

\subsection{Push-pull sequence of probes}
\label{sec:variances-push-pull}

In a push-pull sequence of probes, the property of Eq.~\eqref{eq:probe-covariance-pp} holds  to use the push-pull calibration method. 

\Paragraph{Push-pull method.} In the \emph{push-pull} method, the interaction matrix is
calibrated on the differences between successive WFS frames acquired with opposite probe
commands:
\begin{align}
  y^\Tag{PP}_{t,t'}
  &\bydef d_{t'} - d_{t'-1} = G_t\,x^\Tag{PP}_{t,t'} + z^\Tag{PP}_{t,t'},\\
  x^\Tag{PP}_{t,t'} &= u_{t'-ℓ} - u_{t'-ℓ-1} = 2\,u_{t'-ℓ},\\
  z^\Tag{PP}_{t,t'} &= n_{t'} - n_{t'-1},
\end{align}
for $t'$ even. Hence,
\begin{align}
  ω^\Tag{PP}_j
  &\bydef \Var_U\Paren[\big]{x^\Tag{PP}_{t,t'}[j]}
  = \Var\Paren[\big]{2\,u_{t'-ℓ}[j]} = 4\,ϵ^2_j,
    \label{eq:omega^PP_j}\\
  χ^\Tag{PP}_i
  &\bydef \Var\Paren[\big]{z^\Tag{PP}_{t,t'}[i]}\notag\\
  &= \Var\Paren[\big]{n_{t'}[i]} + \Var\Paren[\big]{n_{t'-1}[i]}
  - 2\,\Cov\Paren[\big]{n_{t'}[i], n_{t'-1}[i]}\notag\\
  &= 2\,(1 - ζ_i)\,σ^2_i.
    \label{eq:chi^PP_i}
\end{align}
with
\begin{equation}
  \label{eq:gamma_i}
  ζ_i = \frac{\Cov\Paren[\big]{n_{t'}[i], n_{t'-1}[i]}}{σ^2_i}
\end{equation}
the coefficient of correlation for successive nuisances in the $i$-th WFS data.




\section{Updating the cross-products}
\label{sec:updating-cross-products}

In this Appendix, we derive the updating rules for the sums of cross products $A^\Tag{D}_t$
and $B^\Tag{D}_t$ used in the \emph{direct} calibration method and defined in
Eqs.~\eqref{eq:A^D_t} and \eqref{eq:B^D_t}. We start by deriving an update rule for
$B^\Tag{D}_t$ which combines the calibration data and the corresponding perturbations. The
update rule for $A^\Tag{D}_t$, which only combines the perturbations, is directly deduced
from that of $B^\Tag{D}_t$.

Considering the definition of $B^\Tag{D}_t$ in Eq.~\eqref{eq:B^D_t} and applying the
constraint in Eq.~\eqref{eq::rho(t,t)} and the recurrences in Eq.~\eqref{eq:rho-recurrence}
and \eqref{eq:eta-recurrence}, $B^\Tag{D}_{t+1}$ can be expanded as follows:
\begin{align}
  B^\Tag{D}_t
  &= ρ_{t,t}\,\Paren{d_t - \mean{d}_t}\,\Paren{u_{t-ℓ} - \mean{u}_t}\T
    \nonumber\\
  &\quad + γ_t\,\sum\nolimits_{t'=ℓ+1}^{t-1} ρ_{t-1,t'}\,
    \Paren{d_{t'} - \mean{d}_t}\,\Paren{u_{t'-ℓ} - \mean{u}_t}\T.
    \label{eq:B^D_t:update:expanded}
\end{align}
Then, using Eqs.~\eqref{eq:mean(d)_t:update}, \eqref{eq:mean(u)_t:update}, and
\eqref{eq:alpha}, we can develop and simplify the sum in the above right-hand side:
\begin{align}
   \sum\nolimits_{t'=ℓ+1}^{t-1}
  & ρ_{t-1,t'}\,\Paren{d_{t'} - \mean{d}_t}\,
    \Paren{u_{t'-ℓ} - \mean{u}_t}\T\nonumber\\
  & = \sum\nolimits_{t'=ℓ+1}^{t-1} ρ_{t-1,t'}\,
    \Paren[\big]{d_{t'} - \mean{d}_{t-1} - α_t\,\Paren{d_t - \mean{d}_{t-1}}}\,
    \nonumber\\[-2ex]
  & \phantom{{} = \sum_{t'=ℓ+1}^{t-1} ρ_{t-1,t'}}\,
    \Paren[\big]{u_{t'-ℓ} - \mean{u}_{t-1} - α_t\,\Paren{u_{t-ℓ} - \mean{u}_{t-1}}}\T
    \nonumber\\
  & = \underbrace{\sum\nolimits_{t'=ℓ+1}^{t-1} ρ_{t-1,t'}\,
    \Paren{d_{t'} - \mean{d}_{t-1}}\,
    \Paren{u_{t'-ℓ} - \mean{u}_{t-1}}\T}_{\text{$B^\Tag{D}_{t-1}$, from Eq.~\eqref{eq:B^D_t}}}
    \nonumber\\
  & \quad - α_t\,\Paren[\Big]{\underbrace{
    \sum\nolimits_{t'=ℓ+1}^{t-1} ρ_{t-1,t'}\,\Paren{d_{t'} - \mean{d}_{t-1}}
    }_{\text{$0$, from Eq.~\eqref{eq:mean(d)_t}}}}
    \,\Paren{u_{t-ℓ} - \mean{u}_{t-1}}\T
    \nonumber\\
  & \quad - α_t\,\Paren{d_t - \mean{d}_{t-1}}\,
    \Paren[\Big]{\underbrace{
    \sum\nolimits_{t'=ℓ+1}^{t-1} ρ_{t-1,t'}\,
    \Paren{u_{t'-ℓ} - \mean{u}_{t-1}}}_{\text{$0$, from Eq.~\eqref{eq:mean(u)_t}}}}\T
    \nonumber\\
  & \quad + α_t^2\,\Paren[\Big]{\underbrace{
    \sum\nolimits_{t'=ℓ+1}^{t-1} ρ_{t-1,t'}}_{\text{$η_{t-1}$, from Eq.~\eqref{eq:sum-of-weights}}}}\,
    \Paren{d_t - \mean{d}_{t-1}}\,
    \Paren{u_{t-ℓ} - \mean{u}_{t-1}}\T
    \nonumber\\
  & = B^\Tag{D}_{t-1} + α_t^2\,η_{t-1}\,
    \Paren{d_t - \mean{d}_{t-1}}\,
    \Paren{u_{t-ℓ} - \mean{u}_{t-1}}\T.
    \label{eq:B^D_t:update:RHS-sum-simplified}
\end{align}
Besides, the updating rules in Eqs.~\eqref{eq:mean(d)_t:update} and
\eqref{eq:mean(u)_t:update} yield the following identities:
\begin{align}
  d_t - \mean{d}_t
  &= (1 - α_t)\,\Paren[\big]{d_t - \mean{d}_{t-1}},\\
  u_{t-ℓ} - \mean{u}_t
  &= (1 - α_t)\,\Paren[\big]{u_{t-ℓ} - \mean{u}_{t-1}}.
\end{align}
Inserting these identities and the last right-hand side of
Eq.~\eqref{eq:B^D_t:update:RHS-sum-simplified}, combining with the right-hand side of
Eq.~\eqref{eq:B^D_t:update:expanded}, and simplifying the factors thanks to
Eq.~\eqref{eq:alpha}, we obtain the updating rule for $B^\Tag{D}_t$:
\begin{align}
  B^\Tag{D}_t
  &= γ_t\,B^\Tag{D}_{t-1} + ρ_{t,t}\,(1 - α_t)\,
    \Paren{d_t - \mean{d}_{t-1}}\,
    \Paren{u_{t-ℓ} - \mean{u}_{t-1}}\T.
    \label{eq:B^D_t:update:simplified}
\end{align}
Replacing the data $d_t$ by the perturbation $u_{t-ℓ}$ directly yields the updating rule for
$A^\Tag{D}_t$:
\begin{align}
  A^\Tag{D}_t
  &= γ_t\,A^\Tag{D}_{t-1} + ρ_{t,t}\,(1 - α_t)\,
    \Paren{u_{t-ℓ} - \mean{u}_{t-1}}\,
    \Paren{u_{t-ℓ} - \mean{u}_{t-1}}\T.
    \label{eq:A^D_t:update:simplified}
\end{align}


\section{Full calibration algorithm}
\label{sec:update-model}

We provide here Algorithm~\ref{alg:update-model} to update all model parameters for any of
the calibration methods under consideration. For the implementation, it is worth noting that
$A_t^{-1}$, $\estim{G}_t$, $\estim{b}$, $\mean{d}_t$, and $\mean{u}_t$ can be \emph{updated
  in-place}, that is to say with the same storage for the quantity before and after the
update.

\begin{algorithm}
  \caption{Update affine model parameters for calibration method $m  \in \Brace{\Tag{D},\Tag{PP},\Tag{SD},\Tag{FD},\Tag{BD}}$. All updated
    quantities are initially set to zero, except $A_t^{-1}$ initialized to
    $A_0^{-1} = (𝔼_{U}\Paren[\big]{x_{t,t'}\,x_{t,t'}\T})^{-1}$ which can easily be
    precomputed from the amplitude of the probes as mentioned at the end of
    Sect.~\ref{sec:updating-imat}.}
  \label{alg:update-model}
  \KwIn{%
    $\estim{b}_{t-1} \in ℝ^{\Ndat}$,
    $\estim{G}_{t-1} \in ℝ^{\Ndat × \Nact}$,
    $A_{t-1}^{-1} \in ℝ^{\Nact × \Nact}$,
    $\Paren{u_{t-ℓ},u_{t-ℓ-1},\mean{u}_{t-1}} \in ℝ^{\Nact\times3}$,
    $\Paren{d_t,d_{t-1},\mean{d}_{t-1}} \in ℝ^{\Ndat\times3}$,
    $η_{t-1} ≥ 0$, $ρ_{t,t} \in \{0,1\}$,
    $γ_t \in [0, 1]$, and
    $m \in \Brace{\Tag{D},\Tag{PP},\Tag{SD},\Tag{FD},\Tag{BD}}$.}

  \KwOut{Updated parameters: $\estim{b}_t$, $\estim{G}_t$, $A_t^{-1}$, $η_t$,
    $\mean{u}_t$, and $\mean{d}_t$.}

  \BlankLine
  \If{$m = \Tag{PP}$ \And $ρ_{t,t} > 0$ \And $u_{t-ℓ} \not= -u_{t-ℓ-1}$}{
    $ρ_{t,t} \gets 0$ \Comment*{Invalid frame for the \emph{push-pull} method.}
  }
  $η_t = ρ_{t,t} + γ_t\,η_{t-1}$ \Comment*{see Eq.~\eqref{eq:eta-recurrence}}
  \uIf{$ρ_{t,t} > 0$}{
    $α_t = ρ_{t,t}/η_t$ \Comment*{see Eq.~\eqref{eq:alpha}}
    $β_t = γ_t\,η_{t-1}/η_t$ \Comment*{$β_t = 1 - α_t$ avoiding rounding errors}
    \uIf{$m = \Tag{D}$}{
      $x_t = u_{t-ℓ} - \mean{u}_{t-1}$ \Comment*{see Eq.~\eqref{eq:x_t}}
      $y_t = d_t - \mean{d}_{t-1}$ \Comment*{see Eq.~\eqref{eq:y_t}}
      $μ_t = ρ_{t,t}\,β_t$ \Comment*{see Eq.~\eqref{eq:mu_t} with $β_t = 1 - α_t$}
    }\Else{
      \uIf{$m = \Tag{PP}$}{
        $x_t = 2\,u_{t-ℓ}$ \Comment*{see Eq.~\eqref{eq:x^PP_(t')}}
      }\uElseIf{$m = \Tag{SD}$}{
        $x_t = u_{t-ℓ} - u_{t-ℓ-1}$ \Comment*{see Eq.~\eqref{eq:x^SD_(t')}}
      }\uElseIf{$m = \Tag{FD}$}{
        $x_t = u_{t-ℓ}$ \Comment*{see Eq.~\eqref{eq:x^FD_(t')}}
      }\ElseIf{$m = \Tag{BD}$}{
        $x_t = -u_{t-ℓ-1}$ \Comment*{see Eq.~\eqref{eq:x^BD_(t')}}
      }
      $y_t = d_t - d_{t-1}$ \Comment*{see Eq.~\eqref{eq:y^dif_(t')}}
      $μ_t = ρ_{t,t}$ \Comment*{see Eq.~\eqref{eq:mu_t}}
    }
    $v_t = A_{t-1}^{-1}\,x_t$ \Comment*{see Eq.~\eqref{eq:v_t}}
    \vspace{0.3ex}
    $θ_t = x_t\T\,v_t$ \Comment*{see Eq.~\eqref{eq:alpha_t}}
    $λ_t = μ_t/\Paren{θ_t\,μ_t + γ_t}$ \Comment*{see Eq.~\eqref{eq:lambda_t}}
    $A_t^{-1} = (A_{t-1}^{-1} - λ_t\,v_t\,v_t\T)/γ_t$ \Comment*{see Eq.~\eqref{eq:A_t^{-1}:update}}
    $e_t = y_t - \estim{G}_{t-1}\,x_t$ \Comment*{see Eq.~\eqref{eq:e_t}}
    $\estim{G}_t = \estim{G}_{t-1} + λ_t\,e_t\,v_t\T$ \Comment*{see Eq.~\eqref{eq:G_t:update}}
    $\mean{d}_t = α_t\,d_t + β_t\,\mean{d}_{t-1}$
    \Comment*{see Eqs.~\eqref{eq:mean(d)_t:update} with $β_t = 1 - α_t$}
    $\mean{u}_t = α_t\,u_{t-ℓ} + β_t\,\mean{u}_{t-1}$
    \Comment*{see Eqs.~\eqref{eq:mean(u)_t:update}  with $β_t = 1 - α_t$}
    $\estim{b}_t = \mean{d}_t - \estim{G}_{t} \, \mean{u}_t$ \Comment*{see Eq.~\eqref{eq:b_t}}
  }\Else{
    $A_t^{-1} = A_{t-1}^{-1}/γ_t$\;
    $\estim{G}_t = \estim{G}_{t-1}$\;
    $\mean{d}_t = \mean{d}_{t-1}$\;
    $\mean{u}_t = \mean{u}_{t-1}$\;
    $\estim{b}_t = \estim{b}_{t-1}$\;
  }
\end{algorithm}


\section{Interaction matrix amplitude and working conditions}
\label{sec:working-conditions}

\begin{figure}
  \centering
  \includegraphics[width=0.9\columnwidth]{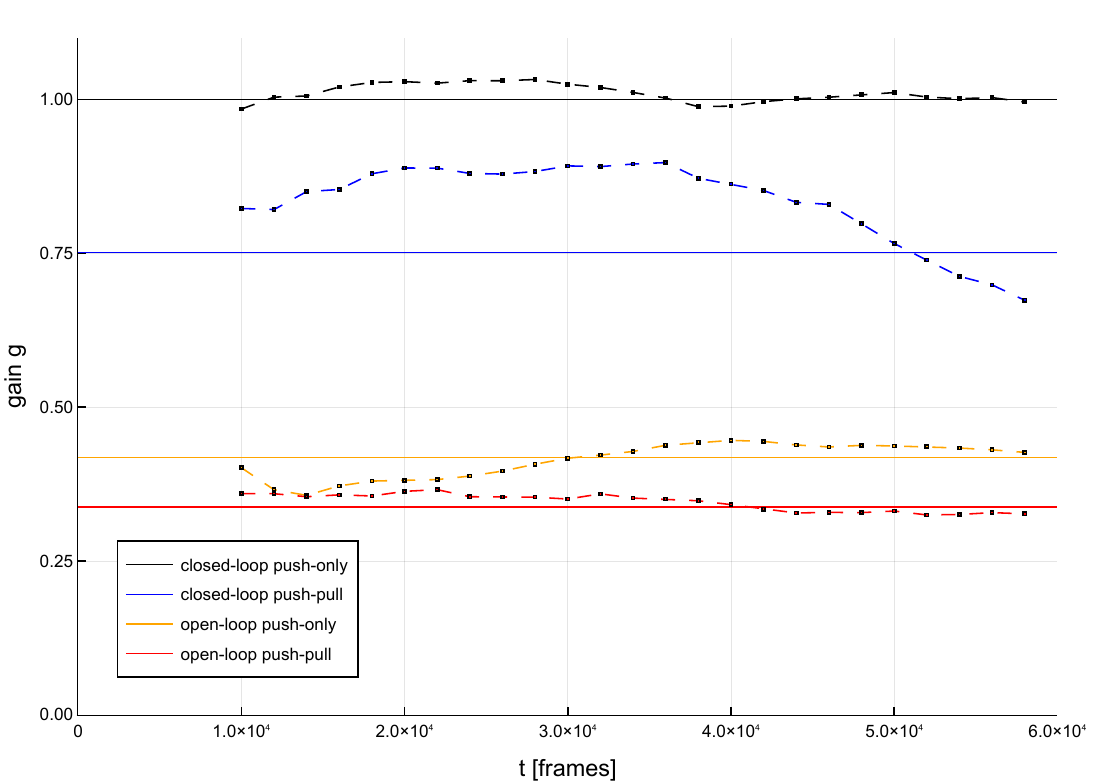}
  \caption{Dashed lines: gain $g$ as a function of $t$ [frames], when using a forgetting
    factor $\gamma_{t}$ defined by Eq.~\eqref{eq:forgetting-factor-gain} and the
    \emph{simple differences} (SD) method on each of the 4 considered sequences (see
    legend). The interaction matrix is constantly updated using
    Algorithm~\ref{alg:update-model} and the gain is assessed every 2\,000 frames.}
  \label{fig:gain-vs-time}
\end{figure}

It has been noted that the amplitude of the interaction matrix coefficients varies
significantly from one telemetry sequence to another, revealing significant changes in the
working conditions. To support this claim, we quantify here these changes through a scalar
gain $g$ representing the average projection of an estimated interaction matrix
$\estim{G}_{t}$ onto another estimated interaction matrix taken as the reference. We
arbitrarily select as a reference for all, the estimated matrix obtained with \emph{simple
  differences} method on the closed-loop \emph{push-pull} sequence, since it has the largest
coefficients amplitudes and the least errors among the closed-loop matrices. This reference
interaction matrix $\estim{G}_{\textrm{ref}}$ is the one considered in the bottom graph of
Fig.~\ref{fig:closed-loop:G:maps-pp}. For an interaction matrix $\estim{G}$ estimated on any
calibration sequence and length, we can compute its projection on the reference matrix in
the least squares sense and determine a scalar gain value $g$, defined as
\begin{equation}
  g = \sum_{i,j} \frac{\estim{G}[i,j] \, \estim{G}_{\textrm{ref}}[i,j]}{(\estim{G}_{\textrm{ref}}[i,j])^{2}}
\end{equation}
The 4 interaction matrices estimated with \emph{simple differences} (SD) methods in
Sec.~\ref{sec:results} --- open-loop on \emph{push-only}, open-loop on \emph{push-pull}, closed-loop on
\emph{push-only} and closed-loop on \emph{push-pull} --- are all obtained with forgetting factors of 1 and
using the full sequences of $60\,000$ frames. Figure~\ref{fig:gain-vs-time} represent the
gain $g$ computed for each of these matrices as an horizontal solid line of the
corresponding color (see the legend).

\newcommand{\ForgettingFactor}{0.999965}

In addition, for each calibration sequence, the estimation of interaction matrix is
repeated, still with \emph{simple differences} method, but with a constant forgetting
factor defined such that the weight of calibration data $20\,000$-frame old is half the
weight of the current data in the weighted least-squares criterion of
Eq.~\eqref{eq:G_t:WLS}, in other words:
\begin{equation}
  \label{eq:forgetting-factor-gain}
  \rho_{t, t - 20\,000} = \gamma_{t}^{20\,000} = 0.5 \quad \forall t\geq20\,000.
\end{equation}
This corresponds to apply $\gamma_{t} \approx \ForgettingFactor$, for all $t$. The gain is
recomputed every $2\,000$ frames, in order to show its evolution with $t$ in the dashed
curves. Firstly, the very different levels of the curves confirm an increased sensitivity of
the system by a factor 2 from open-loop to closed-loop regime. Furthermore, the variations
of $g$ along the sequences of 1-minute telemetry evidences the fast evolution of the working
conditions.


\section{Data Format for the Telemetry}

The software to perform the calibration as proposed in this paper is freely available on
GitHub\footnote{\url{https://github.com/emmt/AOSelfCal.jl}}. The
telemetry data (commands sent to the deformable mirror and wavefront sensor measurements)
are expected to be stored in FITS Binary Tables. This appendix describes these tables.

The telemetry of the deformable mirror is stored in a FITS extension named
\texttt{DM\_TELEMETRY} and whose format is summarized in Table~\ref{tab:dm-telemetry}. The
telemetry of the wavefront sensor is stored in a FITS extension named
\texttt{WFS\_TELEMETRY} and whose format is summarized in Table~\ref{tab:wfs-telemetry}.
These tables may be stored in separate FITS files or in the same FITS file. The columns
data are retrieved by their name; hence, order of columns is irrelevant. As far as no
significant digits are lost, it is acceptable to store column values with a smaller number
of bits, e.g. floating-point values in single precision (format \texttt{E} instead of
\texttt{D}) or integer values as 32-bit signed integers (format \texttt{J} instead of
\texttt{K}).

Each row of Tables~\ref{tab:dm-telemetry} and \ref{tab:wfs-telemetry} stores the telemetry
of a given frame of the device under consideration. The \texttt{MARK} column is a global
index used to synchronize the frames in the telemetry of different devices, it shall be
strictly increasing (although some frames may be missed). Any deformable mirror (DM)
telemetry frame is assumed to result from the commands sent by the controller in response
to a frame of the wavefront sensor (WFS) measurements, two such corresponding frames (the
WFS and the DM ones) must share the same \texttt{MARK}. The \texttt{MARK} is usually
chosen to be the frame index of the wavefront sensor camera (the index $t$ in this paper).

\begin{table}[t]
  \centering
  \resizebox{\columnwidth}{!}{ 
    \setlength{\extrarowheight}{1.5pt}
    \begin{tabular}{lll}
      \hline
      \multicolumn{3}{c}{Header part}\\
      Keyword & Value / Type & Description (notation) \\
      \hline
      \texttt{XTENSION}         & \texttt{'BINTABLE'}      & FITS Binary Table \\
      \texttt{EXTNAME}          & \texttt{'DM\_TELEMETRY'} & Deformable mirror telemetry \\
      \texttt{EXTVER}           & 2                        & Version of this extension \\
      \texttt{NACTS}$^\star$    & Integer                  & Number of controlled degrees of freedom ($\Nact$) \\
      \hline
      \hline
      \multicolumn{3}{c}{Data part (columns)}\\
      Column & Format & Description [units] \\
      \hline
      \texttt{MARK}            & \texttt{K}                & WFS camera frame serial number ($t$) \\
      \texttt{TIME}$^\star$    & \texttt{D}                & Time stamp [s] \\
      \texttt{DM\_CTRL}        & $\Nact$\,\texttt{D}       & Control commands ($c_{t - ℓ}$) \\
      \texttt{DM\_PROBE}       & $\Nact$\,\texttt{D}       & Perturbation commands ($u_{t - ℓ}$) \\
      \texttt{DM\_REF}$^\star$  & $\Nact$\,\texttt{D}      & Reference commands \\
      \texttt{DM\_DEV}$^\star$ & $\Nact$\,\texttt{D}       & Device commands \\
      \hline
    \end{tabular}
  }
  \caption{FITS extension \texttt{DM\_TELEMETRY} to store deformable mirror telemetry.
    Entries marked with $^\star$ are optional. $\Nact$ denotes the number of
    controlled actuators. The \emph{Format} follows the FITS convention for the
    \texttt{TFORM}$j$ keyword (with $j$ the column number) to have the value type
    specified as a capital letter (\texttt{K} for 64-bit signed integer, \texttt{D} for
    64-bit floating point) possibly preceded by a repeat count.}
  \label{tab:dm-telemetry}
\end{table}

\begin{table}
  \centering
  \resizebox{\columnwidth}{!}{ 
    \setlength{\extrarowheight}{1.5pt}
    \begin{tabular}{lll}
      \hline
      \multicolumn{3}{c}{Header part}\\
      Keyword & Value / Type & Description (notation) \\
      \hline
      \texttt{XTENSION}      & \texttt{'BINTABLE'}       & FITS Binary Table \\
      \texttt{EXTNAME}       & \texttt{'WFS\_TELEMETRY'} & Wavefront sensor telemetry \\
      \texttt{EXTVER}        & 2                         & Version of this extension \\
      \texttt{NDATA}$^\star$ & Integer                   & Measurements per frame ($\Ndat$) \\
      \hline
      \hline
      \multicolumn{3}{c}{Data part (columns)}\\
      Column & Format & Description [units] \\
      \hline
      \texttt{MARK}             & \texttt{K}             & WFS camera frame serial number ($t$) \\
      \texttt{TIME}$^\star$     & \texttt{D}             & Time stamp [s] \\
      \texttt{WFS\_DATA}        & $\Ndat$\,\texttt{D}    & Measurements ($d_t$) \\
      \texttt{WFS\_WGT}$^\star$ & $\Nwgt$\,\texttt{D}    & Precision of measurements \\
      \texttt{WFS\_REF}$^\star$ & $\Ndat$\,\texttt{D}    & Reference measurements \\
      \hline
    \end{tabular}
  }
  \caption{FITS extension \texttt{WFS\_TELEMETRY} to store wavefront sensor telemetry.
    Entries marked with $^\star$ are optional. $\Ndat$ denotes the number of WFS
    measurements per frame and $\Nwgt$ denotes the number of associated weights. See
    caption of Table~\ref{tab:dm-telemetry} for more information.}
  \label{tab:wfs-telemetry}
\end{table}

In a \texttt{DM\_TELEMETRY} extension, the requested commands, \Field{dm\_ctrl}, are
relative to the reference and probing commands, \Field{dm\_ref} and \Field{dm\_probe}. The
commands effectively sent to the deformable mirror device are:
\begin{equation}
  \label{eq:dm-commands}
  \Field{dm\_dev}_{\!j,t} =
  \psi_{\!j}(\Field{dm\_ctrl}_{j,t} + \Field{dm\_ref}_{j,t} + \Field{dm\_probe}_{j,t})
\end{equation}
where the mapping $\psi_{j}\from\mathbb{R}\to\mathbb{R}$ accounts for the actual behavior
(clamping, gain, etc.) of the deformable mirror. Index $j \in 1:n_\text{act}$ is the
command vector component number, $t$ is the frame index. In Table~\ref{tab:dm-telemetry},
these data are saved in columns with corresponding names converted to uppercase (due to
FITS conventions). If optional column \texttt{DM\_REF} is missing, \Field{dm\_ref} is
assumed to be zero in all frames. If optional column \texttt{DM\_DEV} is missing,
\Field{dm\_dev} is computed from the other columns assuming $\psi_{\!j}$ is the identity.
Hence, only columns \texttt{DM\_CTRL} and \texttt{DM\_PROBE} are mandatory in a
\texttt{DM\_TELEMETRY} extension.

In a \texttt{WFS\_TELEMETRY} extension, if optional column \texttt{WFS\_REF} is missing,
\Field{wfs\_ref} is assumed to be zero for all measurements in all frames. For a
Shack-Hartmann WFS with $N_\text{sub}$ sub-pupils, a cell of \texttt{WFS\_DATA} or
\texttt{WFS\_REF} contains $2 × N_\text{sub}$ values, a cell of \texttt{WFS\_WGT} contains
$3 × N_\text{sub}$.

\end{document}